\documentclass[a4paper,twocolumn,preprintnumbers,superscriptaddress,accepted=2022-08-03]{quantumarticle}
\pdfoutput=1
\usepackage{amsmath}
\usepackage{amssymb}
\usepackage{graphicx}
\usepackage{hyperref}
\usepackage{physics}
\usepackage{xcolor}
\usepackage{xspace,slashed}
\usepackage{qcircuit}
\usepackage{multirow}
\usepackage[utf8]{inputenc}
\usepackage{subfigure}
\usepackage{amsthm}
\usepackage[numbers,sort&compress]{natbib}

\usepackage[left]{lineno}

\newtheorem{theorem}{Theorem}

\usepackage{array}

\renewcommand{\vec}[1]{\boldsymbol{#1}}

\newcommand{\CERNaff}{Theoretical Physics Department, CERN, CH-1211
  Geneva 23, Switzerland.}

\newcommand{\NYCU}{Institute of Physics, National Yang Ming Chiao Tung University, Hsinchu 30010, Taiwan.}

\newcommand{\tii}{Quantum Research Centre, Technology Innovation Institute, Abu Dhabi, UAE}

\newcommand{\UB}{Departament de F\'isica Qu\`antica i Astrof\'isica and Institut de Ci\`encies del Cosmos (ICCUB), Universitat de Barcelona, Barcelona, Spain.}

\newcommand{\MIaff}{TIF Lab, Dipartimento di Fisica, Universit\`a degli Studi di
  Milano and INFN Sezione di Milano, Milan, Italy.}

\begin{document}

\title{Style-based quantum generative adversarial networks for Monte Carlo events}

\author{Carlos Bravo-Prieto}
\affiliation{\tii}
\affiliation{\UB}

\author{Julien Baglio}
\affiliation{\CERNaff}

\author{Marco C\`e}
\affiliation{\CERNaff}

\author{Anthony Francis}
\affiliation{\NYCU}
\affiliation{\CERNaff}

\author{Dorota M. Grabowska}
\affiliation{\CERNaff}

\author{Stefano Carrazza}
\affiliation{\MIaff}
\affiliation{\CERNaff}
\affiliation{\tii}
\maketitle

\begin{abstract}
We propose and assess an alternative quantum generator architecture in the
context of generative adversarial learning for Monte Carlo event generation,
used to simulate particle physics processes at the Large Hadron Collider (LHC).
We validate this methodology by implementing the quantum network on artificial
data generated from known underlying distributions. The network is then applied
to Monte Carlo-generated datasets of specific LHC scattering processes. The new
quantum generator architecture leads to a generalization of the state-of-the-art implementations, achieving smaller
Kullback-Leibler divergences even with shallow-depth networks. Moreover, the quantum
generator successfully learns the underlying distribution functions even if
trained with small training sample sets; this is particularly interesting for
data augmentation applications. We deploy this novel methodology on two
different quantum hardware architectures, trapped-ion and superconducting
technologies, to test its hardware-independent viability.

\end{abstract}

\section{Introduction}

Quantum computing is a new paradigm whereby quantum phenomena are harnessed to perform computations. The current availability of noisy intermediate-scale quantum (NISQ) computers ~\cite{nisq}, and recent advances towards quantum computational supremacy~\cite{supremacy, zhong2020quantum}, have led to a growing interest in these devices to perform computational tasks faster than classical machines. Among many of the near-term applications~\cite{cerezo2021variational, bharti2021noisy}, the field of Quantum Machine Learning (QML)~\cite{biamonte2017quantum, schuld2018supervised} is held as one promising approach to make use of NISQ computers.

Early work in QML was mostly focused on speeding up linear algebra subroutines~\cite{wiebe2012quantum, lloyd:2013ml, Rebentrost:2014svm, kerenidis2020quantum}, widely used in classical machine learning, by leveraging the Harrow-Hassidim-Lloyd
algorithm~\cite{harrow2009quantum}. This approach is promising, though its utility depends on the existence of large-scale quantum computers with low gate errors and enough qubits to perform quantum error correction. More recent proposals focus on defining a quantum neural network (QNN), or parameterized quantum circuit~\cite{benedetti2019parameterized, sim2019expressibility, bravo2020scaling, larocca2021theory}, which then can be trained to implement a function class~\cite{schuld2021effect, goto2021universal, perez2021one}; these proposals can be implemented on current NISQ-era devices. For example, several QNNs have been proposed for pattern classification~\cite{havlivcek2019supervised, Schuld:2020circuit, perezsalinas:2020reuploading, dutta2021realization} or data compression~\cite{romero2017quantum, pepper2019experimental, bravo2021quantum, cao2021noise}. This QML approach to quantum computing is a research topic that can be adapted, improved, and tested on many research problems in disparate scientific fields. Motivated by this idea, we propose to investigate the possibility of using QNNs for generative modeling~\cite{benedetti2019generative, hamilton2019generative, coyle2020born}. More specifically, we explore the uses of QNNs for the generation of Monte Carlo events through quantum generative adversarial networks (qGANs)~\cite{dallaire2018quantum, lloyd2018quantum}.

The generative adversarial framework employs two competing networks, the generator and the discriminator, that are
trained alternatingly~\cite{goodfellow2014generative}. The generator produces candidates
while the discriminator evaluates them. The objective of the discriminator is to
distinguish the real samples from the generated ones. That is, the discriminator
plays the role of the generator's adversary, and therefore, their competition is
a zero-sum two-player game. By substituting either the discriminator, the
generator, or both with quantum systems, the scheme can be translated to quantum
computing~\cite{dallaire2018quantum}.

In recent months, the spreading interest in QML has led to different qGAN
implementations~\cite{zoufal2019quantum, zeng2019learning, situ2020quantum, hu2019quantum, benedetti2019adversarial, romero2021variational, niu2021entangling}. Our contribution here can be summarized in three distinct aspects. (1) Previous proposals employed toy data for their qGAN training. In contrast, we test our model using data for a quantum scattering process. In particular, we first train and validate our qGAN model with artificial data from known underlying probability density functions. Then, in order to test our model in a realistic set-up, we use as training sets simulated Monte Carlo events for particle physics processes at the Large Hadron Collider (LHC) at CERN. (2) We propose an alternative quantum generator architecture. Traditionally, the prior noise distribution, or latent variables in the language of generative models, is provided to the quantum generator through its first quantum gates. We instead repeatedly encode the latent variables over all the quantum network. This allows us to obtain a generalized version of the state-of-the-art qGAN implementations; when working with a realistic dataset, we achieve significantly smaller Kullback-Leibler (KL) divergences. A similar concept was introduced in the classical context~\cite{karras2019style}, coined as style-based generative adversarial network (GAN), which was proven to be useful in facial recognition tasks. In the classical approach, however, a preprocessing of the latent variables is done. Although our approach do not incorporate this preprocessing step of the latent variables, given the analogy of repeatedly encoding them into the network, from now on we refer to our qGAN model as style-qGAN. (3) We validate and assess our style-qGAN in quantum hardware. Specifically, we successfully implement our model in two different quantum architectures, namely ion traps and superconducting qubits.

\begin{figure}[t!]
\centering
  \includegraphics[width=0.45\textwidth]{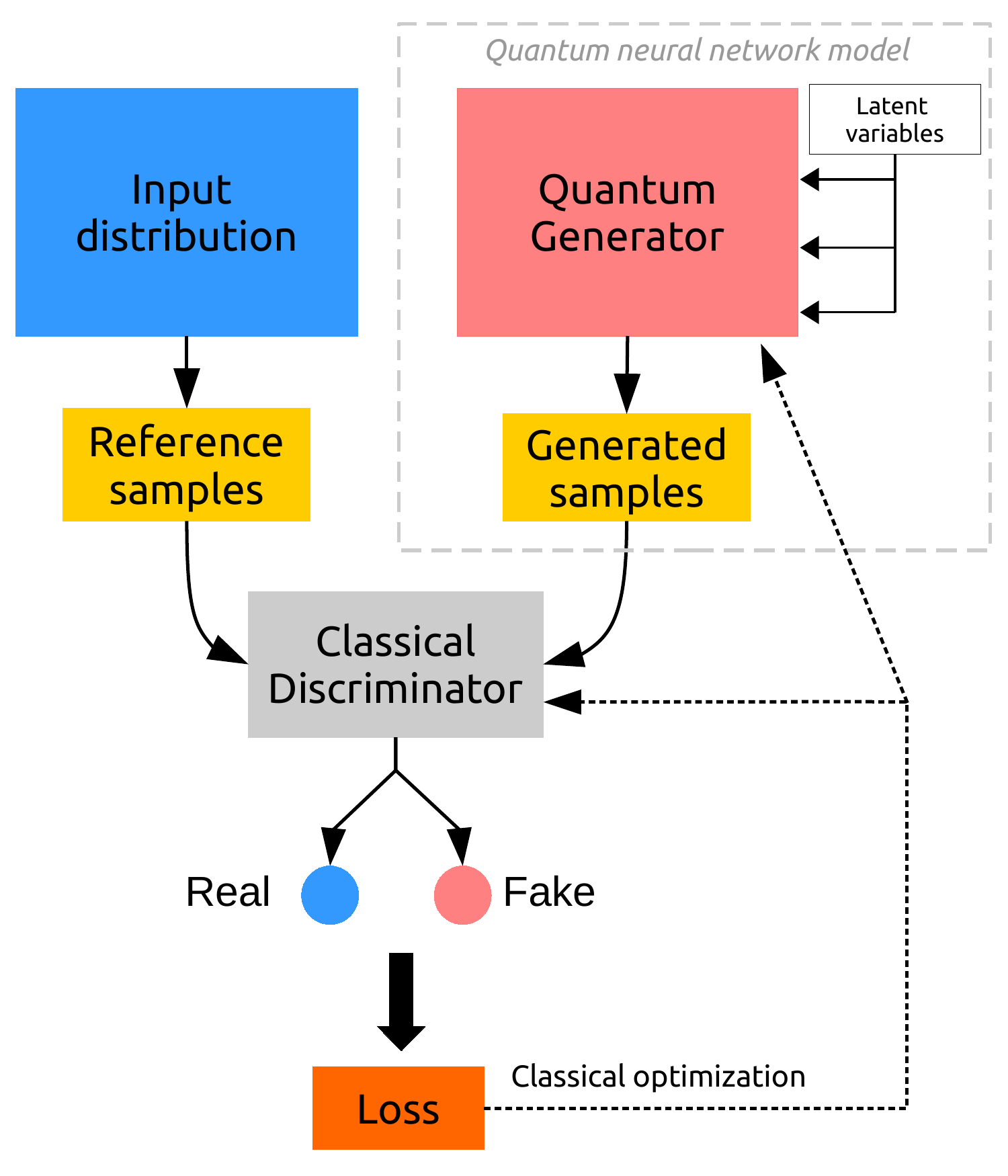}
  \caption{\label{fig:scheme} Schematic steps involved in the style-qGAN training. Reference samples from a known input distribution are initially prepared. Simultaneously, we inject the latent variables over the whole quantum generator to produce the fake generated samples. Then, both set of samples are used to train the classical discriminator. The quality of the training is measured by an appropriate loss function, which is then also used to update the quantum generator and the classical discriminator. This procedure is repeated following an adversarial approach.}
\end{figure}

It is important to highlight that several research groups from the high-energy
physics (HEP) community are investigating potential applications of quantum
technologies in HEP applications and obtaining interesting
results~\cite{P_rez_Salinas_2021,Guan_2021,chang2021quantum,Chang_2021,Belis_2021,khattak2021fast}.
The study presented in this manuscript should be considered both as an improvement over the state-of-the-art from the algorithmic point of view, and as a
proof-of-concept for a future quantum hardware deployment, providing a robust
and reproducible starting point for future investigations. In particular, the
introduction of GAN models in HEP Monte Carlo simulation has been discussed
extensively in the last years, see
Refs.~\cite{baldi2021gan,Backes_2021,butter2020generative,Butter_2021,Butter_2020,Bellagente_2020,Butter_2019}.
In this work, we consider the possibility to use a qGAN model in a data
augmentation context, where the model is trained with a small amount of input
samples and it learns how to sample the underlying distribution.

The paper is structured as follows. In Sec.~\ref{sec:implementation} we define
our style-qGAN model. In Sec.~\ref{sec:validation} we validate the style-qGAN
model using toy data. Then, in Sec.~\ref{sec:lhc} we train the style-qGAN
generator on simulated LHC events. Finally, in Sec.~\ref{sec:deployment} we test
our final model on real quantum hardware, in Sec.~\ref{sec:comparison} we
compare our model to previous qGAN architectures, and in
Sec.~\ref{sec:conclusion} we present our conclusion and future development
directions. In addition, we include in Appendix~\ref{sec:appendixnoise} a noise
simulation of the algorithm using the simplified noise model provided by IBM~Q
for the target machine used in Sec.~\ref{sec:deployment}.

\section{Implementation}
\label{sec:implementation}

\subsection{Workflow design}

The classical implementation of a GAN model~\cite{goodfellow2014generative}
involves at least three components: the discriminator model, the
generator model, and the adversarial training procedure. In this work we
consider a hybrid quantum-classical system, where the generator model has a
quantum representation through a QNN while the
discriminator is a classical neural network model. This choice is motivated by
the practical positive implications of a hardware-based generative model, in
particular the possibility to obtain performance improvements in a real quantum
device. The idea of using a quantum device for the generation of
samples is very attractive because the complicated aspects of density modeling and sampling are delegated to a hardware architecture.

There are alternative approaches where both models could be represented by a
QNN~\cite{dallaire2018quantum, hu2019quantum, benedetti2019adversarial, romero2021variational, niu2021entangling}. However, after testing some prototype architectures, we have observed faster convergence when using a classical discriminator.

In Figure~\ref{fig:scheme} we schematically show the steps involved in the style-qGAN
presented here. The procedure starts from the preparation of reference samples from a
known distribution function that we would like to encode in the quantum
generator model. At the same time, we define a QNN model where
we inject stochastic noise in all quantum gates of the network. The generator model is then used to extract fake generated samples
that, after the training procedure, should match the quality of the known input
distribution. Lastly, both sets of samples are used to train the discriminator
model. The quality of the training is measured by an appropriate loss function
which is monitored and optimized classically by a minimization algorithm based on
the adversarial approach. The training process consists of simultaneous
stochastic gradient descent for both models which, after reaching convergence,
delivers a quantum generator model with realistic sampling.

In the following paragraphs, we first introduce the optimization procedure and the quantum
generator network, and validate the procedure by using reference samples from known
distribution functions to train the model on a quantum simulator. We then
train our style-qGAN model with Monte Carlo-generated LHC events using again a quantum
simulator. Finally, the best-trained model is deployed on real quantum hardware
devices based on superconducting and trapped-ion technologies.

All calculations involving quantum circuit simulation are performed using
{\tt Qibo} v0.1.6~\cite{efthymiou2020qibo,stavros_efthymiou_2021_5088103} on
classical hardware. For this particular project, we have used the {\tt
tensorflow}~\cite{tensorflow2015-whitepaper} backend which provides the
possibility to use gradient descent optimizers during the training step. The
style-qGAN model is publicly available through the {\tt Qibo}
framework and the code repository in~\cite{afrancis_heplat_2021_5567077}.

\subsection{Optimization procedure}

Our style-qGAN comprises of a QNN for the generator $G(\vec{\phi_g}, \vec{z})$ and a classical network for the discriminator $D(\vec{\phi_d}, \vec{x})$, where $\vec{\phi_g}$ and $\vec{\phi_d}$ are the parameters of the corresponding networks. The quantum generator transforms samples from a prior standard Gaussian noise distribution $\vec{z} \sim p_{\mathrm{prior}}(\vec{z})$, also called latent vector, into samples generated by $G(\vec{\phi_g})$, thus mapping $p_{\mathrm{prior}}(\vec{z})$ to a different distribution $p_{\mathrm{fake}}$ of generated data. On the other hand, the discriminator takes as input samples $\vec{x}$ and tries to distinguish between fake data from the generator and real data from the reference input distribution $p_{\mathrm{real}}$. The training corresponds to an adversarial game, where we alternate between improving the discriminator to distinguish fake and real data, and the generator to cheat the discriminator with new fake data.

In this work, we consider the binary cross-entropy for the optimization objective. The generator's loss function can be defined as
\begin{equation}
   \mathcal{L}_G(\vec{\phi_g},\vec{\phi_d}) = -\mathbb{E}_{\vec{z} \sim p_{\mathrm{prior}}(\vec{z})}[\log D(\vec{\phi_d},G(\vec{\phi_g},\vec{z}))]  \,,
\end{equation}
while the discriminator's loss function can be defined as
\begin{equation}
\begin{split}
   \mathcal{L}_D(\vec{\phi_g},\vec{\phi_d}) = \mathbb{E}_{\vec{x} \sim p_{\mathrm{real}}(\vec{x})}[\log D(\vec{\phi_d},\vec{x})] \\+\, \mathbb{E}_{\vec{z} \sim p_{\mathrm{prior}}(\vec{z})}[\log (1-D(\vec{\phi_d},G(\vec{\phi_g},\vec{z})))]\,.
\end{split}
\end{equation}
Notice that the adversarial training corresponds to a minimax two-player game,
\begin{equation}
 \underset{\vec{\phi_g}}{\min}\,\,\mathcal{L}_G(\vec{\phi_g},\vec{\phi_d})  \,,
\end{equation}
\begin{equation}
 \underset{\vec{\phi_d}}{\max}\,\,\mathcal{L}_D(\vec{\phi_g},\vec{\phi_d})  \,,
\end{equation}
where the optimum uniquely corresponds to the Nash equilibrium between the loss functions.

The optimization of the parameters $\vec{\phi_g}$ and $\vec{\phi_d}$ is done alternatingly by updating the quantum generator and classical discriminator. The optimizer used to update the steps in this work is the ADADELTA~\cite{zeiler2012adadelta}, which is a stochastic gradient descent method that monotonically decreases its learning rate. The starting learning rates utilized are $0.1$ for the classical discriminator and $0.5$ for the quantum generator.

\subsection{Quantum generator architecture}
\begin{figure}[t!]
  \includegraphics[width=1.0\columnwidth]{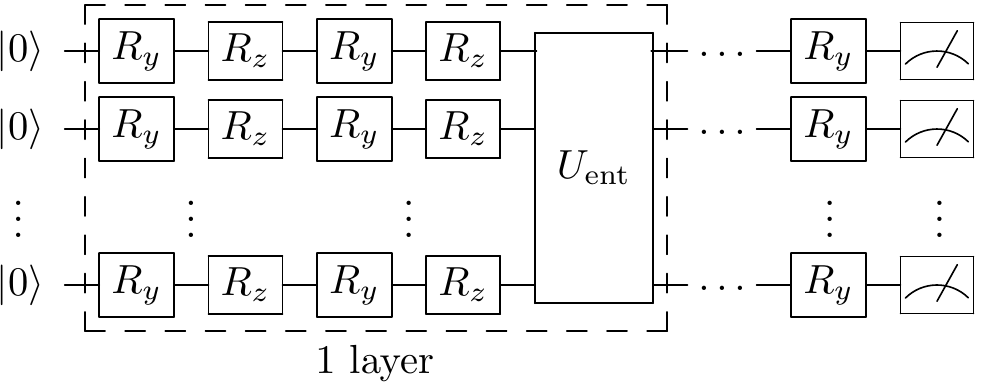}
  \caption{\label{fig:circuit}Quantum neural network employed for the qGAN model. As indicated by the dashed box, each layer is composed of a set of entangling gates $U_{\rm ent}$, to be specified for each example, preceded by two alternating $R_y$ and $R_z$ single-qubit rotations. After implementing the layered network, a final layer of $R_y$ gates is applied.}
\end{figure}

The goal of the quantum generator is to implement a quantum feature map to encode the latent variables and then produce the fake samples. Let $\mathcal{H}$ be a Hilbert space. The quantum feature map $\Psi_{\phi_g} : \vec{z} \rightarrow \mathcal{H}$ encodes the latent variables into a quantum state $\left|\Psi_{\phi_g}(\vec{z})\right\rangle$. This action is equivalent to applying a quantum circuit $\mathcal{U}_{\phi_g}(\vec{z})$ to the initial state $\left|0\right\rangle^n$, where $n$ is the number of qubits. In particular, we employ the architecture shown in Figure~\ref{fig:circuit}. We consider a layered QNN, where each layer is composed of a set of entangling gates $U_{\rm ent}$ preceded by two alternating $R_y$ and $R_z$ single-qubit rotations. After implementing the layered network, a final layer of $R_y$ gates is applied. Note that $U_{\rm ent}$ is specific to each example and $R_j(\theta_k) = e^{-i\theta_k \sigma_j /2}$, where $\sigma_j$ are the Pauli operators. The number of layers can be modified to tune the capacity of the quantum generator.

The action of the quantum feature map is dictated by each qubit rotation. That is, each gate is parameterized by the set of trainable parameters $\vec{\phi_g}$ and, most importantly, the latent vector $\vec{z}$. We encode them by using a linear function as
\begin{equation}
    \label{eq:rotation} R^{l,m}_{y,z}\left(\vec{\phi_g}, \vec{z}\right) = R_{y,z}\left(\phi_g^{(l)} z^{(m)} + \phi_g^{(l-1)}\right)\,,
\end{equation}
where $l=\{2,4,6,\hdots\}$ takes different even values on each rotation up to the total number of trainable parameters, and $m=\{1,2,\hdots,D_{\mathrm{latent}}\}$, being $D_{\mathrm{latent}}$ the latent dimension to be specified for each implementation. The encoding in Eq.~\ref{eq:rotation} is analogous to that used in classical neural networks. Namely, $\phi^{(l)}_g$ and $\phi^{(l-1)}_g$ play the role
of the weights and biases, respectively, while the rotation gate plays the role of the non-linear activation function.

The novelty of the quantum generator architecture used for the style-qGAN model is the embedding of the latent variables into every quantum gate of the network, in contrast to previous qGAN proposals which introduced them only once at the beginning of the network. Our style-based generator contains sequences of layers with latent variable encoding and trainable parameters, much in the spirit of recent developments in the QML field. For instance, for univariate functions $f(z)$, Ref.~\cite{schuld2021effect} shows that large enough QNNs with sequences of quantum gates with data encoding and trainable parameters are universal function approximators. In contrast, encoding data only once restricts the function class that can be learned~\cite{ostaszewski2021structure}. Although much less is known for multivariate functions $f(\vec{z})$, Ref.~\cite{goto2021universal} shows that two particular schemes with repetition blocks of data encoding are also universal function approximators. Hence, the style-based approach is the intuitive step for qGAN implementations toward quantum feature maps that may be classically intractable and, as we numerically explore in Sections~\ref{sec:validation} and~\ref{sec:lhc}, also reliable function approximators even with shallow QNNs.

In the following, let us show that the style-based quantum generator generalises the standard quantum generator approach.

\begin{theorem}
\label{theorem:style}
 Let be a standard quantum generator $\mathcal{U}_{\phi^\ast_g}(\vec{z})$ as a unitary operation where the latent variables $\vec{z}$ are encoded once in the circuit, and $\mathcal{S}^\ast_{\Psi} = \{\mathcal{U}_{\phi^\ast_g}(\vec{z})\}$ be its set of achievable quantum feature maps for any set of parameters $\vec{\phi^\ast_g}$. Equivalently, let $\mathcal{S}_{\Psi} = \{\mathcal{U}_{\phi_g}(\vec{z})\}$ be the set of achievable quantum feature maps by a style-based quantum generator for any set of parameters $\vec{\phi_g}$. Then, the following relation holds
 \[\mathcal{S}^\ast_{\Psi} \subset \mathcal{S}_{\Psi}\,.\]
 Moreover, for any set of parameters $\vec{\phi^\ast_g}$ there exists a set of parameters $\vec{\phi_g}$ such that
\[\mathcal{S}^\ast_{\Psi} = \mathcal{S}_{\Psi}\,.\]
\end{theorem}
\begin{proof}
Assume any standard quantum generator $\mathcal{U}_{\phi^\ast_g}(\vec{z})$ where the latent variables $\vec{z}$ are encoded once in the circuit. Then, $\mathcal{U}_{\phi^\ast_g}(\vec{z})$ can be split as the action of an encoder of the latent variables $E_{\phi^{\ast1}_g}(\vec{z})$ and a variational circuit $V_{\phi^{\ast2}_g}$ as $\mathcal{U}_{\phi^\ast_g}(\vec{z}) = V_{\phi^{\ast2}_g}\,\,E_{\phi^{\ast1}_g}(\vec{z})$, where $\vec{\phi^{\ast}_g} = (\vec{\phi^{\ast1}_g}, \vec{\phi^{\ast2}_g})$. In contrast, the style-based approach utilises the same circuit architecture $\mathcal{U}_{\phi_g}(\vec{z})$ but repeatedly performs a linear encoding  as in Eq.~\ref{eq:rotation}. Equivalently, the action of the style-based quantum generator can be seen as  $\mathcal{U}_{\phi_g}(\vec{z}) = V_{\phi^2_g}(\vec{z})\,\,E_{\phi^1_g}(\vec{z})$, where $\vec{\phi_g} = (\vec{\phi^1_g}, \vec{\phi^2_g})$.

By a proper tuning of the parameters $\vec{\phi^2_g}$ in $V_{\phi^2_g}(\vec{z})$, that is, by setting $\phi_g^{(l)}$ to zero and $\phi_g^{(l-1)}$ equal to each component of $\vec{\phi^{\ast2}_g}$ in Eq.~\ref{eq:rotation}, we obtain $V_{\phi^2_g}(\vec{z}) = V_{\phi^{\ast2}_g}$. Similarly, by setting $\vec{\phi^1_g} = \vec{\phi^{\ast1}_g}$, we obtain $E_{\phi^1_g}(\vec{z}) = E_{\phi^{\ast1}_g}(\vec{z})$. Hence, we recover the standard quantum generator from the style-based approach, namely, $\mathcal{U}_{\phi^\ast_g}(\vec{z}) = \mathcal{U}_{\phi_g}(\vec{z})$ and therefore $\mathcal{S}^\ast_{\Psi} = \mathcal{S}_{\Psi}$. Given the extra degrees of freedom in the style-based approach, it also holds that $\mathcal{S}^\ast_{\Psi} \subset \mathcal{S}_{\Psi}$ for any set of parameters.
\end{proof}
From the previous result, the following observation can be made. The training of a style-qGAN might be initialized with the optimal parameters of a pretrained qGAN model. This way, the style-qGAN improves the performance of the latter, given the extra degrees of freedom, yet using the same quantum circuit. This is particularly interesting for the NISQ era, where quantum resources are limited. In the worst case scenario, the style-qGAN remains with the same optimal parameters, and therefore, with an equal performance.

Recall that the quantum generator's task is creating fake samples to fool the classical discriminator. The fake samples are prepared by acting with the parameterized QNN on the initial $n$-qubit state $\ket{0}^{\otimes n}$, and then measuring in the computational basis. For our implementations, each qubit delivers one sample component. That is, the sample $\vec{x} \in \mathbb{R}^n$ is generated as
\begin{equation}
    \label{eq:samples} \vec{x} = \left(-\left\langle\sigma_z^1\right\rangle,-\left\langle\sigma_z^2\right\rangle,\hdots,-\left\langle\sigma_z^n\right\rangle\right)\,,
\end{equation}
with
\begin{equation}
    \label{eq:expectation}\left\langle\sigma_z^i\right\rangle = \left\langle\Psi_{\phi_g}(\vec{z})\left|\,\sigma_z^i\,\right|\Psi_{\phi_g}(\vec{z})\right\rangle \,,
\end{equation}
where $\left|\Psi_{\phi_g}(\vec{z})\right\rangle$ is the output state from the quantum generator. Notice, however, that for other models, more sophisticated ways of generating fake samples could be more convenient to implement. For instance, one could generate a sample component by computing expectation values involving several qubits. Finally, let us briefly comment that we used a deep convolutional neural network for the discriminator. More details about the classical discriminator implementation can be found in the code~\cite{afrancis_heplat_2021_5567077}.

\section{Validation examples}
\label{sec:validation}

In this section, we show examples of style-qGAN models obtained for known prior
distribution functions in one and three dimensions. The results presented here
have been obtained after a systematic process of fine-tuning and manual
hyper-optimization of the training and quantum generator model.

\subsection{1D Gamma distribution}
\label{sec:gamma}

\begin{figure}
\centering
  \includegraphics[width=0.49\textwidth]{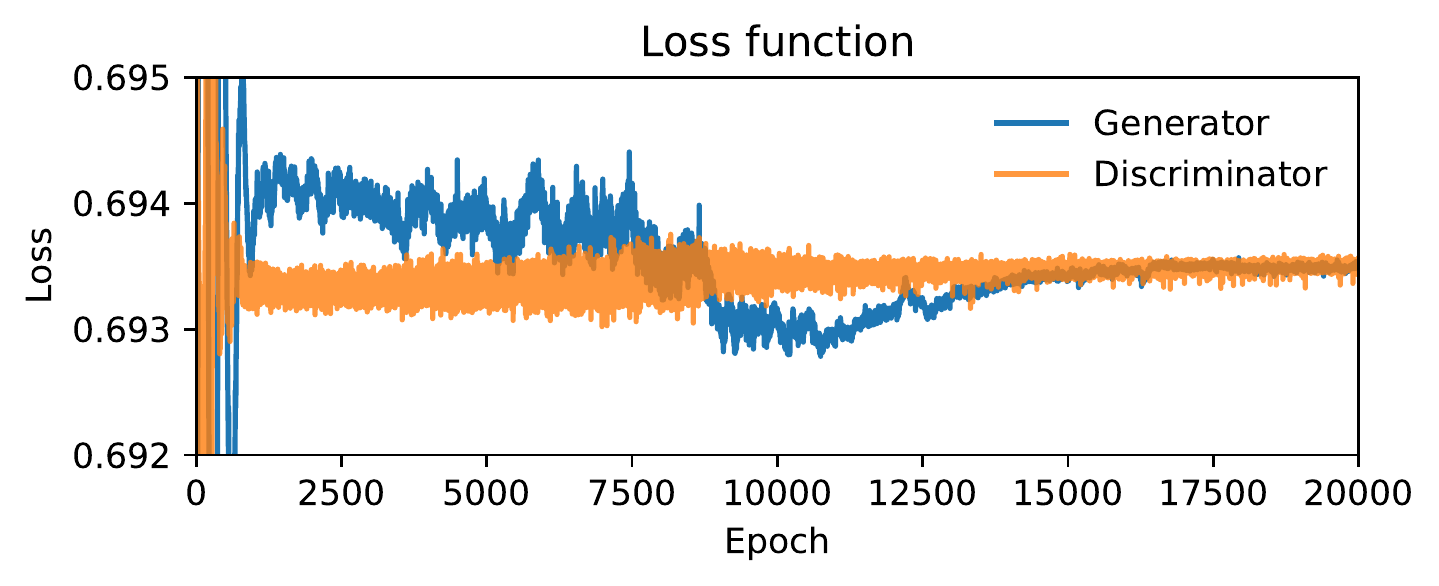}
  \caption{\label{fig:loss}Example of loss function convergence. After an initial warm-up phase, the loss function of both models converges. This indicates that the style-qGAN has been successfully trained.}
\end{figure}

In order to test the framework proposed above, we
consider the sampling of a 1D gamma distribution with probability density
function given by
\begin{equation}
  p_\gamma (x, \alpha, \beta) = x^{\alpha-1} \frac{e^{-x/\beta}}{\beta^\alpha \Gamma(\alpha)}\,,
\end{equation}
where $\Gamma$ is the Gamma function. In this example we take $p_\gamma (x, 1,
1)$ as the input distribution and train a style-qGAN with 1 qubit, 1 latent dimension and
1 layer using $10^4$ samples from the input distribution. The total number of trainable parameters is 10. We perform a linear pre-processing of the data to fit the samples within $x \in [-1, 1]$. We undo this transformation after the training. Note that this single-qubit validation example is useful to test the quality of the style-qGAN model, as it has been shown that a single qubit is enough to approximate univariate functions~\cite{perez2021one}.
In Figure~\ref{fig:loss} we show the evolution of the loss function for the
generator and discriminator models in terms of the number of epochs. We observe the
typical behavior of GAN training and a convergence region after 15000 epochs.
The style-qGAN is trained with batch sizes of 128 samples.

A necessary property of this framework is that the style-qGAN model learns the underlying distribution from a small data set. To demonstrate this, we train a style-qGAN model with a set number of reference samples and then use it to generate two sample sets of different size. In particular, we train the style-qGAN with $10^4$ reference samples and then use it to generate sets of $10^4$ and $10^5$ samples.

\begin{figure}[t!]
  \includegraphics[width=0.45\textwidth]{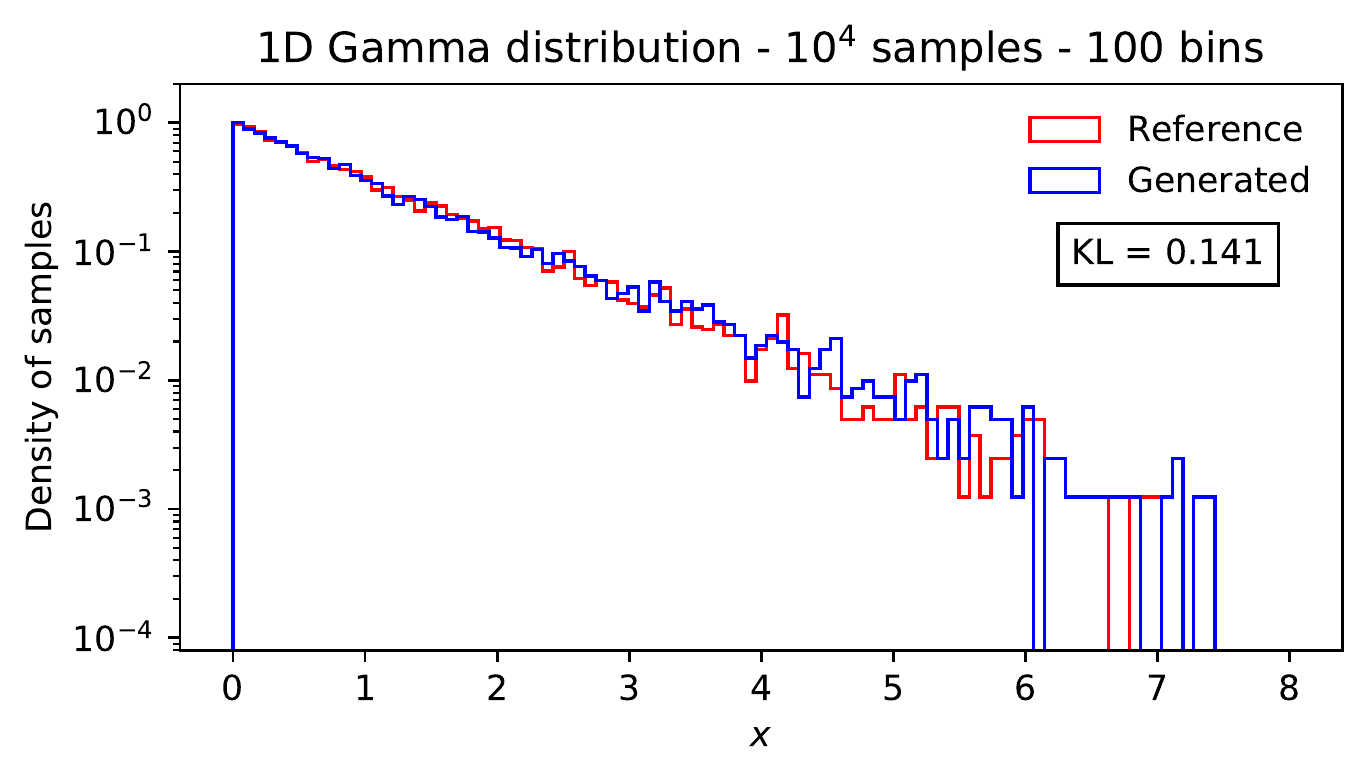}
  \includegraphics[width=0.45\textwidth]{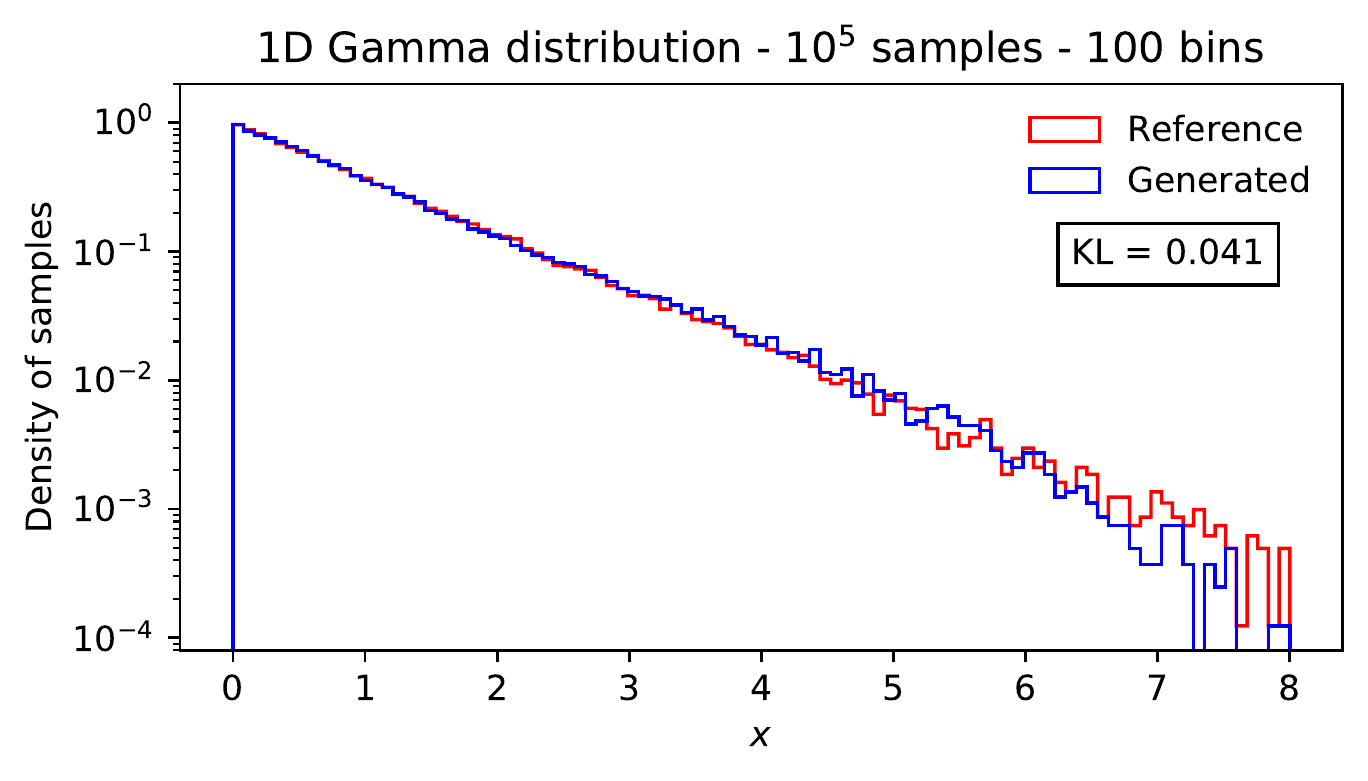}
    \includegraphics[width=0.45\textwidth]{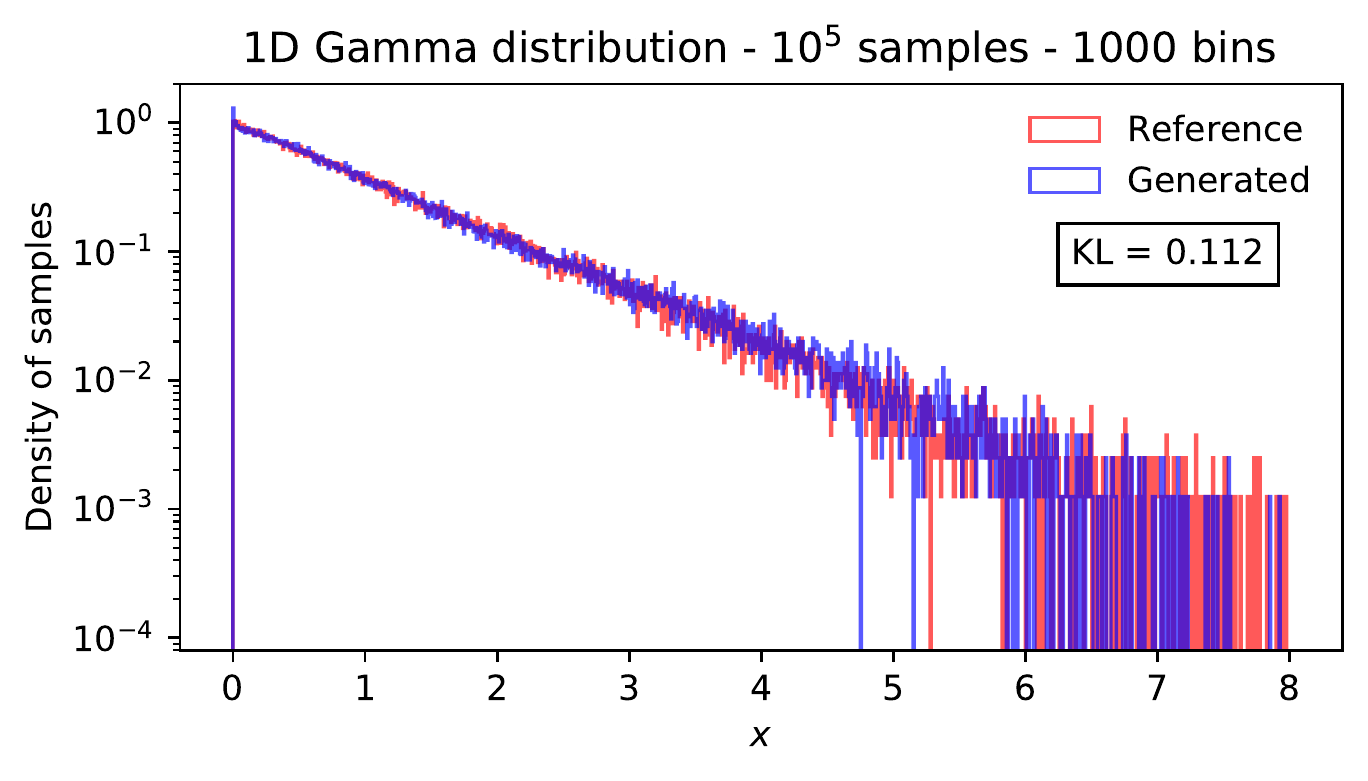}
  \caption{\label{fig:gamma} Examples of 1D gamma distribution sampling for the
  reference underlying distribution (red) and a style-qGAN model (blue) that has been
  trained with $10^4$ reference samples. The top plot compares $10^4$ generated samples using 100 bins. The middle plot compares $10^5$ generated samples using 100 bins.
Finally, the bottom plot compares $10^5$ samples using 1000 bins, which enables a direct comparison of the KL divergences.
  We observe a good level of agreement between both distributions with low values of the KL distance, despite the model being trained on a small training set in addition to an almost constant KL in the data augmentation regime.}
\end{figure}

The top of Figure~\ref{fig:gamma} shows the smaller sample distribution generated by the style-qGAN model in blue and a sampling of the reference distribution of the same size in red. This enables a comparison also using the Kullback-Leibler divergence (KL)~\cite{kullback1951information}. In both cases, the $10^4$ samples have been transformed into histograms with 100 bins linearly spaced on the $x$-axis of the figure. We observe that the distributions are statistically similar even for this high-density binning choice. The KL divergence of two displayed distributions is 0.141, which entails a high degree of similarity.

Going further in the middle of Figure~\ref{fig:gamma} we show the same results as in the top on the larger set containing $10^5$ samples
. Again, we use for comparison a re-sampling of the reference distribution at the same size as the generated set and show both distributions on a grid with 100 linearly spaced bins.
Having generated an order of magnitude more samples than the training set we observe that the style-qGAN model performs well. Both distributions are visually very close to each other and the KL divergence of 0.041 signals a high degree of similarity.

In order to compare the two KL divergences, note that they are computed on discrete histograms. Therefore, for a direct comparison, the number of bins for the larger sample set has to be increased proportionally to the increase in generated sample size, i.e. to compare with 100 bins for the $10^4$ sample size we need to set 1000 bins for the $10^5$ sample size. In this case, for a style-qGAN that provides an equally good description of the underlying distribution function, the KL divergence will stay constant or decrease. Here, with this change in binning, we find the KL divergence is 0.112. The results are shown in the bottom panel of Figure~\ref{fig:gamma}.
This behavior confirms that the style-qGAN model is able to learn the underlying
distribution function even if trained with a small training sample set. Such a
feature is particularly interesting in the context of data augmentation
applications~\cite{frid2018synthetic,tanaka2019data}, where few samples are
available, nonetheless the style-qGAN model can generalize and learn the underlying distribution
with satisfactory outcome.

\subsection{3D correlated Gaussian distribution}

The previous test shows that a style-qGAN model implemented on a single qubit can be
trained and produce acceptable results. However, this particular set-up does not
include entanglement between qubits. In order to study the impact of the
entanglement term $U_{\rm ent}$ in the considered QNN, we select as an
underlying distribution a 3D correlated Gaussian distribution centered at
$\overline{\mu}=(0,0,0)$ with covariance matrix
\begin{equation}
\label{eq:covmat}
  \sigma =
\begin{pmatrix}
  0.5 & 0.1 & 0.25\\
  0.1 & 0.5 & 0.1\\
  0.25 & 0.1 & 0.5\\
  \end{pmatrix}.
\end{equation}

For this specific set-up, we consider a 3-qubit model with 3 latent dimensions and
1 layer. The $U_{\rm ent}$ consists of two controlled-$R_{y}$ gates acting sequentially on the 3 qubits. The total number of trainable parameters is 34. As in the previous example, we perform a linear pre-processing of the data to fit the samples within $x \in [-1, 1]$, and then we undo this transformation after the training. In Table~\ref{table:summary}
we summarize the style-qGAN configurations obtained for both examples discussed in
this section.

\begin{table}[h]
\centering
  \begin{tabular}{l|c|c}
     & {\bf 1D gamma} & {\bf 3D Gaussian} \tabularnewline
    \hline
    Qubits & 1 & 3 \tabularnewline
    $D_{\mathrm{latent}}$ & 1 & 3 \tabularnewline
    Layers & 1 & 1 \tabularnewline
    Epochs & $3\times10^4$ & $1.3\times10^4$ \tabularnewline
    Training set & $10^4$ & $10^4$ \tabularnewline
    Batch size & 128 & 128 \tabularnewline
    Parameters & 10 & 34 \tabularnewline
    $U_{\rm ent}$ & None & 2 sequential C$R_y$ \tabularnewline
    \hline
  \end{tabular}

  \caption{\label{table:summary} Summary of the style-qGAN set-up for the 1D
  gamma distribution and the 3D correlated Gaussian distribution.}
\end{table}

\begin{figure*}[t!]

  \hspace{2.0em}\includegraphics[width=0.29\textwidth]{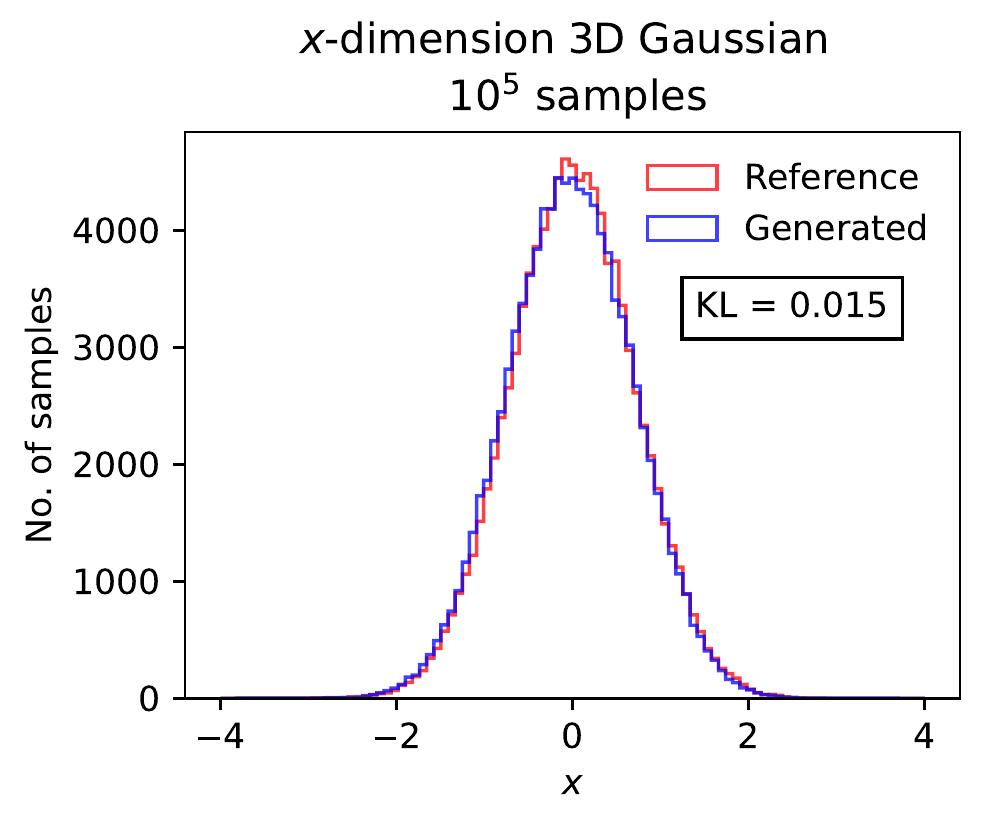}%
  \includegraphics[width=0.29\textwidth]{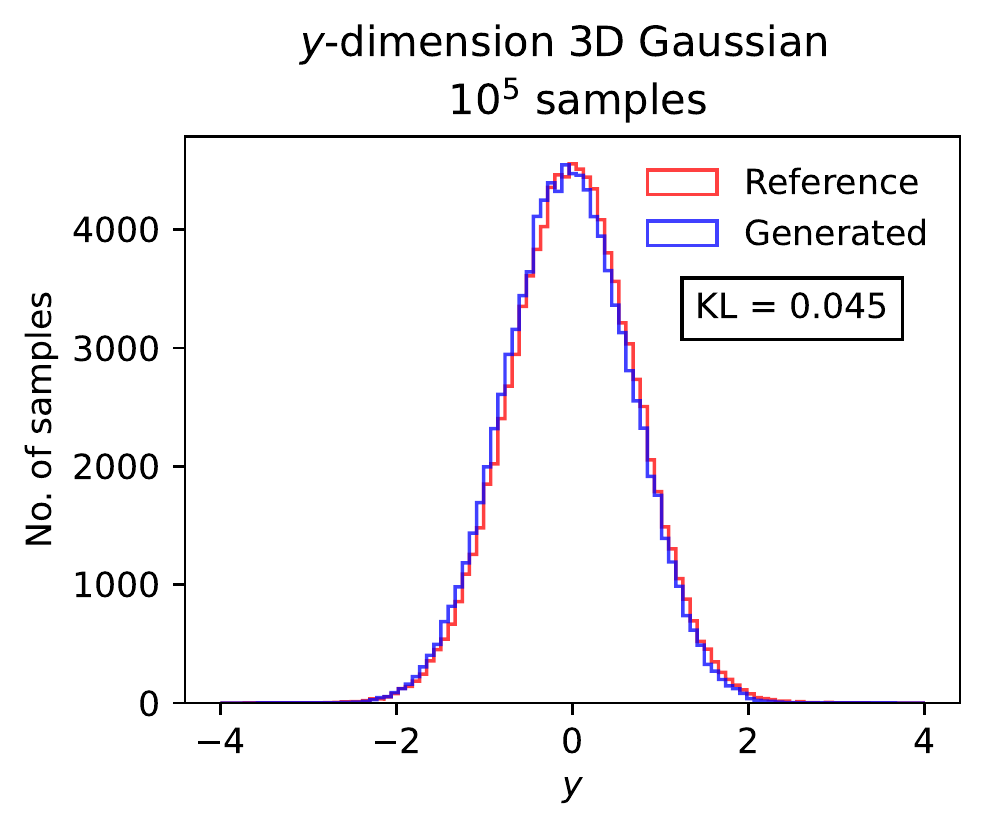}%
  \includegraphics[width=0.29\textwidth]{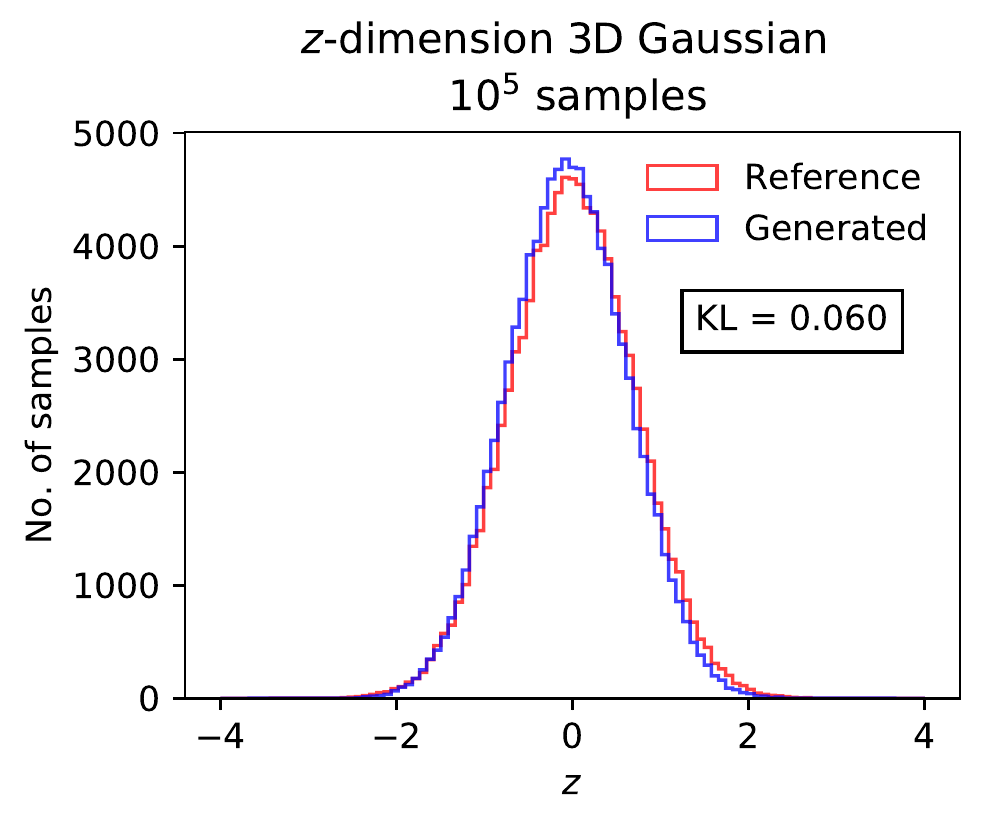}

  \hspace{2.15em}\includegraphics[width=0.305\textwidth]{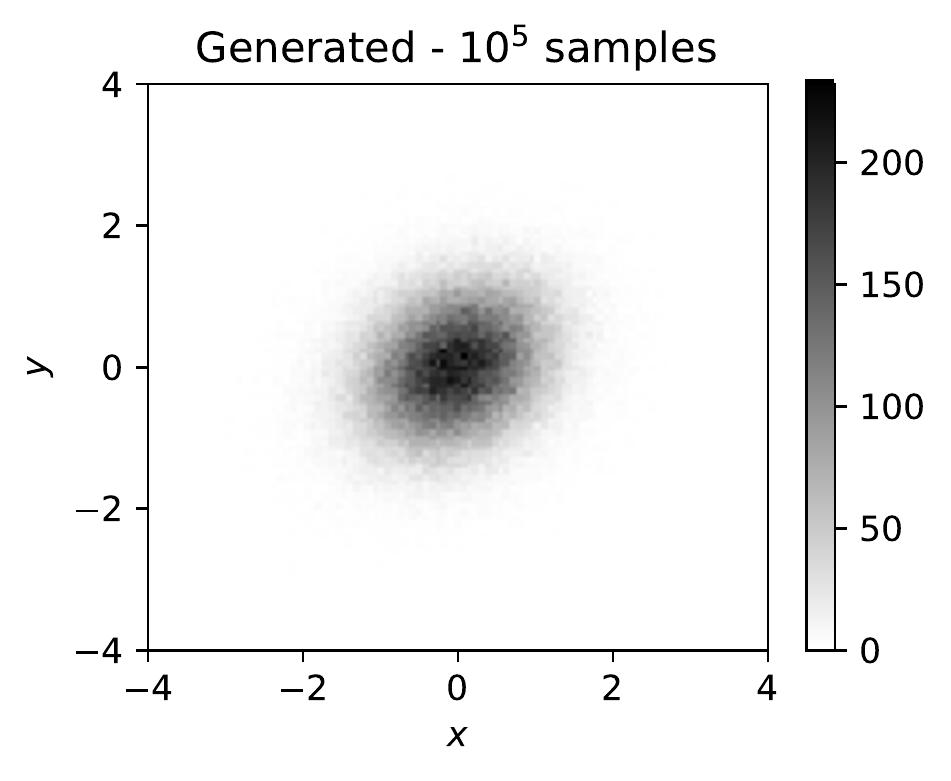}%
  \includegraphics[width=0.305\textwidth]{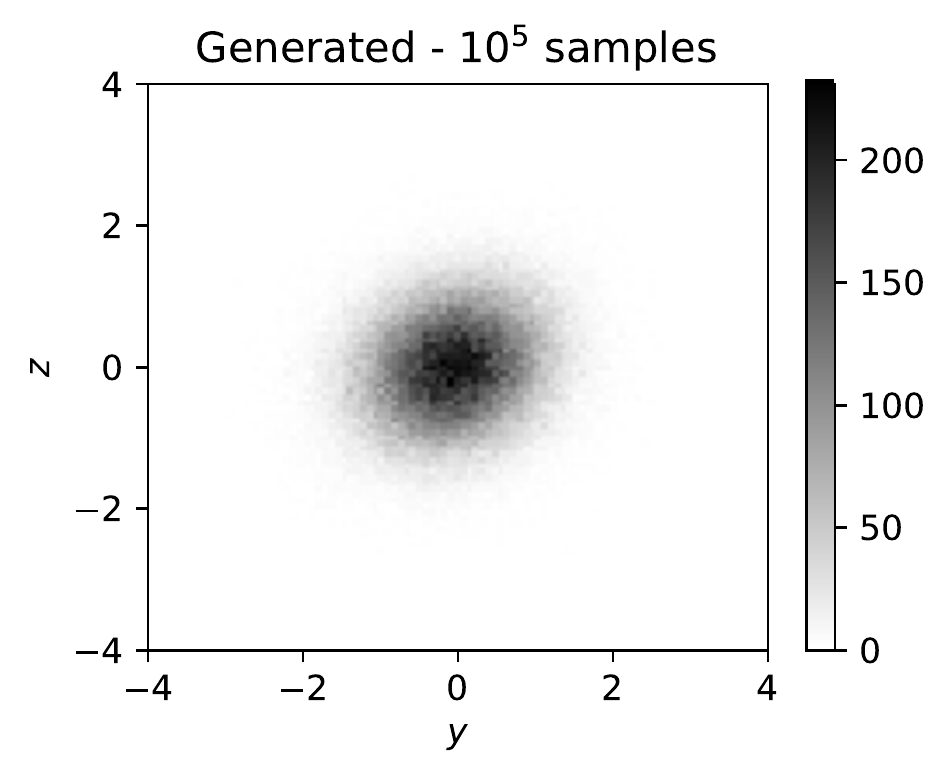}%
  \includegraphics[width=0.305\textwidth]{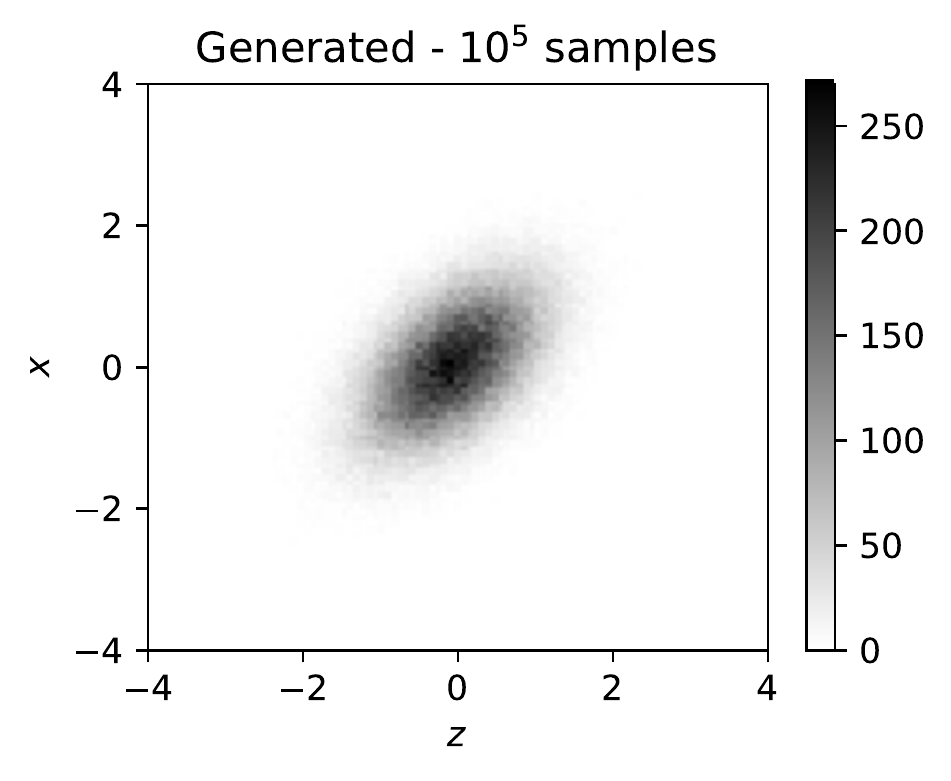}

   \hspace{2.0em}\includegraphics[width=0.3\textwidth]{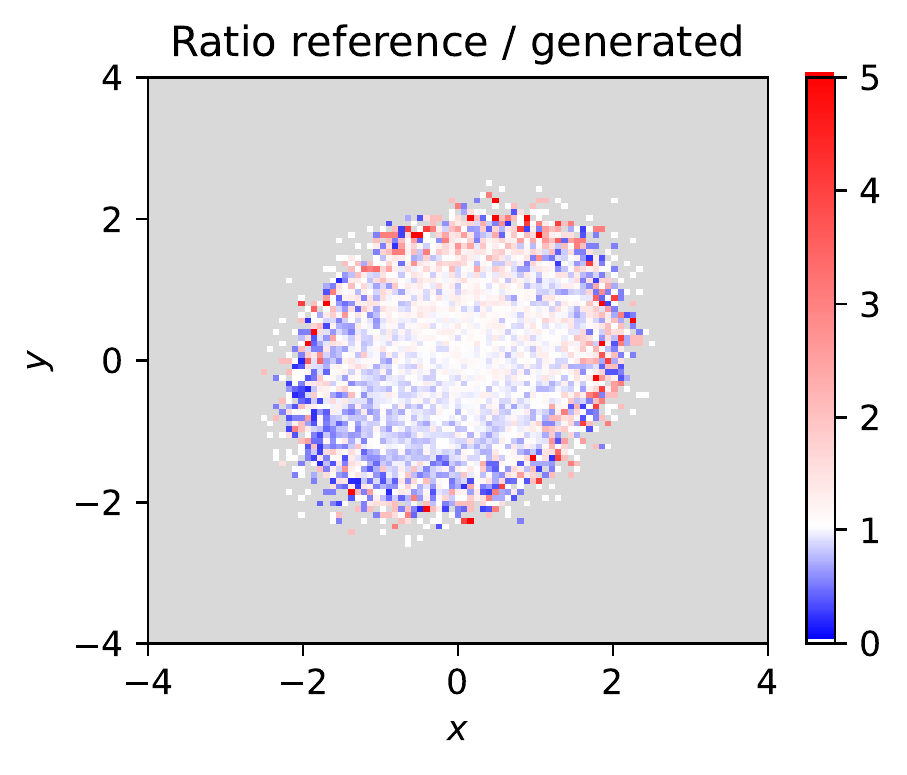}%
  \includegraphics[width=0.3\textwidth]{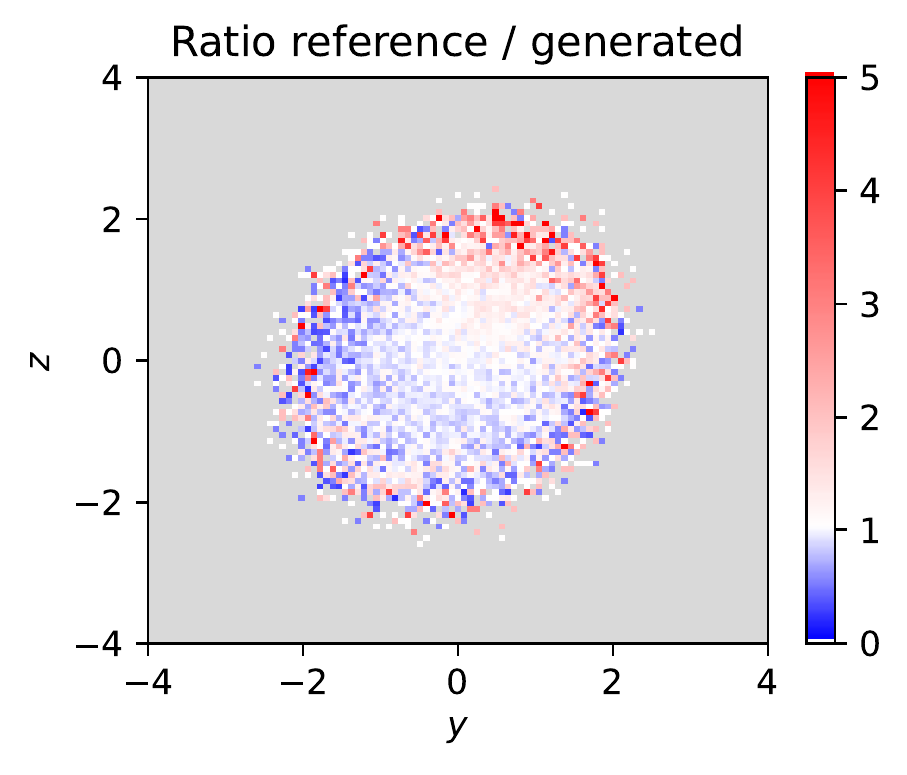}%
  \includegraphics[width=0.3\textwidth]{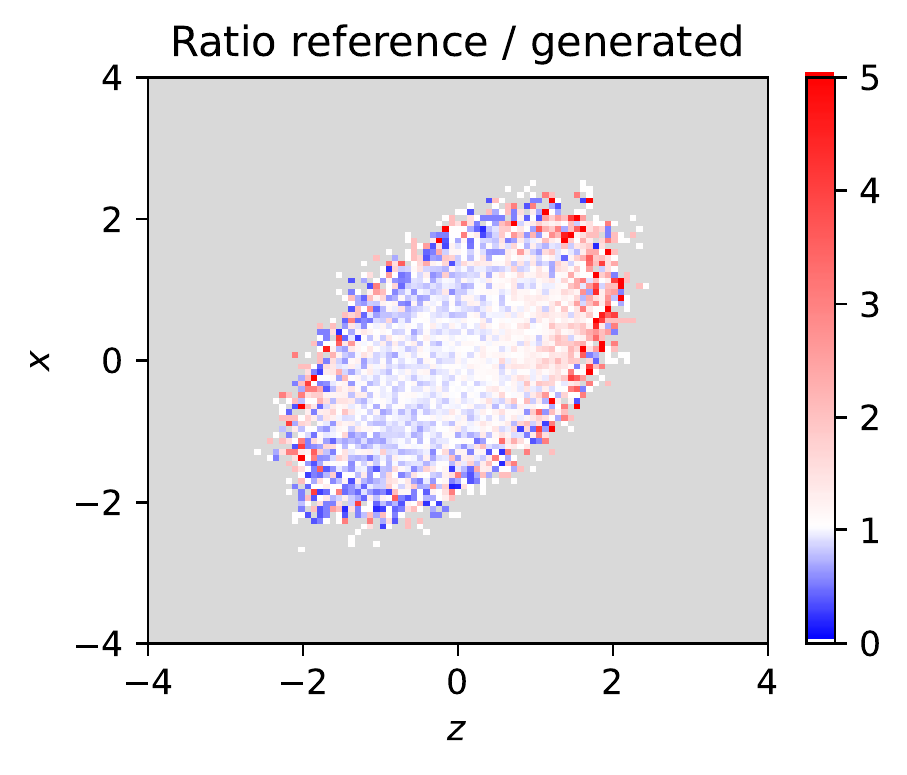}

  \caption{\label{fig:3dgauss}Marginal samples distributions for each dimension $x,y,z$
    of the 3D correlated Gaussian distribution for the style-qGAN model trained
    with $10^4$ samples (top row), together with the corresponding
    two-dimensional sampling projections (middle row) and the ratio to the reference underlying
    prior distribution (bottom row). The style-qGAN generator model learns the correlations and
    provides acceptable samples when compared to the reference distribution. Note that we choose
    a grey background for the plots at the bottom row to more clearly highlight a ratio of one between reference and generated samples, indicated by white.}
\end{figure*}

Following the same training procedure employed in Section~\ref{sec:gamma} and again using $10^4$ reference samples to train the style-qGAN model, we test how well our model samples this specific 3D correlated Gaussian distribution. The results are shown in Figure~\ref{fig:3dgauss}. In the top row, we compare the one-dimensional cumulative projections of samples generated by the style-qGAN model with the reference input distribution function for $10^5$ samples. We again use a grid of 100 linearly spaced bins per dimension in order to highlight small differences between the prior reference distribution and the artificial samples.  For this example, we also observe that the distributions are statistically similar as the corresponding KL distances are quite small and close to each other. In the second row, we show $10^5$ samples produced by the style-qGAN model in two-dimensional projections.

To further study the features of the style-qGAN model in the third row of plots in Figure~\ref{fig:3dgauss} we show the two-dimensional projections of the ratio between samples generated from the prior reference distribution and the style-qGAN model.
In this way, we can visualize how well the model learns not only the distributions but also the correlations between the dimensions of the problem. A ratio of one, given by a white coloring of the corresponding bin in the figure, would imply the reference and generated samples are identical. Note that we aim to generate unseen samples, not an identical copy of the reference set. However, at the same time, the model should not diverge significantly, depicted by deep blue and red in the figure, nor occupy space in the grey area of the figure.  We observe a good level of agreement, in particular in those regions where the sampling frequency is higher. The largest deviations are seen at the edges of the distributions, where the sampling frequency is lower. These deviations are evidence of the limitations in our model, common to the GAN method; however, their severe appearance is an artifact of visualization due to data augmentation.

To better quantify how well the correlations have been learned, we study the covariance matrices defined by the reference and the generated samples. The summed eigenvalues of the reference and generated covariance matrices give a means to estimate the similarity between the learned underlying correlations. We find agreement between the reference and generated eigenvalues to the better of $10\%$ for style-qGAN set-ups with equal and more than 3 latent dimensions. Recall that the latent variables are introduced in every gate of the circuit, including the entangling ones $U_{\mathrm{ent}}$.  With $D_{\mathrm{latent}}< 3$ we observe significant deviations of factors of one order of magnitude while for $D_{\mathrm{latent}}\geq 3$ no further significant improvement is seen. The same holds for increasing the number of layers in the style-qGAN model. This suggests that the number of latent dimensions introduced is a key hyperparameter once the number of layers allows a sufficient complexity. However, training success also depends on the convergence of the GAN parameters through optimization. This means that, in practice, having more layers and parameters than the minimal set might be a better choice.

Since the eigenvalues are known also exactly through Eq.~\ref{eq:covmat} we furthermore can compare the performance of the style-qGAN with increased generation sample size. We find that the style-qGAN with 3 latent dimensions and 1 layer (shown here) generates sets that reproduce the exact eigenvalues of the input covariance matrix to better than $\lesssim 6\%$ for $10^3$, $\lesssim 1.3\%$ for $5\times10^3$ and $\lesssim 0.8\%$ for $2\times10^4$ samples.

This analysis demonstrates a key property of a functioning GAN model -- that the larger set of generated samples more closely agrees with the reference input distribution function. The observation that our style-qGAN fulfils this property confirms its viability as a functioning quantum implementation of the generative adversarial network idea for multi-dimensional correlated data.

\section{Generating LHC events}
\label{sec:lhc}

After the validation of the style-qGAN model presented in the previous section,
let us consider a training dataset from HEP. One of the big challenges involving
Monte Carlo (MC) event generation for modelling physics processes in HEP is the large number of data points required in order to compare predictions of
physical observables to experimental data.

In this context, we have generated $10^5$ MC events for $pp\rightarrow t\bar{t}$
production at LHC with $\sqrt{s} = 13$ TeV with MadGraph ({\tt
MG5\_aMC}~\cite{Alwall:2014hca,Frederix:2018nkq}) at leading order in the strong coupling constant. From this simulated events we
sample the Mandelstam variables $(s,t)$ and the rapidity. Here, $s$
and $t$ are understood as the local partonic variables,
$s=(p_1^{}+p_2^{})^2$, $t=(p_1^{}-p_3^{})_{}^2$, where $p_1^{}$ and
$p_2^{}$ are the four-momenta of the incoming quarks within the proton that collide to produce a top quark with four-momentum $p_3^{}$ and an anti-top quark with four-momentum $p_4^{}$. Note that all momenta are given in the center-of-mass frame.

We consider a 3-qubit model with 5 latent dimensions and 2 layers. Again,
$U_{\rm ent}$ consists of two controlled-$R_{y}$ gates acting sequentially on
the 3 qubits. The total number of trainable parameters is 62. The style-qGAN
model has been trained on $10^4$ samples. See Table~\ref{table:summary_lhc} for more details. In this case, we perform a linear preprocessing of the data
to fit the samples within $x \in [-1, 1]$ after a power
transform~\cite{yeo2000new} from the Python package {\tt Scikit Learn}~\cite{scikit-learn}. As previously, we undo this transformation after
the training.

\begin{figure*}[t!]

  \hspace{2.0em}\includegraphics[width=0.29\textwidth]{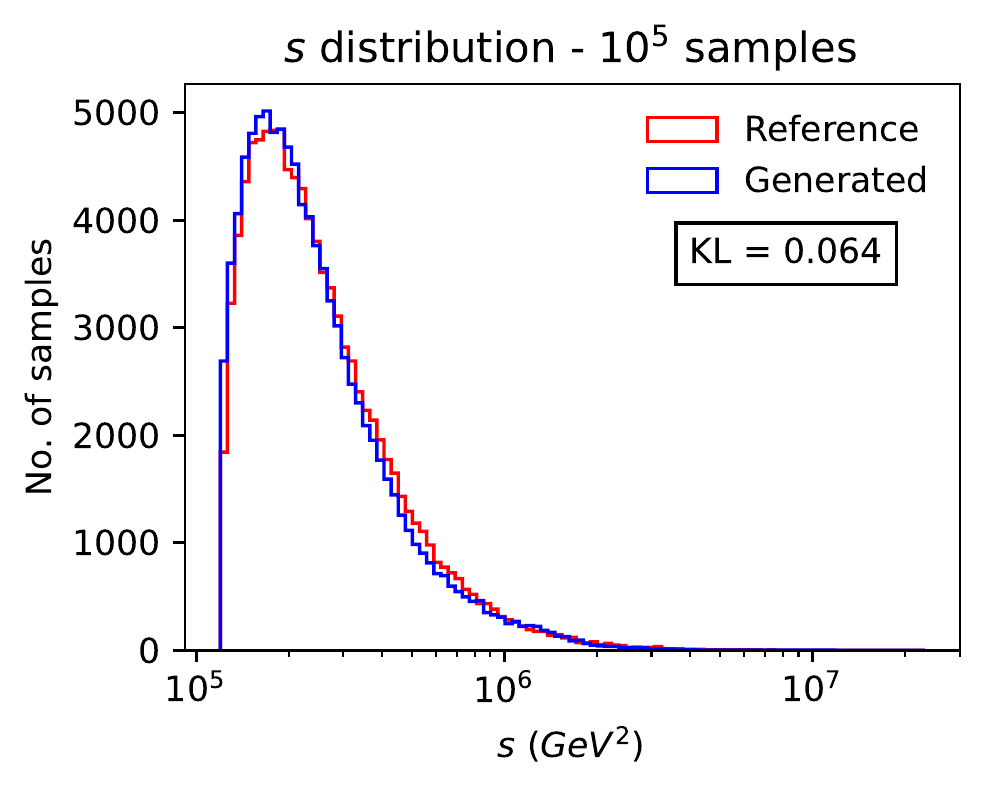}%
  \includegraphics[width=0.29\textwidth]{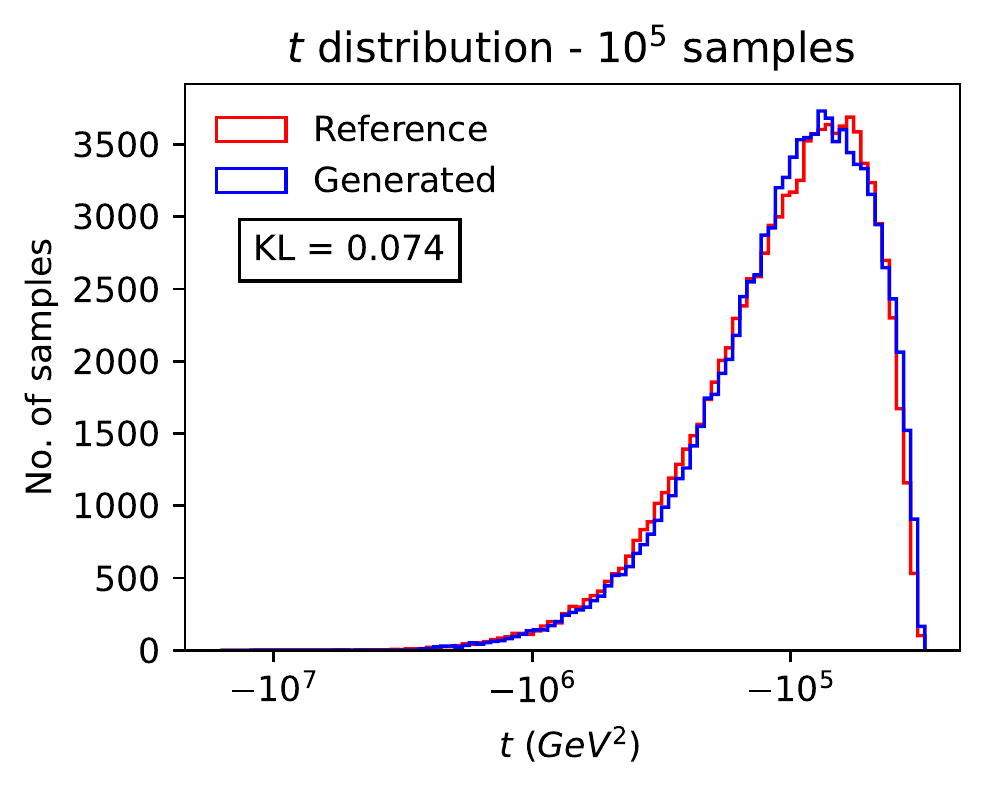}%
  \includegraphics[width=0.29\textwidth]{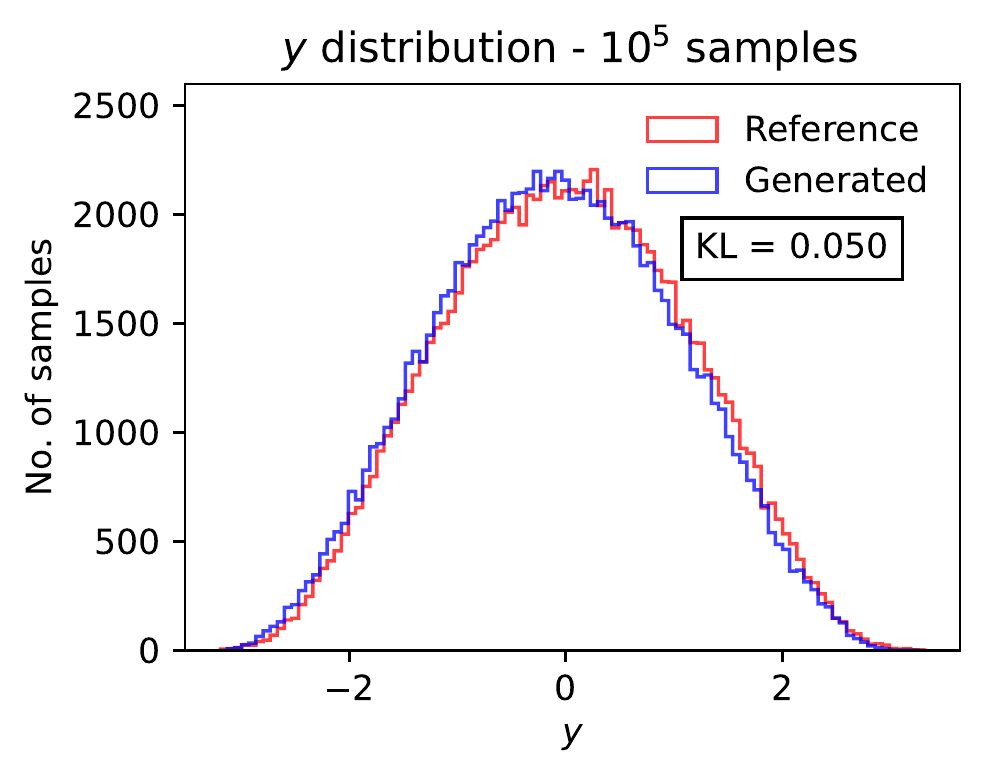}

  \hspace{1.5em}\includegraphics[width=0.315\textwidth]{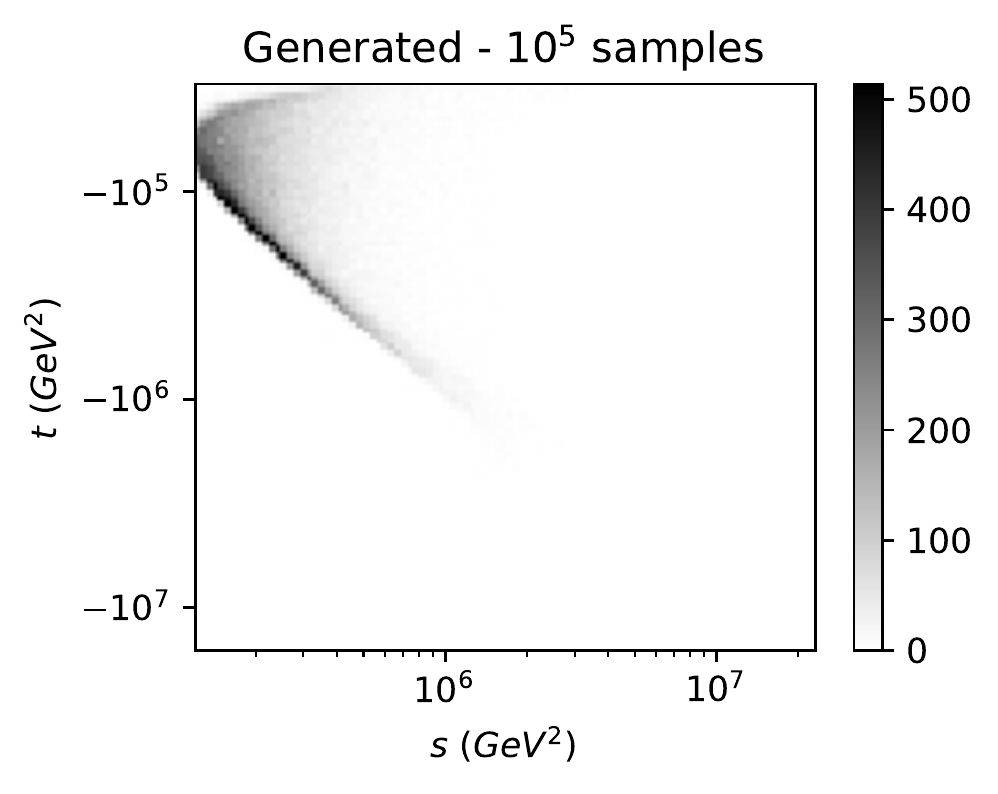}%
  \includegraphics[width=0.295\textwidth]{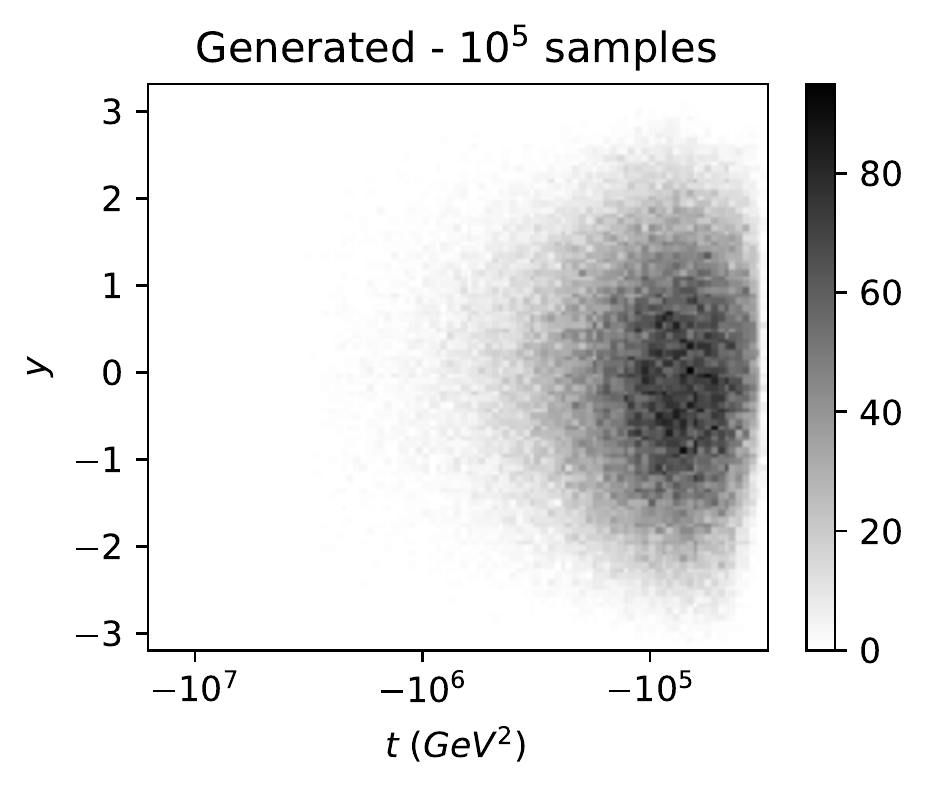}%
  \includegraphics[width=0.313\textwidth]{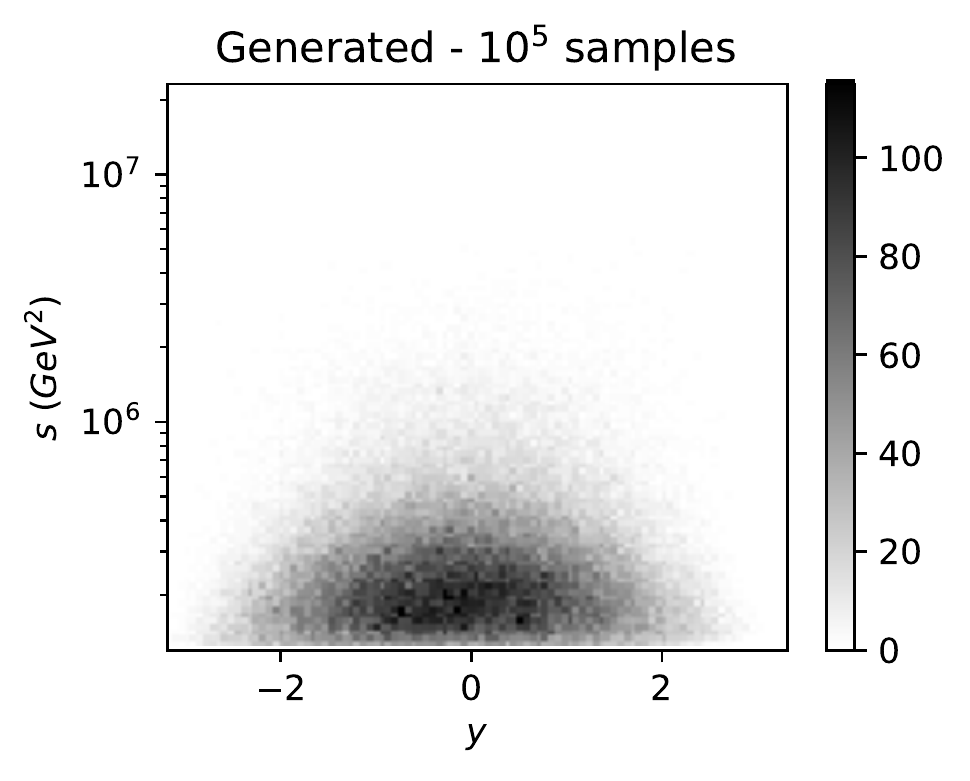}

  \hspace{1.5em}\includegraphics[width=0.308\textwidth]{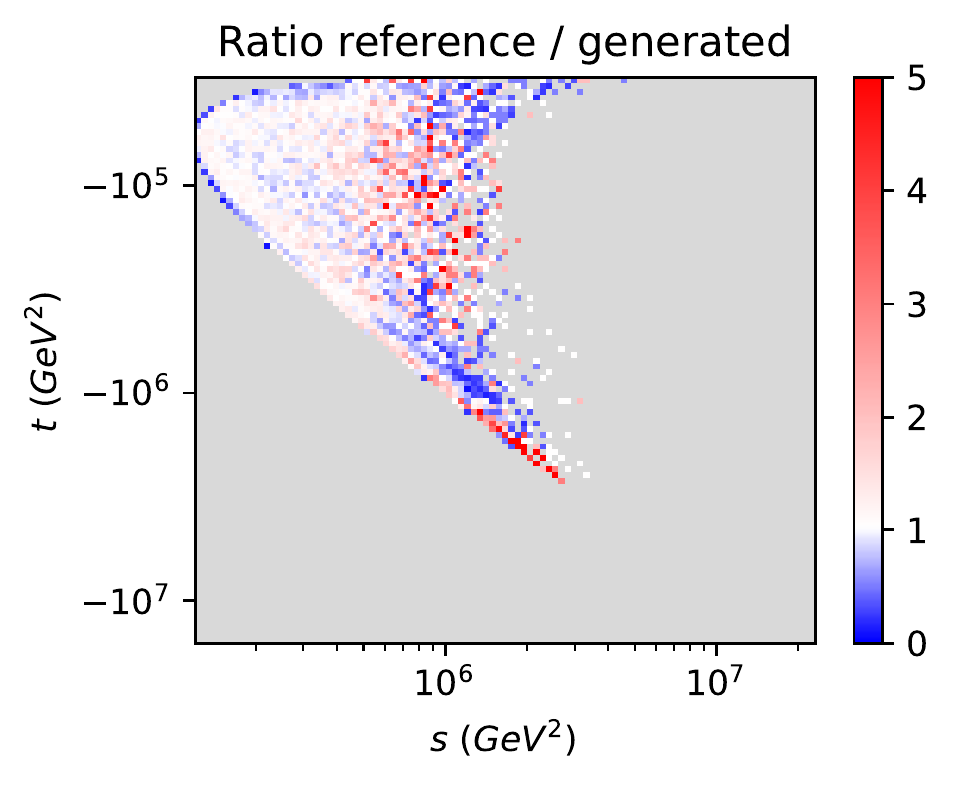}%
  \includegraphics[width=0.295\textwidth]{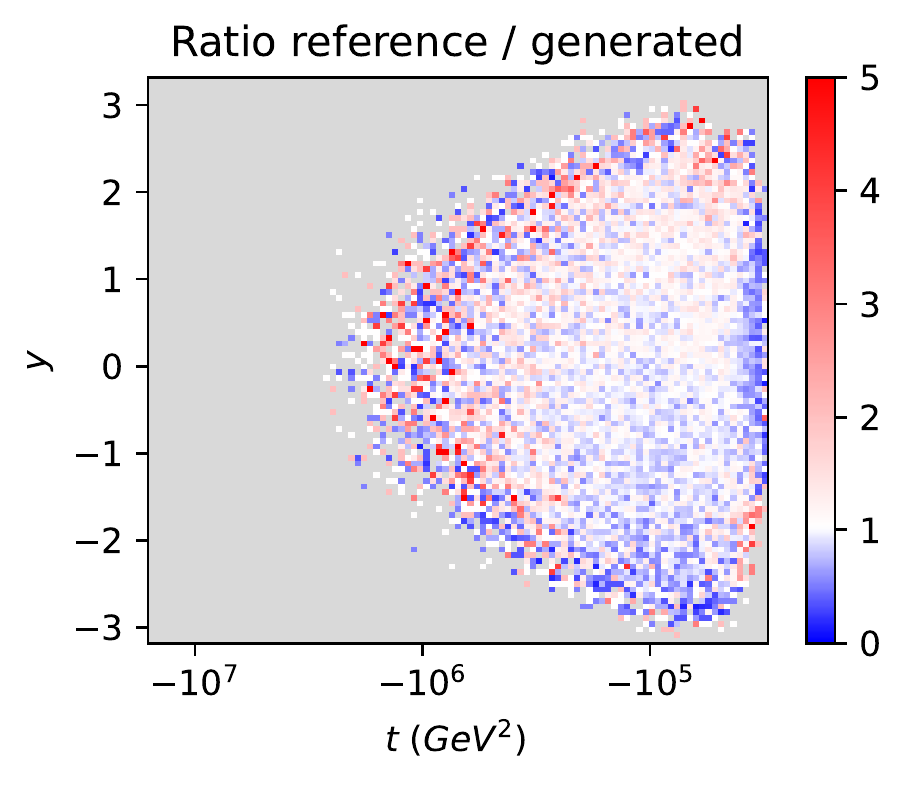}%
  \includegraphics[width=0.305\textwidth]{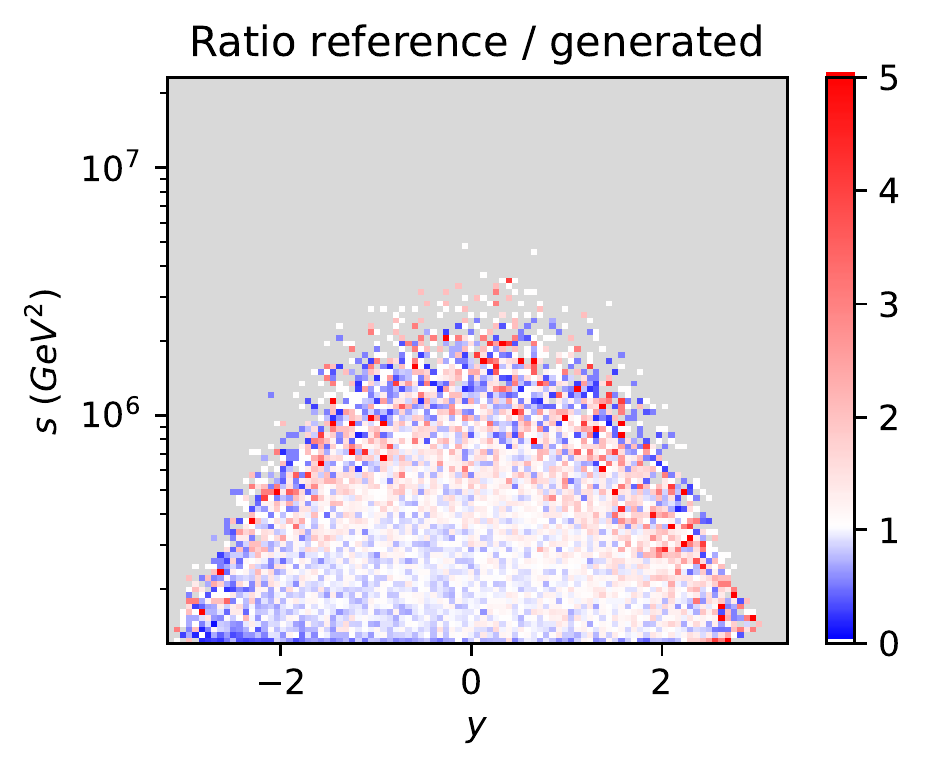}

  \caption{\label{fig:ttbar}
    Marginal samples distributions for the physical observables $s,t,y$
    in $pp\rightarrow t\bar{t}$ production at the LHC for the style-qGAN model trained
    with $10^4$ samples (top row), together with the corresponding
    two-dimensional sampling projections (middle row) and the ratio to the reference underlying
    prior MC distribution (bottom row). The style-qGAN generator model learns the correlations and
    provides acceptable samples when compared to the reference distribution. Note that we choose
    a grey background for the plots at the bottom row to more clearly highlight a ratio of one between reference and generated samples, indicated by white.
}
\end{figure*}

Following the same training procedure employed in the previous section, in the top row of Figure~\ref{fig:ttbar} we compare the one-dimensional cumulative projections of samples generated by the style-qGAN model with the reference input distribution function for $10^5$ samples. We use a grid of 100 linearly spaced bins for $y$ and 100 log-spaced bins for $s$ and $t$.  For this example, the distributions are again statistically similar, with the corresponding KL distance being small and close to each other. In the second row of Figure~\ref{fig:ttbar} we show $10^5$ samples produced by the style-qGAN model in two-dimensional projections.

The bottom row of plots in Figure~\ref{fig:ttbar} shows the ratio between
samples generated from the prior original MC distribution and the style-qGAN
model. Again, even for this physically-realistic model, we observe a remarkable level of agreement, especially in those regions where the sampling frequency is higher. Most importantly, we observe that the style-qGAN learns
the correlations between the three dimensions.

Applying the same reasoning as in the previous section we compute the eigenvalues of the covariance matrices derived from the reference and generated data sets. To this extent we use the larger sized reference data set calculated previously using MadGraph ({\tt
MG5\_aMC}). We find that the summed eigenvalues of the covariance matrices derived from samples generated by the shown style-qGAN with 5 latent dimensions and 2 layers agree with the corresponding reference to $\sim 9-13\%$ for $10^3$, $\sim 8-15\%$ for $5\times10^3$ and $\sim 7-14\%$ for $2\times10^4$ samples. Here the quoted range originates from comparing different samples of the reference data. Furthermore, we suppress effects from the inverse transformation that converts the generated sample and instead focus on the learning capability of the style-qGAN model by estimating the covariances on the transformed reference data sets.
It should be stressed that this test is slightly different from the one in the previous section since the exact eigenvalues are not known. As a result, the sampling error of the reference enters and an agreement at the previous level should not be expected as too close of an agreement would indicate the model is overfitted.  However, our model exhibits the expected and necessary behavior, even when applied to realistic data.

\begin{table}
\centering
  \begin{tabular}{l|c}
     & {\bf $pp \rightarrow t\bar{t}$ LHC events} \tabularnewline
    \hline
    Qubits & 3  \tabularnewline
    $D_{\mathrm{latent}}$ & 5 \tabularnewline
    Layers & 2  \tabularnewline
    Epochs & $3\times10^4$ \tabularnewline
    Training set & $10^4$ \tabularnewline
    Batch size & 128 \tabularnewline
    Parameters & 62 \tabularnewline
    $U_{\rm ent}$ & 2 sequential C$R_y$ \tabularnewline
    \hline
  \end{tabular}

  \caption{\label{table:summary_lhc} Summary of the style-qGAN set-up for the LHC events distribution.}
\end{table}

\section{Sampling from quantum hardware}
\label{sec:deployment}

In order to benchmark our style-qGAN model on real quantum hardware, we performed several runs on two different types of
architectures. This allows us to qualitatively assess the impact of
decoherence and noise, issues that are typical for NISQ computers, and
to check whether the model can already give good results without
waiting for error-corrected machines. The first quantum architecture we used is based on
superconducting transmon qubits as provided by IBM~Q
quantum computers~\footnote{\href{https://research.ibm.com/blog/ibm-quantum-roadmap}{IBM's roadmap for scaling quantum technology}, Sept. 2020.}.
The second is based on trapped ion technology as provided by IonQ quantum computers~\footnote{\href{https://IonQ.com/posts/december-09-2020-scaling-quantum-computer-roadmap}{Scaling IonQ's Quantum Computers: The Roadmap}, Dec. 2020.}
and accessible to us using cloud resources from Amazon Web Services (AWS).

Implementing our style-qGAN onto real quantum hardware introduces a new parameter into the model: the number of shots done for each calculation. Specifically,  we now perform a quantum experiment each time we measure the three-qubit state, and we collect the results after a set number of experiments (shots) have been carried out. These then build up expectation values which are the generated samples.
In this work we typically perform a number of 1000 shots per sample.

 Prior to running on actual quantum hardware, we performed noise simulations using the
IBM~Q simplified noise model, which provides an approximation of the properties of real device backends, and enables us to test how well the results presented in Section~\ref{sec:lhc} would be preserved in the noisy environment. Results are provided in Appendix~\ref{sec:appendixnoise} and show that the impact of the noise is expected to be visible to a degree. We leave noise mitigation to further work. For the noise simulation as well as the actual runs on IBM~Q quantum devices,
we have selected in particular the {\tt ibmq\_santiago} 5-qubit Falcon r4L quantum processor. For our circuit, we need only three qubits with at least one directly connected to the other two. We use a translation layer written in \texttt{Qiskit}~\cite{gadi_aleksandrowicz_2019_2562111} to implement the circuit in Figure~\ref{fig:circuit} and automatically select the three qubits out of the five that have the best noise properties. Note that this also allows us to test the impact of potential
interference between qubits, as IonQ qubits are fully connected while those of IBM~Q are not.

\begin{figure*}[t!]

  \hspace{1.2em}\includegraphics[width=0.30\textwidth]{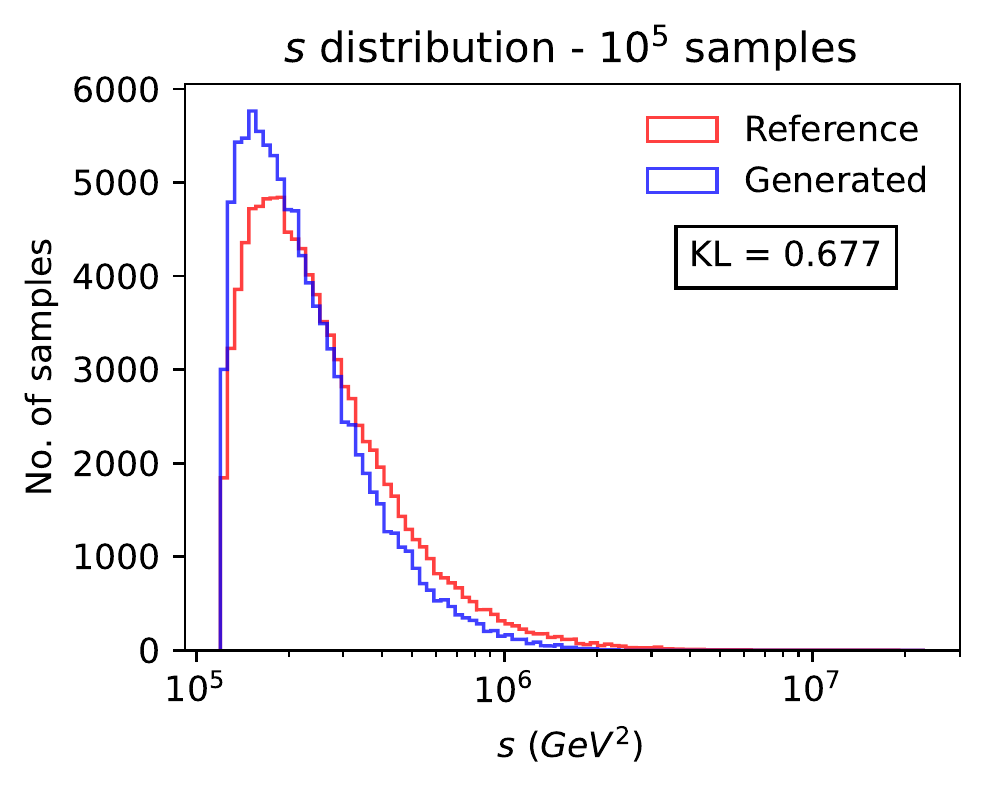}%
  \includegraphics[width=0.30\textwidth]{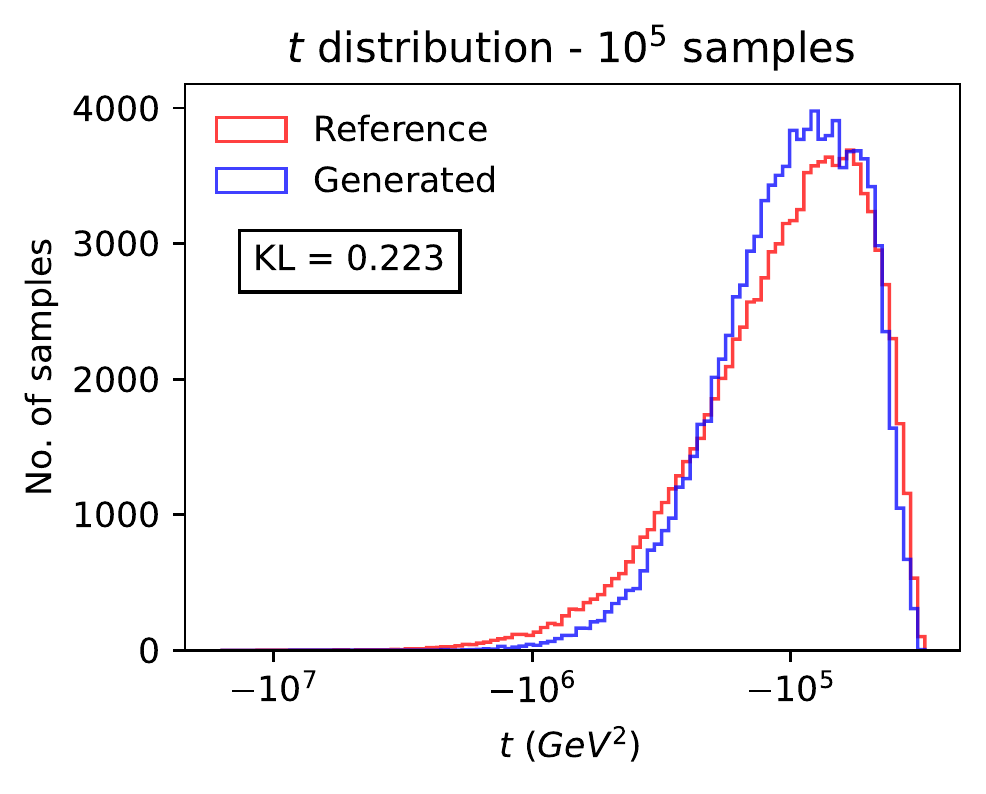}%
  \includegraphics[width=0.30\textwidth]{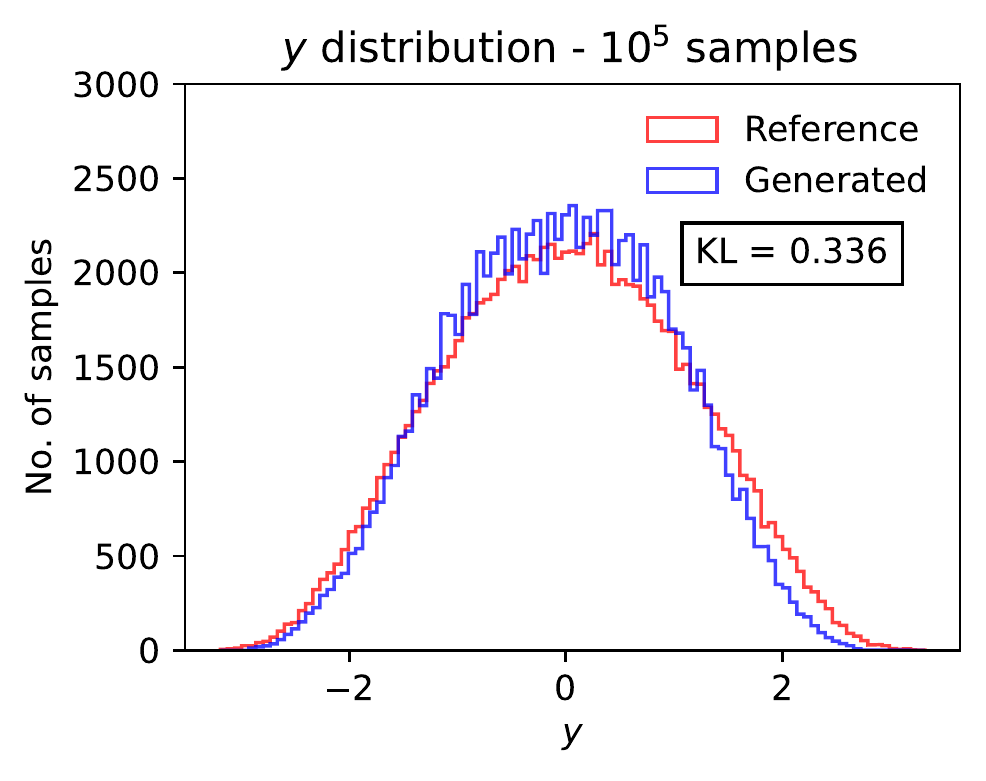}

  \hspace{0.7em}\includegraphics[width=0.335\textwidth]{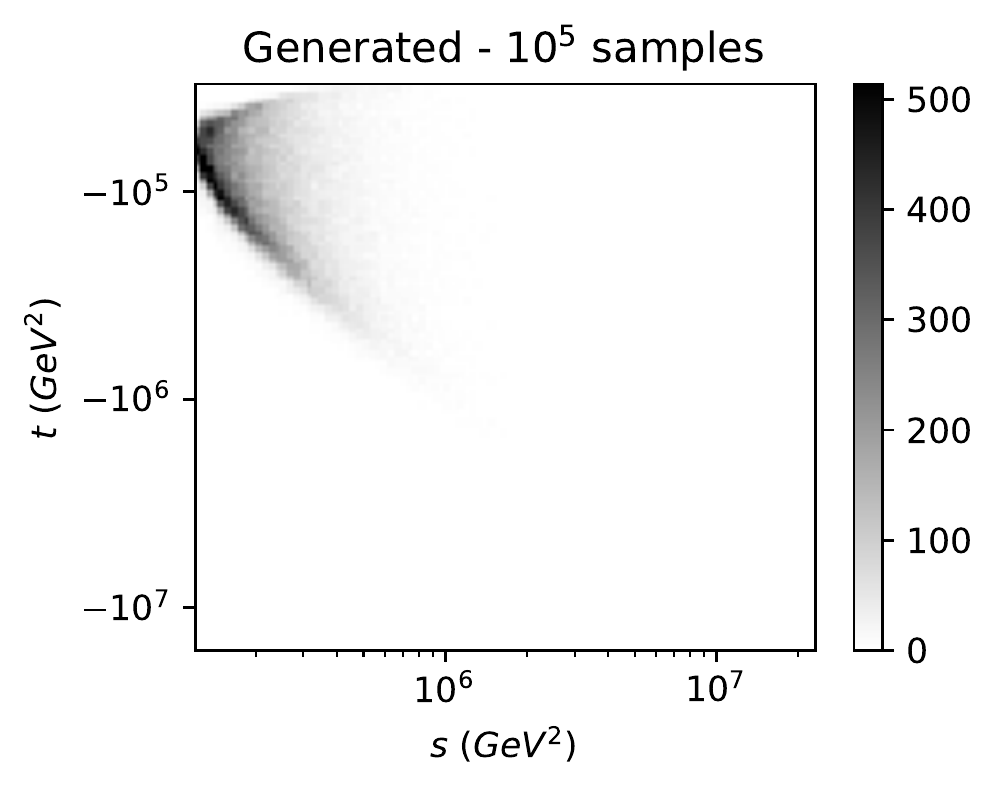}%
  \includegraphics[width=0.310\textwidth]{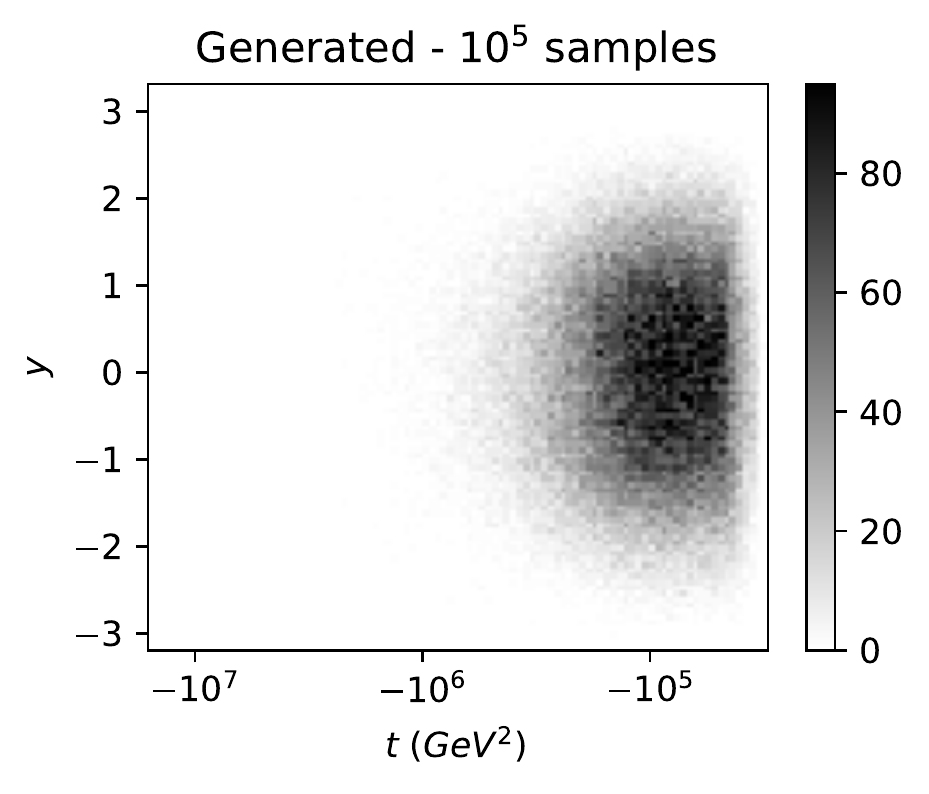}%
  \includegraphics[width=0.325\textwidth]{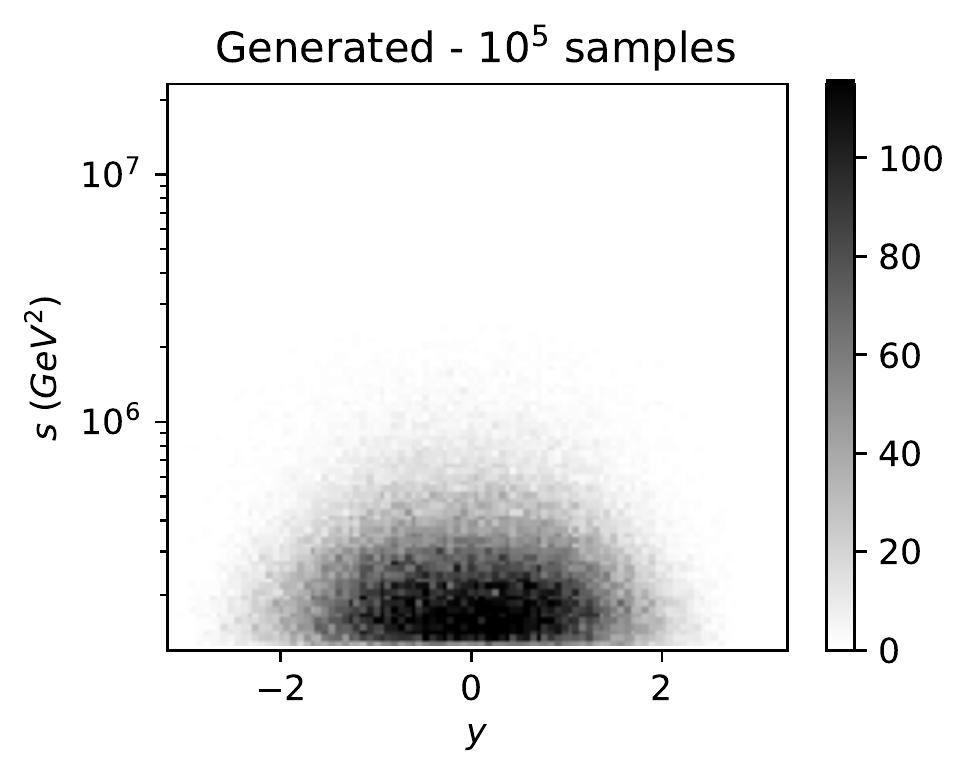}

  \hspace{0.8em}\includegraphics[width=0.325\textwidth]{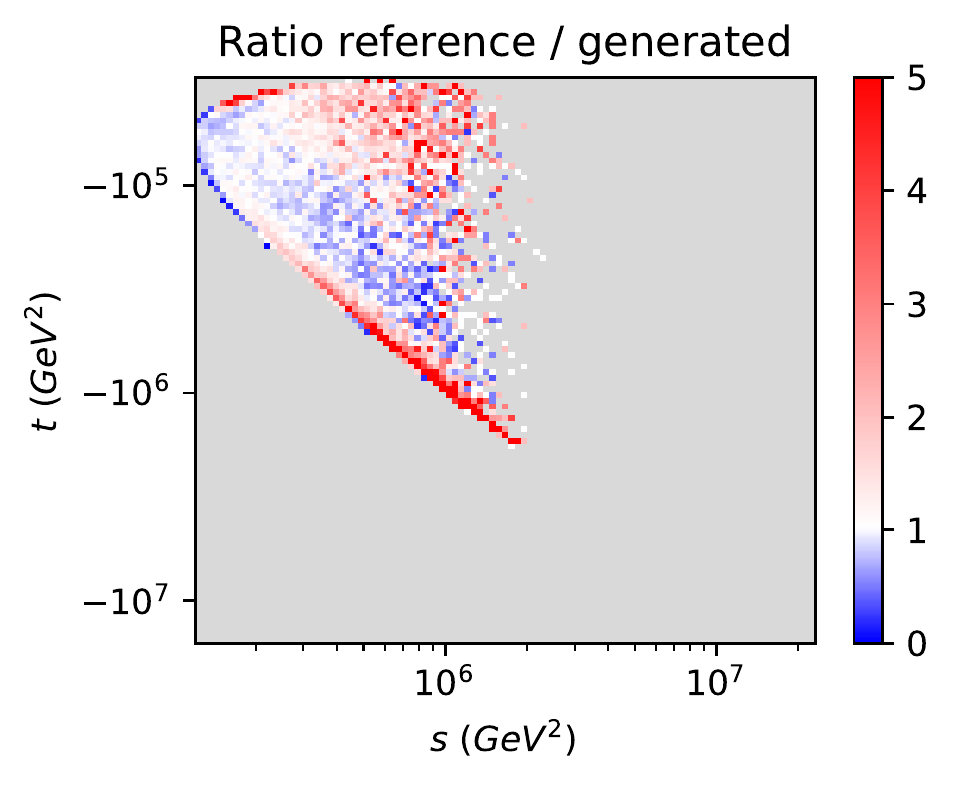}%
  \includegraphics[width=0.305\textwidth]{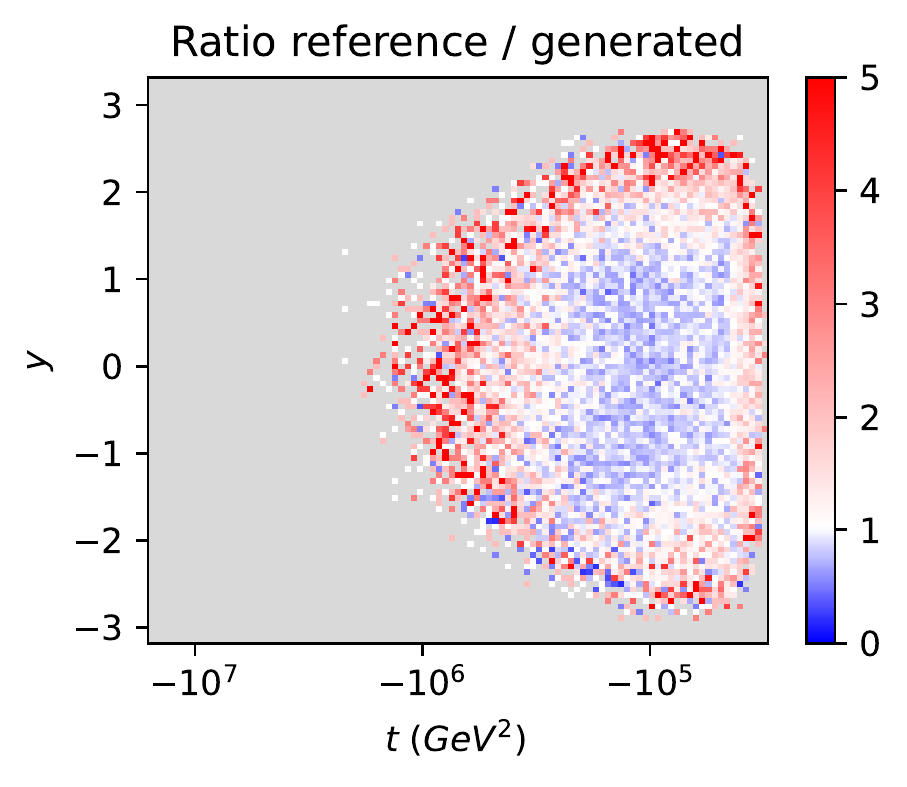}%
  \includegraphics[width=0.32\textwidth]{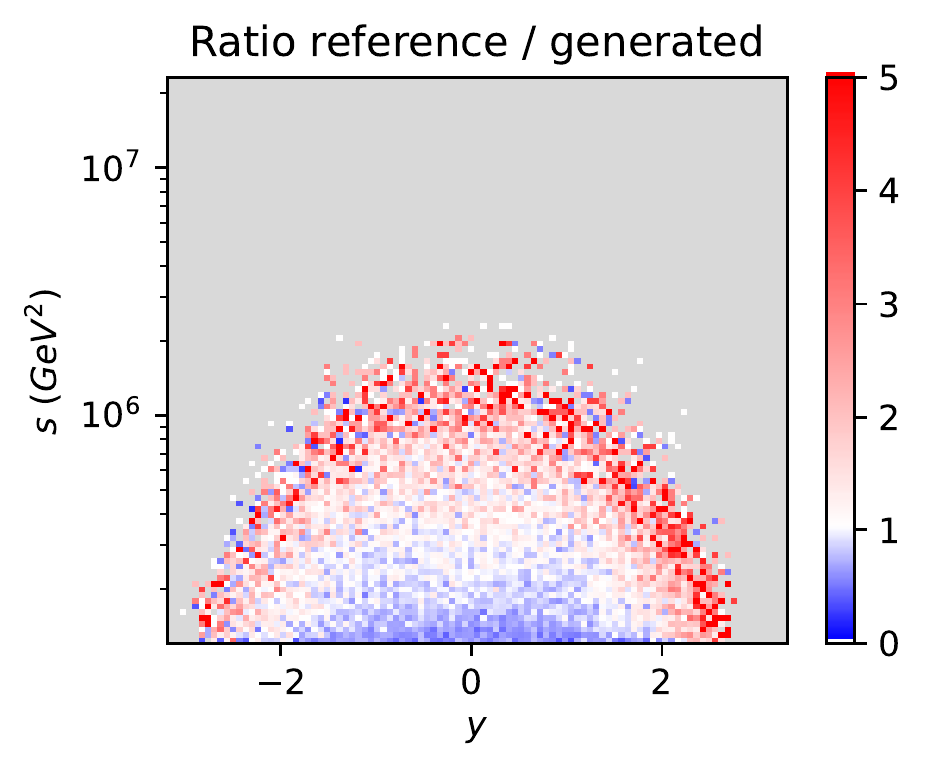}

  \caption{\label{fig:ibm}
    Marginal samples distributions for the physical observables $s,t,y$
    in $pp\rightarrow t\bar{t}$ production at the LHC using the style-qGAN generator model trained
    with $10^4$ samples on {\tt ibmq\_santiago} (top row), together with the corresponding
    two-dimensional sampling projections (middle row) and the ratio to the reference underlying
    prior MC distribution (bottom row). Note that we choose
    a grey background for the plots at the bottom row to more clearly highlight a ratio of one between reference and generated samples, indicated by white.}
\end{figure*}

We present in Figure~\ref{fig:ibm} examples of samples that have been generated using the {\tt ibmq\_santiago} machine on IBM~Q. We use a 3-qubit model with 5 latent dimensions and 1 layer and for which the hyperparameters are the same as the ones used in Section~\ref{sec:lhc} and trained on $10^4_{}$ samples. In contrast to the previous Sec. \ref{sec:lhc}, for this implementation in the quantum hardware we reduced the number of layers to one. This means that we have trained a different style-qGAN with only one layer and then deployed the model to the quantum architecture. This change is motivated by the desire to diminish the effect of noise by reducing the depth of the circuit. Note, the analysis presented in Appendix~\ref{sec:appendixnoise} shows little deviation between the one- and two-layer result ratios, further strengthening this choice. To compute each fake sample, we have performed 1000 shots on the quantum circuits. In the top row of Figure~\ref{fig:ibm}, we compare the one-dimensional cumulative projections of samples generated by the style-qGAN model with the reference input distribution functions for $10^5$ samples. The binning choice is equivalent to that used in Figure \ref{fig:ttbar}. In the middle row, we display the generation of $10^5$ samples in two-dimensional projections. In the bottom row of plots in Figure~\ref{fig:ibm}, we show again the ratio between the reference samples, generated using the MC event generator, and the samples generated by the style-qGAN on the {\tt ibmq\_santiago} quantum hardware. As expected, the agreement is worse than in Figure~\ref{fig:ttbar} because of the noise and reduced capacity of the quantum generator, nevertheless the results are reasonable. The style-qGAN generator model deployed in this NISQ hardware still manages to capture the correlations and provides reasonably good samples when compared to the reference distribution. The  KL distances reported in the top row of plots are still relatively small, at most one order of magnitude larger than the KL distances reported in Figure~\ref{fig:ttbar}.

\begin{figure*}
  \includegraphics[width=0.32\textwidth]{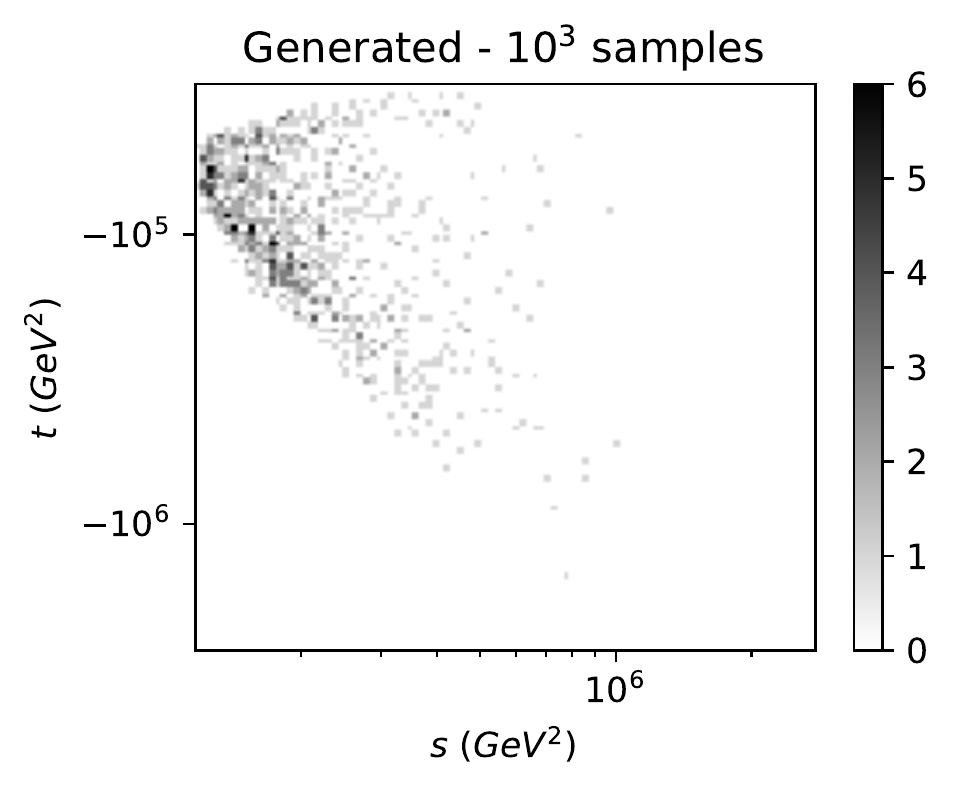}%
  \includegraphics[width=0.32\textwidth]{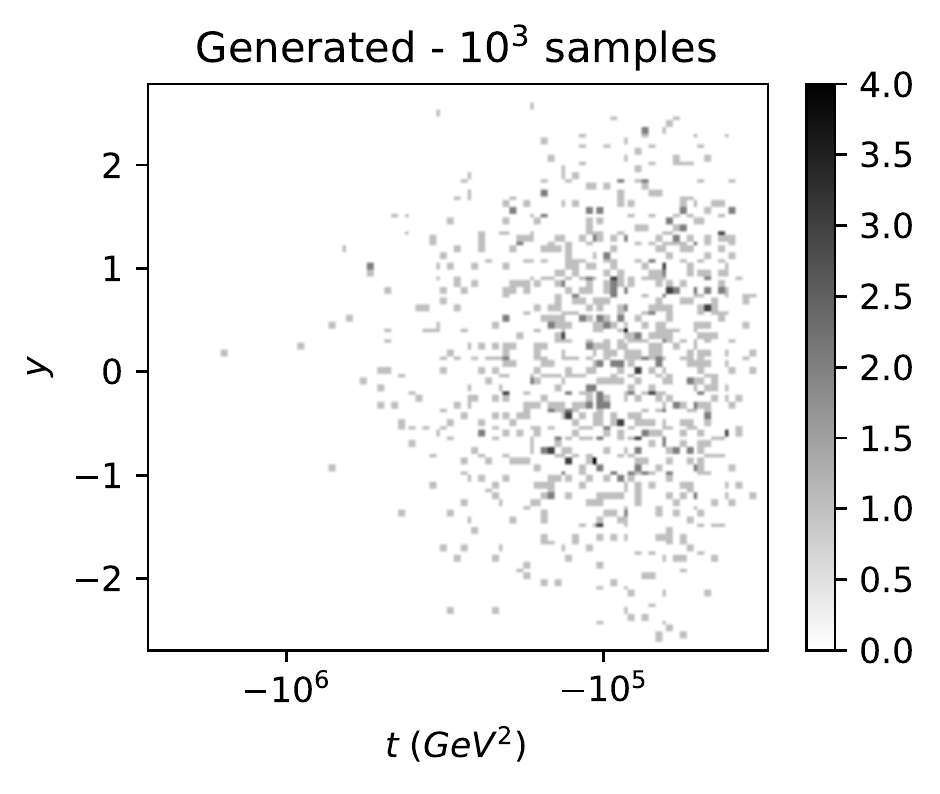}%
  \includegraphics[width=0.32\textwidth]{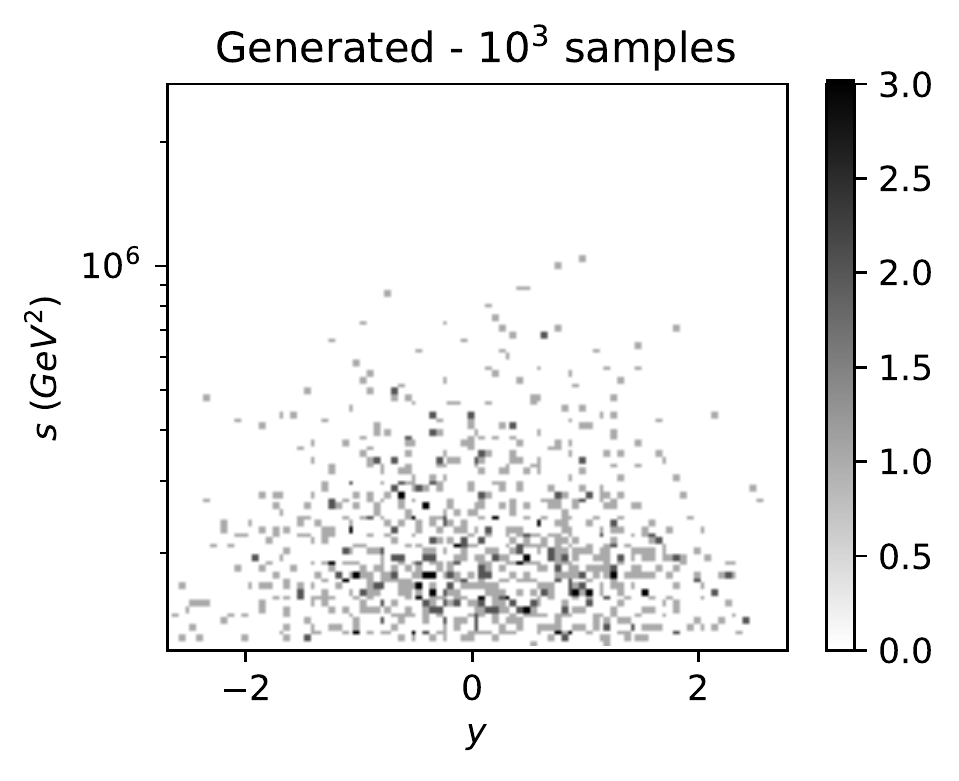}

  \includegraphics[width=0.32\textwidth]{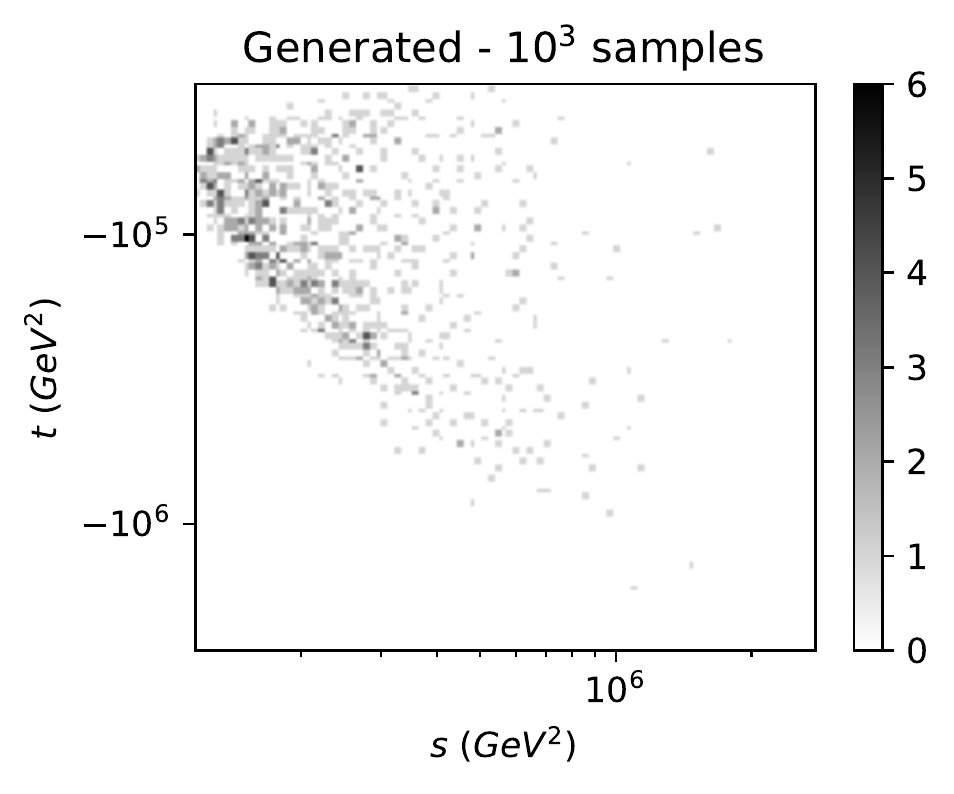}%
  \includegraphics[width=0.32\textwidth]{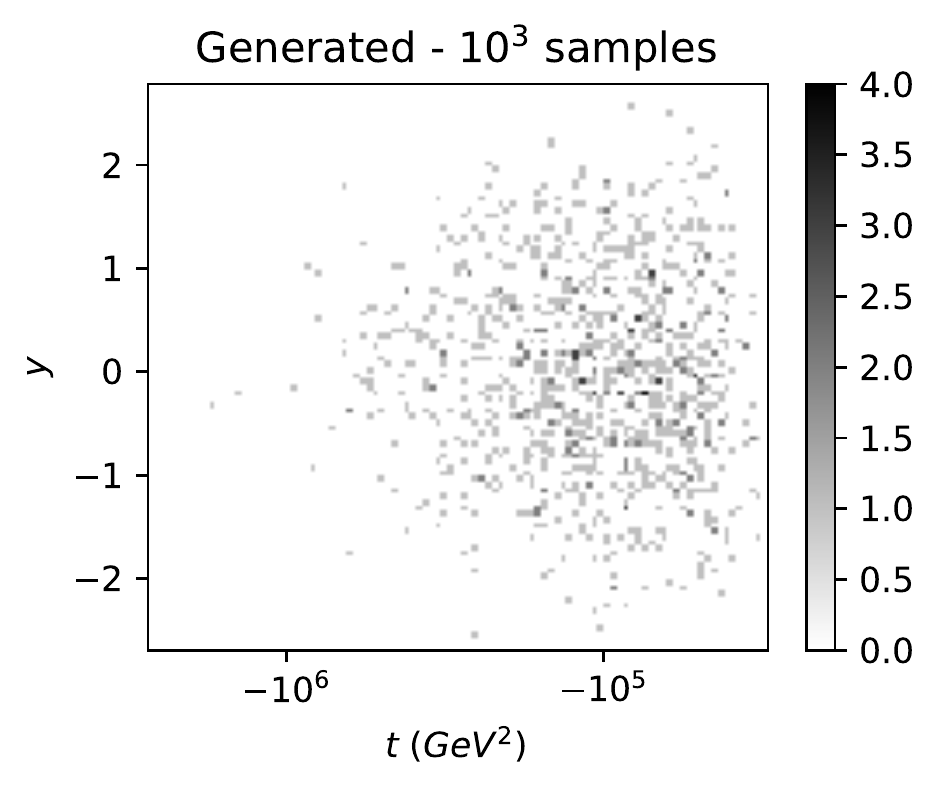}%
  \includegraphics[width=0.32\textwidth]{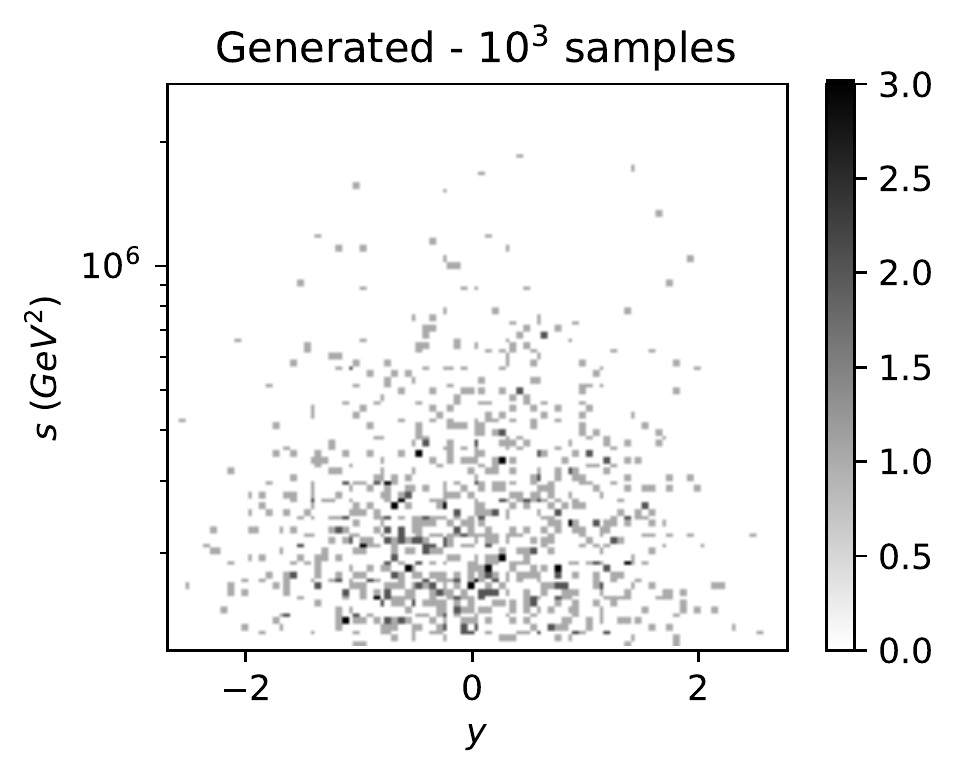}

  \caption{\label{fig:ionq}Example of two-dimensional sampling projections for
  $pp \rightarrow t\bar{t}$ production using the style-qGAN generator
  model on {\tt ibmq\_santiago} (top row) and IonQ (bottom row) trained with $10^4$ samples.}
\end{figure*}

During the current NISQ-era, the different quantum hardware architectures are not standardized and can have limits
on the potential applications of the machines. As part of the implementation of our model onto quantum hardware, we were also able to study how the style-qGAN performs across different platforms.
The aim is to understand whether and to what extent the style-qGAN's performance is hardware-dependent and also its potential hardware transferability.
In view of this study of different quantum technologies, we have also performed a run
with $10^3$ samples only, on IonQ machines and separately on IBM~Q
machines. We have selected this fairly small amount of samples
mainly due to external constraints on IonQ machine access on AWS.
Note that the purpose of these tests is not to compare the two different hardware technologies,
but instead to test whether the style-qGAN model works well on different quantum architectures.
We use again a translation layer, written in Python with the
\texttt{Braket SDK} from Amazon, between our circuit and the quantum hardware, and we
have also performed around 1000 shots for the measurement of the generated samples.
We stress again that although the amount of samples is quite low, the purpose of this test is to assess
how the algorithm performs on different quantum technologies using the same amount of
samples, not to obtain fine-grained results.

We show in Figure~\ref{fig:ionq} the two-dimensional projections using
IBM~Q  {\tt ibmq\_santiago} machine (upper row) and IonQ machine (lower row).
It is clear that the sampling is sparser than in Figure~\ref{fig:ibm}, due to the lower number of
samples; nonetheless, the style-qGAN captures the underlying
distribution and correlations. This is particularly visible on the left plots for the $t-s$ correlation.
The comparison between the upper row and the lower row also indicates that both architectures
obtain similar results. This demonstrates that the style-qGAN can give good results on two different quantum hardware architectures.

\section{Comparison with previous models}
\label{sec:comparison}

In order to quantify the quality of the generalized style-qGAN model described
in the previous sections, we implement and compare the results obtained with the
standard qGAN architecture presented in Refs.~\cite{zoufal2019quantum,
zeng2019learning, situ2020quantum} and adapted to the continuous real sampling
space in Ref.~\cite{romero2021variational}.

\begin{table}[h]
  \begin{tabular}{l|c|c}
     & {\bf style-qGAN} & {\bf qGAN~\cite{romero2021variational}} \tabularnewline
    \hline
    Qubits & 3 & 3 \tabularnewline
    $D_{\mathrm{latent}}$ & 5 & 5 \tabularnewline
    Layers & 2 & 2 \tabularnewline
    Epochs & $3\times10^4$ & $3\times10^4$ \tabularnewline
    Training set & $10^4$ & $10^4$ \tabularnewline
    Batch size & 128 & 128 \tabularnewline
    Parameters & 62 & 36 \tabularnewline
    $U_{\rm ent}$ & 2 C$R_y$ gates & 2 C$R_y$ gates \tabularnewline
    \hline
    KL $s$ & 0.064 & 0.338 \tabularnewline
    KL $t$ & 0.074 & 11.171 \tabularnewline
    KL $y$ & 0.050 & 0.049\tabularnewline
    \hline
  \end{tabular}

  \caption{\label{table:comparison} Summary of the style-qGAN and qGAN
  configuration for the training setup in Section~\ref{sec:lhc}. The bottom rows show the KL divergences for
$(s,t,y)$ samples obtained with both models after training.}
\end{table}

In Table~\ref{table:comparison} we summarize the differences between both
architectures in the context of LHC event generation presented in
Section~\ref{sec:lhc}. In order to provide a fair comparison we have tested both
models keeping identical the number of qubits, layers, latent space
dimension, training epochs, batch size and entangling gates. With this
setup, the style-qGAN model provides 62 parameters to be trained while the qGAN
model 36 parameters. This is an artifact of the linear encoding in
Eq.~\ref{eq:rotation} requiring two trainable parameters in each rotation,
instead of one. Note that in order to increase the number of parameters of the
qGAN model we have to include extra layers and thus modify radically its
architecture.
At the bottom of Table~\ref{table:comparison} we compute the KL divergences for
$(s,t,y)$ samples obtained with both models after training using the simulation
setup presented in Section~\ref{sec:lhc}. While the KL divergence for
Gaussian-like distributions, such as $y$, are similar between both models, there
is a remarkable difference in terms of quality for non-Gaussian distributions,
$(s,t)$. This translates the lack of flexibility of the original qGAN which has
been addressed in its generalization through the style-qGAN procedure.

\section{Conclusion}
\label{sec:conclusion}
In this work, we explore the use of quantum neural networks (QNNs) for Monte Carlo event generation, specifically for scattering processes at the Large Hadron Collider (LHC). We focus specifically on quantum generative adversarial networks (qGANs), which employ two competing networks, the generator and the discriminator, that are trained alternatingly. Here we propose a novel quantum generator model that does not follow the traditional path where the prior noise distribution is provided to the quantum generator through its first quantum gates. We instead choose to embed it on every single-qubit and entangling gate of the network. This allows for a generalization of the state-of-the-art qGAN implementations, achieving
smaller Kullback-Leibler divergences even with shallow-depth networks.  As a similar concept has been utilized in the classical context, coined as style-GANs, we choose to call our novel architecture a style-qGAN. A future interesting development of this work could be introducing additional features from the classical style-GAN, such as the preprocessing of the latent variables.

As this is a new quantum generative architecture, the body of this work focused on validating and assessing our methodology on various data sets and hardware architectures. In particular, we not only trained our model on toy data, namely 1D gamma and 3D correlated Gaussian distributions, but also on data for realistic quantum processes at the LHC, generated via MadGraph. For both toy data and realistic data, we saw strong evidence that the style-qGAN model could be used for data augmentation, as it was able to reproduce known reference distributions from small sample sets. Additionally, we deployed the models on two different quantum hardware architectures -- superconducting qubits (IBM~Q) and trapped ions (IonQ). Despite working with a small sample set, we observed that the style-qGAN works well, at least qualitatively, on different hardware architectures. This points to its hardware-independent viability.

The results presented here should be considered both as an improvement over the state-of-the-art from the algorithmic point of view, and as a
proof-of-concept for a
future quantum hardware deployment, providing a robust and reproducible starting
point for future investigations. Nevertheless, this is a first attempt to bridge
the power of quantum machine learning algorithms into the complexity of Monte
Carlo simulation in HEP. We wish that the approach presented here will inspire
new HEP applications that may benefit from quantum computing.

\vspace*{0.5cm}

\paragraph*{Code availability:}
The code is available in Ref.~\cite{afrancis_heplat_2021_5567077}.

\acknowledgments

C.B.-P. acknowledges Stavros Efthymiou for useful discussions about the
realization of the manuscript's code. S.C.~thanks Marco Zaro and Luigi Favaro for
discussions about classical GAN models applied to Monte Carlo events. This
project is supported by CERN's Quantum Technology Initiative.
M.C.\ is supported by the European Union’s Horizon 2020 research and innovation programme under the Marie Skłodowska-Curie grant agreement number 843134.
S.C. is supported by the European Research
Council under the European Union's Horizon 2020 research and innovation
Programme (grant agreement number 740006). The authors acknowledge the support of the CloudBank EU as a pilot brokering cloud service at CERN, allowing
 for access to Amazon Web Services in order to run our algorithm on IonQ hardware. The authors also acknowledge the use of IBM Quantum services for this work.

\bibliographystyle{apsrev4-2}
\bibliography{qgan}

\begin{thebibliography}{68}%
\makeatletter
\providecommand \@ifxundefined [1]{%
 \@ifx{#1\undefined}
}%
\providecommand \@ifnum [1]{%
 \ifnum #1\expandafter \@firstoftwo
 \else \expandafter \@secondoftwo
 \fi
}%
\providecommand \@ifx [1]{%
 \ifx #1\expandafter \@firstoftwo
 \else \expandafter \@secondoftwo
 \fi
}%
\providecommand \natexlab [1]{#1}%
\providecommand \enquote  [1]{``#1''}%
\providecommand \bibnamefont  [1]{#1}%
\providecommand \bibfnamefont [1]{#1}%
\providecommand \citenamefont [1]{#1}%
\providecommand \href@noop [0]{\@secondoftwo}%
\providecommand \href [0]{\begingroup \@sanitize@url \@href}%
\providecommand \@href[1]{\@@startlink{#1}\@@href}%
\providecommand \@@href[1]{\endgroup#1\@@endlink}%
\providecommand \@sanitize@url [0]{\catcode `\\12\catcode `\$12\catcode
  `\&12\catcode `\#12\catcode `\^12\catcode `\_12\catcode `\%12\relax}%
\providecommand \@@startlink[1]{}%
\providecommand \@@endlink[0]{}%
\providecommand \url  [0]{\begingroup\@sanitize@url \@url }%
\providecommand \@url [1]{\endgroup\@href {#1}{\urlprefix }}%
\providecommand \urlprefix  [0]{URL }%
\providecommand \Eprint [0]{\href }%
\providecommand \doibase [0]{https://doi.org/}%
\providecommand \selectlanguage [0]{\@gobble}%
\providecommand \bibinfo  [0]{\@secondoftwo}%
\providecommand \bibfield  [0]{\@secondoftwo}%
\providecommand \translation [1]{[#1]}%
\providecommand \BibitemOpen [0]{}%
\providecommand \bibitemStop [0]{}%
\providecommand \bibitemNoStop [0]{.\EOS\space}%
\providecommand \EOS [0]{\spacefactor3000\relax}%
\providecommand \BibitemShut  [1]{\csname bibitem#1\endcsname}%
\let\auto@bib@innerbib\@empty
\bibitem [{\citenamefont {Preskill}(2018)}]{nisq}%
  \BibitemOpen
  \bibfield  {author} {\bibinfo {author} {\bibfnamefont {J.}~\bibnamefont
  {Preskill}},\ }\href {https://doi.org/10.22331/q-2018-08-06-79} {\bibfield
  {journal} {\bibinfo  {journal} {Quantum}\ }\textbf {\bibinfo {volume} {2}},\
  \bibinfo {pages} {79} (\bibinfo {year} {2018})}\BibitemShut {NoStop}%
\bibitem [{\citenamefont {Arute}\ \emph {et~al.}(2019)\citenamefont {Arute},
  \citenamefont {Arya}, \citenamefont {Babbush}, \citenamefont {Bacon},
  \citenamefont {Bardin}, \citenamefont {Barends}, \citenamefont {Biswas},
  \citenamefont {Boixo}, \citenamefont {G.~S. L.~Brandao}, \citenamefont
  {Buell} \emph {et~al.}}]{supremacy}%
  \BibitemOpen
  \bibfield  {author} {\bibinfo {author} {\bibfnamefont {F.}~\bibnamefont
  {Arute}}, \bibinfo {author} {\bibfnamefont {K.}~\bibnamefont {Arya}},
  \bibinfo {author} {\bibfnamefont {R.}~\bibnamefont {Babbush}}, \bibinfo
  {author} {\bibfnamefont {D.}~\bibnamefont {Bacon}}, \bibinfo {author}
  {\bibfnamefont {J.~C.}\ \bibnamefont {Bardin}}, \bibinfo {author}
  {\bibfnamefont {R.}~\bibnamefont {Barends}}, \bibinfo {author} {\bibfnamefont
  {R.}~\bibnamefont {Biswas}}, \bibinfo {author} {\bibfnamefont
  {S.}~\bibnamefont {Boixo}}, \bibinfo {author} {\bibfnamefont
  {F.}~\bibnamefont {G.~S. L.~Brandao}}, \bibinfo {author} {\bibfnamefont
  {D.~A.}\ \bibnamefont {Buell}}, \emph {et~al.},\ }\href
  {https://doi.org/10.1038/s41586-019-1666-5} {\bibfield  {journal} {\bibinfo
  {journal} {Nature}\ }\textbf {\bibinfo {volume} {574}},\ \bibinfo {pages}
  {505} (\bibinfo {year} {2019})}\BibitemShut {NoStop}%
\bibitem [{\citenamefont {Zhong}\ \emph {et~al.}(2020)\citenamefont {Zhong},
  \citenamefont {Wang}, \citenamefont {Deng}, \citenamefont {Chen},
  \citenamefont {Peng}, \citenamefont {Luo}, \citenamefont {Qin}, \citenamefont
  {Wu}, \citenamefont {Ding}, \citenamefont {Hu} \emph
  {et~al.}}]{zhong2020quantum}%
  \BibitemOpen
  \bibfield  {author} {\bibinfo {author} {\bibfnamefont {H.-S.}\ \bibnamefont
  {Zhong}}, \bibinfo {author} {\bibfnamefont {H.}~\bibnamefont {Wang}},
  \bibinfo {author} {\bibfnamefont {Y.-H.}\ \bibnamefont {Deng}}, \bibinfo
  {author} {\bibfnamefont {M.-C.}\ \bibnamefont {Chen}}, \bibinfo {author}
  {\bibfnamefont {L.-C.}\ \bibnamefont {Peng}}, \bibinfo {author}
  {\bibfnamefont {Y.-H.}\ \bibnamefont {Luo}}, \bibinfo {author} {\bibfnamefont
  {J.}~\bibnamefont {Qin}}, \bibinfo {author} {\bibfnamefont {D.}~\bibnamefont
  {Wu}}, \bibinfo {author} {\bibfnamefont {X.}~\bibnamefont {Ding}}, \bibinfo
  {author} {\bibfnamefont {Y.}~\bibnamefont {Hu}}, \emph {et~al.},\ }\href
  {https://doi.org/10.1126/science.abe8770} {\bibfield  {journal} {\bibinfo
  {journal} {Science}\ }\textbf {\bibinfo {volume} {370}},\ \bibinfo {pages}
  {1460} (\bibinfo {year} {2020})}\BibitemShut {NoStop}%
\bibitem [{\citenamefont {Cerezo}\ \emph {et~al.}(2021)\citenamefont {Cerezo},
  \citenamefont {Arrasmith}, \citenamefont {Babbush}, \citenamefont {Benjamin},
  \citenamefont {Endo}, \citenamefont {Fujii}, \citenamefont {McClean},
  \citenamefont {Mitarai}, \citenamefont {Yuan}, \citenamefont {Cincio} \emph
  {et~al.}}]{cerezo2021variational}%
  \BibitemOpen
  \bibfield  {author} {\bibinfo {author} {\bibfnamefont {M.}~\bibnamefont
  {Cerezo}}, \bibinfo {author} {\bibfnamefont {A.}~\bibnamefont {Arrasmith}},
  \bibinfo {author} {\bibfnamefont {R.}~\bibnamefont {Babbush}}, \bibinfo
  {author} {\bibfnamefont {S.~C.}\ \bibnamefont {Benjamin}}, \bibinfo {author}
  {\bibfnamefont {S.}~\bibnamefont {Endo}}, \bibinfo {author} {\bibfnamefont
  {K.}~\bibnamefont {Fujii}}, \bibinfo {author} {\bibfnamefont {J.~R.}\
  \bibnamefont {McClean}}, \bibinfo {author} {\bibfnamefont {K.}~\bibnamefont
  {Mitarai}}, \bibinfo {author} {\bibfnamefont {X.}~\bibnamefont {Yuan}},
  \bibinfo {author} {\bibfnamefont {L.}~\bibnamefont {Cincio}}, \emph
  {et~al.},\ }\href {https://doi.org/10.1038/s42254-021-00348-9} {\bibfield
  {journal} {\bibinfo  {journal} {Nature Reviews Physics}\ }\textbf {\bibinfo
  {volume} {3}},\ \bibinfo {pages} {625–644} (\bibinfo {year}
  {2021})}\BibitemShut {NoStop}%
\bibitem [{\citenamefont {Bharti}\ \emph {et~al.}(2022)\citenamefont {Bharti},
  \citenamefont {Cervera-Lierta}, \citenamefont {Kyaw}, \citenamefont {Haug},
  \citenamefont {Alperin-Lea}, \citenamefont {Anand}, \citenamefont {Degroote},
  \citenamefont {Heimonen}, \citenamefont {Kottmann}, \citenamefont {Menke},
  \citenamefont {Mok}, \citenamefont {Sim}, \citenamefont {Kwek},\ and\
  \citenamefont {Aspuru-Guzik}}]{bharti2021noisy}%
  \BibitemOpen
  \bibfield  {author} {\bibinfo {author} {\bibfnamefont {K.}~\bibnamefont
  {Bharti}}, \bibinfo {author} {\bibfnamefont {A.}~\bibnamefont
  {Cervera-Lierta}}, \bibinfo {author} {\bibfnamefont {T.~H.}\ \bibnamefont
  {Kyaw}}, \bibinfo {author} {\bibfnamefont {T.}~\bibnamefont {Haug}}, \bibinfo
  {author} {\bibfnamefont {S.}~\bibnamefont {Alperin-Lea}}, \bibinfo {author}
  {\bibfnamefont {A.}~\bibnamefont {Anand}}, \bibinfo {author} {\bibfnamefont
  {M.}~\bibnamefont {Degroote}}, \bibinfo {author} {\bibfnamefont
  {H.}~\bibnamefont {Heimonen}}, \bibinfo {author} {\bibfnamefont {J.~S.}\
  \bibnamefont {Kottmann}}, \bibinfo {author} {\bibfnamefont {T.}~\bibnamefont
  {Menke}}, \bibinfo {author} {\bibfnamefont {W.-K.}\ \bibnamefont {Mok}},
  \bibinfo {author} {\bibfnamefont {S.}~\bibnamefont {Sim}}, \bibinfo {author}
  {\bibfnamefont {L.-C.}\ \bibnamefont {Kwek}},\ and\ \bibinfo {author}
  {\bibfnamefont {A.}~\bibnamefont {Aspuru-Guzik}},\ }\href
  {https://doi.org/10.1103/RevModPhys.94.015004} {\bibfield  {journal}
  {\bibinfo  {journal} {Reviews of Modern Physics}\ }\textbf {\bibinfo {volume}
  {94}},\ \bibinfo {pages} {015004} (\bibinfo {year} {2022})}\BibitemShut
  {NoStop}%
\bibitem [{\citenamefont {Biamonte}\ \emph {et~al.}(2017)\citenamefont
  {Biamonte}, \citenamefont {Wittek}, \citenamefont {Pancotti}, \citenamefont
  {Rebentrost}, \citenamefont {Wiebe},\ and\ \citenamefont
  {Lloyd}}]{biamonte2017quantum}%
  \BibitemOpen
  \bibfield  {author} {\bibinfo {author} {\bibfnamefont {J.}~\bibnamefont
  {Biamonte}}, \bibinfo {author} {\bibfnamefont {P.}~\bibnamefont {Wittek}},
  \bibinfo {author} {\bibfnamefont {N.}~\bibnamefont {Pancotti}}, \bibinfo
  {author} {\bibfnamefont {P.}~\bibnamefont {Rebentrost}}, \bibinfo {author}
  {\bibfnamefont {N.}~\bibnamefont {Wiebe}},\ and\ \bibinfo {author}
  {\bibfnamefont {S.}~\bibnamefont {Lloyd}},\ }\href
  {https://doi.org/10.1038/nature23474} {\bibfield  {journal} {\bibinfo
  {journal} {Nature}\ }\textbf {\bibinfo {volume} {549}},\ \bibinfo {pages}
  {195} (\bibinfo {year} {2017})}\BibitemShut {NoStop}%
\bibitem [{\citenamefont {Schuld}\ and\ \citenamefont
  {Petruccione}(2018)}]{schuld2018supervised}%
  \BibitemOpen
  \bibfield  {author} {\bibinfo {author} {\bibfnamefont {M.}~\bibnamefont
  {Schuld}}\ and\ \bibinfo {author} {\bibfnamefont {F.}~\bibnamefont
  {Petruccione}},\ }\href {https://doi.org/10.1007/978-3-319-96424-9} {\emph
  {\bibinfo {title} {Supervised learning with quantum computers}}},\
  Vol.~\bibinfo {volume} {17}\ (\bibinfo  {publisher} {Springer},\ \bibinfo
  {year} {2018})\BibitemShut {NoStop}%
\bibitem [{\citenamefont {Wiebe}\ \emph {et~al.}(2012)\citenamefont {Wiebe},
  \citenamefont {Braun},\ and\ \citenamefont {Lloyd}}]{wiebe2012quantum}%
  \BibitemOpen
  \bibfield  {author} {\bibinfo {author} {\bibfnamefont {N.}~\bibnamefont
  {Wiebe}}, \bibinfo {author} {\bibfnamefont {D.}~\bibnamefont {Braun}},\ and\
  \bibinfo {author} {\bibfnamefont {S.}~\bibnamefont {Lloyd}},\ }\href
  {https://doi.org/10.1103/PhysRevLett.109.050505} {\bibfield  {journal}
  {\bibinfo  {journal} {Physical Review Letters}\ }\textbf {\bibinfo {volume}
  {109}},\ \bibinfo {pages} {050505} (\bibinfo {year} {2012})}\BibitemShut
  {NoStop}%
\bibitem [{\citenamefont {Lloyd}\ \emph {et~al.}(2013)\citenamefont {Lloyd},
  \citenamefont {Mohseni},\ and\ \citenamefont {Rebentrost}}]{lloyd:2013ml}%
  \BibitemOpen
  \bibfield  {author} {\bibinfo {author} {\bibfnamefont {S.}~\bibnamefont
  {Lloyd}}, \bibinfo {author} {\bibfnamefont {M.}~\bibnamefont {Mohseni}},\
  and\ \bibinfo {author} {\bibfnamefont {P.}~\bibnamefont {Rebentrost}},\
  }\href {https://doi.org/10.48550/arXiv.1307.0411} {\bibfield  {journal}
  {\bibinfo  {journal} {arXiv preprint arXiv:1307.0411}\ } (\bibinfo {year}
  {2013})}\BibitemShut {NoStop}%
\bibitem [{\citenamefont {Rebentrost}\ \emph {et~al.}(2014)\citenamefont
  {Rebentrost}, \citenamefont {Mohseni},\ and\ \citenamefont
  {Lloyd}}]{Rebentrost:2014svm}%
  \BibitemOpen
  \bibfield  {author} {\bibinfo {author} {\bibfnamefont {P.}~\bibnamefont
  {Rebentrost}}, \bibinfo {author} {\bibfnamefont {M.}~\bibnamefont
  {Mohseni}},\ and\ \bibinfo {author} {\bibfnamefont {S.}~\bibnamefont
  {Lloyd}},\ }\href {https://doi.org/10.1103/physrevlett.113.130503} {\bibfield
   {journal} {\bibinfo  {journal} {Physical Review Letters}\ }\textbf {\bibinfo
  {volume} {113}},\ \bibinfo {pages} {130503} (\bibinfo {year}
  {2014})}\BibitemShut {NoStop}%
\bibitem [{\citenamefont {Kerenidis}\ and\ \citenamefont
  {Prakash}(2020)}]{kerenidis2020quantum}%
  \BibitemOpen
  \bibfield  {author} {\bibinfo {author} {\bibfnamefont {I.}~\bibnamefont
  {Kerenidis}}\ and\ \bibinfo {author} {\bibfnamefont {A.}~\bibnamefont
  {Prakash}},\ }\href {https://doi.org/10.1103/PhysRevA.101.022316} {\bibfield
  {journal} {\bibinfo  {journal} {Physical Review A}\ }\textbf {\bibinfo
  {volume} {101}},\ \bibinfo {pages} {022316} (\bibinfo {year}
  {2020})}\BibitemShut {NoStop}%
\bibitem [{\citenamefont {Harrow}\ \emph {et~al.}(2009)\citenamefont {Harrow},
  \citenamefont {Hassidim},\ and\ \citenamefont {Lloyd}}]{harrow2009quantum}%
  \BibitemOpen
  \bibfield  {author} {\bibinfo {author} {\bibfnamefont {A.~W.}\ \bibnamefont
  {Harrow}}, \bibinfo {author} {\bibfnamefont {A.}~\bibnamefont {Hassidim}},\
  and\ \bibinfo {author} {\bibfnamefont {S.}~\bibnamefont {Lloyd}},\ }\href
  {https://doi.org/10.1103/PhysRevLett.103.150502} {\bibfield  {journal}
  {\bibinfo  {journal} {Physical Review Letters}\ }\textbf {\bibinfo {volume}
  {103}},\ \bibinfo {pages} {150502} (\bibinfo {year} {2009})}\BibitemShut
  {NoStop}%
\bibitem [{\citenamefont {Benedetti}\ \emph
  {et~al.}(2019{\natexlab{a}})\citenamefont {Benedetti}, \citenamefont {Lloyd},
  \citenamefont {Sack},\ and\ \citenamefont
  {Fiorentini}}]{benedetti2019parameterized}%
  \BibitemOpen
  \bibfield  {author} {\bibinfo {author} {\bibfnamefont {M.}~\bibnamefont
  {Benedetti}}, \bibinfo {author} {\bibfnamefont {E.}~\bibnamefont {Lloyd}},
  \bibinfo {author} {\bibfnamefont {S.}~\bibnamefont {Sack}},\ and\ \bibinfo
  {author} {\bibfnamefont {M.}~\bibnamefont {Fiorentini}},\ }\href
  {https://doi.org/10.1088/2058-9565/ab4eb5} {\bibfield  {journal} {\bibinfo
  {journal} {Quantum Science and Technology}\ }\textbf {\bibinfo {volume}
  {4}},\ \bibinfo {pages} {043001} (\bibinfo {year}
  {2019}{\natexlab{a}})}\BibitemShut {NoStop}%
\bibitem [{\citenamefont {Sim}\ \emph {et~al.}(2019)\citenamefont {Sim},
  \citenamefont {Johnson},\ and\ \citenamefont
  {Aspuru-Guzik}}]{sim2019expressibility}%
  \BibitemOpen
  \bibfield  {author} {\bibinfo {author} {\bibfnamefont {S.}~\bibnamefont
  {Sim}}, \bibinfo {author} {\bibfnamefont {P.~D.}\ \bibnamefont {Johnson}},\
  and\ \bibinfo {author} {\bibfnamefont {A.}~\bibnamefont {Aspuru-Guzik}},\
  }\href {https://doi.org/10.1002/qute.201900070} {\bibfield  {journal}
  {\bibinfo  {journal} {Advanced Quantum Technologies}\ }\textbf {\bibinfo
  {volume} {2}},\ \bibinfo {pages} {1900070} (\bibinfo {year}
  {2019})}\BibitemShut {NoStop}%
\bibitem [{\citenamefont {Bravo-Prieto}\ \emph {et~al.}(2020)\citenamefont
  {Bravo-Prieto}, \citenamefont {Lumbreras-Zarapico}, \citenamefont
  {Tagliacozzo},\ and\ \citenamefont {Latorre}}]{bravo2020scaling}%
  \BibitemOpen
  \bibfield  {author} {\bibinfo {author} {\bibfnamefont {C.}~\bibnamefont
  {Bravo-Prieto}}, \bibinfo {author} {\bibfnamefont {J.}~\bibnamefont
  {Lumbreras-Zarapico}}, \bibinfo {author} {\bibfnamefont {L.}~\bibnamefont
  {Tagliacozzo}},\ and\ \bibinfo {author} {\bibfnamefont {J.~I.}\ \bibnamefont
  {Latorre}},\ }\href {https://doi.org/10.22331/q-2020-05-28-272} {\bibfield
  {journal} {\bibinfo  {journal} {Quantum}\ }\textbf {\bibinfo {volume} {4}},\
  \bibinfo {pages} {272} (\bibinfo {year} {2020})}\BibitemShut {NoStop}%
\bibitem [{\citenamefont {Larocca}\ \emph {et~al.}(2021)\citenamefont
  {Larocca}, \citenamefont {Ju}, \citenamefont {Garc{\'\i}a-Mart{\'\i}n},
  \citenamefont {Coles},\ and\ \citenamefont {Cerezo}}]{larocca2021theory}%
  \BibitemOpen
  \bibfield  {author} {\bibinfo {author} {\bibfnamefont {M.}~\bibnamefont
  {Larocca}}, \bibinfo {author} {\bibfnamefont {N.}~\bibnamefont {Ju}},
  \bibinfo {author} {\bibfnamefont {D.}~\bibnamefont
  {Garc{\'\i}a-Mart{\'\i}n}}, \bibinfo {author} {\bibfnamefont {P.~J.}\
  \bibnamefont {Coles}},\ and\ \bibinfo {author} {\bibfnamefont
  {M.}~\bibnamefont {Cerezo}},\ }\href
  {https://doi.org/10.48550/arXiv.2109.11676} {\bibfield  {journal} {\bibinfo
  {journal} {arXiv preprint arXiv:2109.11676}\ } (\bibinfo {year}
  {2021})}\BibitemShut {NoStop}%
\bibitem [{\citenamefont {Schuld}\ \emph {et~al.}(2021)\citenamefont {Schuld},
  \citenamefont {Sweke},\ and\ \citenamefont {Meyer}}]{schuld2021effect}%
  \BibitemOpen
  \bibfield  {author} {\bibinfo {author} {\bibfnamefont {M.}~\bibnamefont
  {Schuld}}, \bibinfo {author} {\bibfnamefont {R.}~\bibnamefont {Sweke}},\ and\
  \bibinfo {author} {\bibfnamefont {J.~J.}\ \bibnamefont {Meyer}},\ }\href
  {https://doi.org/10.1103/PhysRevA.103.032430} {\bibfield  {journal} {\bibinfo
   {journal} {Physical Review A}\ }\textbf {\bibinfo {volume} {103}},\ \bibinfo
  {pages} {032430} (\bibinfo {year} {2021})}\BibitemShut {NoStop}%
\bibitem [{\citenamefont {Goto}\ \emph {et~al.}(2021)\citenamefont {Goto},
  \citenamefont {Tran},\ and\ \citenamefont {Nakajima}}]{goto2021universal}%
  \BibitemOpen
  \bibfield  {author} {\bibinfo {author} {\bibfnamefont {T.}~\bibnamefont
  {Goto}}, \bibinfo {author} {\bibfnamefont {Q.~H.}\ \bibnamefont {Tran}},\
  and\ \bibinfo {author} {\bibfnamefont {K.}~\bibnamefont {Nakajima}},\ }\href
  {https://doi.org/10.1103/PhysRevLett.127.090506} {\bibfield  {journal}
  {\bibinfo  {journal} {Physical Review Letters}\ }\textbf {\bibinfo {volume}
  {127}},\ \bibinfo {pages} {090506} (\bibinfo {year} {2021})}\BibitemShut
  {NoStop}%
\bibitem [{\citenamefont {P{\'e}rez-Salinas}\ \emph {et~al.}(2021)\citenamefont
  {P{\'e}rez-Salinas}, \citenamefont {L{\'o}pez-N{\'u}{\~n}ez}, \citenamefont
  {Garc{\'\i}a-S{\'a}ez}, \citenamefont {Forn-D{\'\i}az},\ and\ \citenamefont
  {Latorre}}]{perez2021one}%
  \BibitemOpen
  \bibfield  {author} {\bibinfo {author} {\bibfnamefont {A.}~\bibnamefont
  {P{\'e}rez-Salinas}}, \bibinfo {author} {\bibfnamefont {D.}~\bibnamefont
  {L{\'o}pez-N{\'u}{\~n}ez}}, \bibinfo {author} {\bibfnamefont
  {A.}~\bibnamefont {Garc{\'\i}a-S{\'a}ez}}, \bibinfo {author} {\bibfnamefont
  {P.}~\bibnamefont {Forn-D{\'\i}az}},\ and\ \bibinfo {author} {\bibfnamefont
  {J.~I.}\ \bibnamefont {Latorre}},\ }\href
  {https://doi.org/10.1103/PhysRevA.104.012405} {\bibfield  {journal} {\bibinfo
   {journal} {Physical Review A}\ }\textbf {\bibinfo {volume} {104}},\ \bibinfo
  {pages} {012405} (\bibinfo {year} {2021})}\BibitemShut {NoStop}%
\bibitem [{\citenamefont {Havl{\'\i}{\v{c}}ek}\ \emph
  {et~al.}(2019)\citenamefont {Havl{\'\i}{\v{c}}ek}, \citenamefont
  {C{\'o}rcoles}, \citenamefont {Temme}, \citenamefont {Harrow}, \citenamefont
  {Kandala}, \citenamefont {Chow},\ and\ \citenamefont
  {Gambetta}}]{havlivcek2019supervised}%
  \BibitemOpen
  \bibfield  {author} {\bibinfo {author} {\bibfnamefont {V.}~\bibnamefont
  {Havl{\'\i}{\v{c}}ek}}, \bibinfo {author} {\bibfnamefont {A.~D.}\
  \bibnamefont {C{\'o}rcoles}}, \bibinfo {author} {\bibfnamefont
  {K.}~\bibnamefont {Temme}}, \bibinfo {author} {\bibfnamefont {A.~W.}\
  \bibnamefont {Harrow}}, \bibinfo {author} {\bibfnamefont {A.}~\bibnamefont
  {Kandala}}, \bibinfo {author} {\bibfnamefont {J.~M.}\ \bibnamefont {Chow}},\
  and\ \bibinfo {author} {\bibfnamefont {J.~M.}\ \bibnamefont {Gambetta}},\
  }\href {https://doi.org/10.1038/s41586-019-0980-2} {\bibfield  {journal}
  {\bibinfo  {journal} {Nature}\ }\textbf {\bibinfo {volume} {567}},\ \bibinfo
  {pages} {209} (\bibinfo {year} {2019})}\BibitemShut {NoStop}%
\bibitem [{\citenamefont {Schuld}\ \emph {et~al.}(2020)\citenamefont {Schuld},
  \citenamefont {Bocharov}, \citenamefont {Svore},\ and\ \citenamefont
  {Wiebe}}]{Schuld:2020circuit}%
  \BibitemOpen
  \bibfield  {author} {\bibinfo {author} {\bibfnamefont {M.}~\bibnamefont
  {Schuld}}, \bibinfo {author} {\bibfnamefont {A.}~\bibnamefont {Bocharov}},
  \bibinfo {author} {\bibfnamefont {K.~M.}\ \bibnamefont {Svore}},\ and\
  \bibinfo {author} {\bibfnamefont {N.}~\bibnamefont {Wiebe}},\ }\href
  {https://doi.org/10.1103/physreva.101.032308} {\bibfield  {journal} {\bibinfo
   {journal} {Physical Review A}\ }\textbf {\bibinfo {volume} {101}},\ \bibinfo
  {pages} {032308} (\bibinfo {year} {2020})}\BibitemShut {NoStop}%
\bibitem [{\citenamefont {P{\'{e}}rez-Salinas}\ \emph
  {et~al.}(2020)\citenamefont {P{\'{e}}rez-Salinas}, \citenamefont
  {Cervera-Lierta}, \citenamefont {Gil-Fuster},\ and\ \citenamefont
  {Latorre}}]{perezsalinas:2020reuploading}%
  \BibitemOpen
  \bibfield  {author} {\bibinfo {author} {\bibfnamefont {A.}~\bibnamefont
  {P{\'{e}}rez-Salinas}}, \bibinfo {author} {\bibfnamefont {A.}~\bibnamefont
  {Cervera-Lierta}}, \bibinfo {author} {\bibfnamefont {E.}~\bibnamefont
  {Gil-Fuster}},\ and\ \bibinfo {author} {\bibfnamefont {J.~I.}\ \bibnamefont
  {Latorre}},\ }\href {https://doi.org/10.22331/q-2020-02-06-226} {\bibfield
  {journal} {\bibinfo  {journal} {Quantum}\ }\textbf {\bibinfo {volume} {4}},\
  \bibinfo {pages} {226} (\bibinfo {year} {2020})}\BibitemShut {NoStop}%
\bibitem [{\citenamefont {Dutta}\ \emph {et~al.}(2022)\citenamefont {Dutta},
  \citenamefont {P{\'e}rez-Salinas}, \citenamefont {Cheng}, \citenamefont
  {Latorre},\ and\ \citenamefont {Mukherjee}}]{dutta2021realization}%
  \BibitemOpen
  \bibfield  {author} {\bibinfo {author} {\bibfnamefont {T.}~\bibnamefont
  {Dutta}}, \bibinfo {author} {\bibfnamefont {A.}~\bibnamefont
  {P{\'e}rez-Salinas}}, \bibinfo {author} {\bibfnamefont {J.~P.~S.}\
  \bibnamefont {Cheng}}, \bibinfo {author} {\bibfnamefont {J.~I.}\ \bibnamefont
  {Latorre}},\ and\ \bibinfo {author} {\bibfnamefont {M.}~\bibnamefont
  {Mukherjee}},\ }\href {https://doi.org/10.1103/PhysRevA.106.012411}
  {\bibfield  {journal} {\bibinfo  {journal} {Physical Review A}\ }\textbf
  {\bibinfo {volume} {106}},\ \bibinfo {pages} {012411} (\bibinfo {year}
  {2022})}\BibitemShut {NoStop}%
\bibitem [{\citenamefont {Romero}\ \emph {et~al.}(2017)\citenamefont {Romero},
  \citenamefont {Olson},\ and\ \citenamefont
  {Aspuru-Guzik}}]{romero2017quantum}%
  \BibitemOpen
  \bibfield  {author} {\bibinfo {author} {\bibfnamefont {J.}~\bibnamefont
  {Romero}}, \bibinfo {author} {\bibfnamefont {J.~P.}\ \bibnamefont {Olson}},\
  and\ \bibinfo {author} {\bibfnamefont {A.}~\bibnamefont {Aspuru-Guzik}},\
  }\href {https://doi.org/10.1088/2058-9565/aa8072} {\bibfield  {journal}
  {\bibinfo  {journal} {Quantum Science and Technology}\ }\textbf {\bibinfo
  {volume} {2}},\ \bibinfo {pages} {045001} (\bibinfo {year}
  {2017})}\BibitemShut {NoStop}%
\bibitem [{\citenamefont {Pepper}\ \emph {et~al.}(2019)\citenamefont {Pepper},
  \citenamefont {Tischler},\ and\ \citenamefont
  {Pryde}}]{pepper2019experimental}%
  \BibitemOpen
  \bibfield  {author} {\bibinfo {author} {\bibfnamefont {A.}~\bibnamefont
  {Pepper}}, \bibinfo {author} {\bibfnamefont {N.}~\bibnamefont {Tischler}},\
  and\ \bibinfo {author} {\bibfnamefont {G.~J.}\ \bibnamefont {Pryde}},\ }\href
  {https://doi.org/10.1103/PhysRevLett.122.060501} {\bibfield  {journal}
  {\bibinfo  {journal} {Physical Review Letters}\ }\textbf {\bibinfo {volume}
  {122}},\ \bibinfo {pages} {060501} (\bibinfo {year} {2019})}\BibitemShut
  {NoStop}%
\bibitem [{\citenamefont {Bravo-Prieto}(2021)}]{bravo2021quantum}%
  \BibitemOpen
  \bibfield  {author} {\bibinfo {author} {\bibfnamefont {C.}~\bibnamefont
  {Bravo-Prieto}},\ }\href {https://doi.org/10.1088/2632-2153/ac0616}
  {\bibfield  {journal} {\bibinfo  {journal} {Machine Learning: Science and
  Technology}\ }\textbf {\bibinfo {volume} {2}},\ \bibinfo {pages} {035028}
  (\bibinfo {year} {2021})}\BibitemShut {NoStop}%
\bibitem [{\citenamefont {Cao}\ and\ \citenamefont
  {Wang}(2021)}]{cao2021noise}%
  \BibitemOpen
  \bibfield  {author} {\bibinfo {author} {\bibfnamefont {C.}~\bibnamefont
  {Cao}}\ and\ \bibinfo {author} {\bibfnamefont {X.}~\bibnamefont {Wang}},\
  }\href {https://doi.org/10.1103/PhysRevApplied.15.054012} {\bibfield
  {journal} {\bibinfo  {journal} {Physical Review Applied}\ }\textbf {\bibinfo
  {volume} {15}},\ \bibinfo {pages} {054012} (\bibinfo {year}
  {2021})}\BibitemShut {NoStop}%
\bibitem [{\citenamefont {Benedetti}\ \emph
  {et~al.}(2019{\natexlab{b}})\citenamefont {Benedetti}, \citenamefont
  {Garcia-Pintos}, \citenamefont {Perdomo}, \citenamefont {Leyton-Ortega},
  \citenamefont {Nam},\ and\ \citenamefont
  {Perdomo-Ortiz}}]{benedetti2019generative}%
  \BibitemOpen
  \bibfield  {author} {\bibinfo {author} {\bibfnamefont {M.}~\bibnamefont
  {Benedetti}}, \bibinfo {author} {\bibfnamefont {D.}~\bibnamefont
  {Garcia-Pintos}}, \bibinfo {author} {\bibfnamefont {O.}~\bibnamefont
  {Perdomo}}, \bibinfo {author} {\bibfnamefont {V.}~\bibnamefont
  {Leyton-Ortega}}, \bibinfo {author} {\bibfnamefont {Y.}~\bibnamefont {Nam}},\
  and\ \bibinfo {author} {\bibfnamefont {A.}~\bibnamefont {Perdomo-Ortiz}},\
  }\href {https://doi.org/10.1038/s41534-019-0157-8} {\bibfield  {journal}
  {\bibinfo  {journal} {npj Quantum Information}\ }\textbf {\bibinfo {volume}
  {5}},\ \bibinfo {pages} {1} (\bibinfo {year}
  {2019}{\natexlab{b}})}\BibitemShut {NoStop}%
\bibitem [{\citenamefont {Hamilton}\ \emph {et~al.}(2019)\citenamefont
  {Hamilton}, \citenamefont {Dumitrescu},\ and\ \citenamefont
  {Pooser}}]{hamilton2019generative}%
  \BibitemOpen
  \bibfield  {author} {\bibinfo {author} {\bibfnamefont {K.~E.}\ \bibnamefont
  {Hamilton}}, \bibinfo {author} {\bibfnamefont {E.~F.}\ \bibnamefont
  {Dumitrescu}},\ and\ \bibinfo {author} {\bibfnamefont {R.~C.}\ \bibnamefont
  {Pooser}},\ }\href {https://doi.org/10.1103/PhysRevA.99.062323} {\bibfield
  {journal} {\bibinfo  {journal} {Physical Review A}\ }\textbf {\bibinfo
  {volume} {99}},\ \bibinfo {pages} {062323} (\bibinfo {year}
  {2019})}\BibitemShut {NoStop}%
\bibitem [{\citenamefont {Coyle}\ \emph {et~al.}(2020)\citenamefont {Coyle},
  \citenamefont {Mills}, \citenamefont {Danos},\ and\ \citenamefont
  {Kashefi}}]{coyle2020born}%
  \BibitemOpen
  \bibfield  {author} {\bibinfo {author} {\bibfnamefont {B.}~\bibnamefont
  {Coyle}}, \bibinfo {author} {\bibfnamefont {D.}~\bibnamefont {Mills}},
  \bibinfo {author} {\bibfnamefont {V.}~\bibnamefont {Danos}},\ and\ \bibinfo
  {author} {\bibfnamefont {E.}~\bibnamefont {Kashefi}},\ }\href
  {https://doi.org/10.1038/s41534-020-00288-9} {\bibfield  {journal} {\bibinfo
  {journal} {npj Quantum Information}\ }\textbf {\bibinfo {volume} {6}},\
  \bibinfo {pages} {1} (\bibinfo {year} {2020})}\BibitemShut {NoStop}%
\bibitem [{\citenamefont {Dallaire-Demers}\ and\ \citenamefont
  {Killoran}(2018)}]{dallaire2018quantum}%
  \BibitemOpen
  \bibfield  {author} {\bibinfo {author} {\bibfnamefont {P.-L.}\ \bibnamefont
  {Dallaire-Demers}}\ and\ \bibinfo {author} {\bibfnamefont {N.}~\bibnamefont
  {Killoran}},\ }\href {https://doi.org/10.1103/PhysRevA.98.012324} {\bibfield
  {journal} {\bibinfo  {journal} {Physical Review A}\ }\textbf {\bibinfo
  {volume} {98}},\ \bibinfo {pages} {012324} (\bibinfo {year}
  {2018})}\BibitemShut {NoStop}%
\bibitem [{\citenamefont {Lloyd}\ and\ \citenamefont
  {Weedbrook}(2018)}]{lloyd2018quantum}%
  \BibitemOpen
  \bibfield  {author} {\bibinfo {author} {\bibfnamefont {S.}~\bibnamefont
  {Lloyd}}\ and\ \bibinfo {author} {\bibfnamefont {C.}~\bibnamefont
  {Weedbrook}},\ }\href {https://doi.org/10.1103/PhysRevLett.121.040502}
  {\bibfield  {journal} {\bibinfo  {journal} {Physical Review Letters}\
  }\textbf {\bibinfo {volume} {121}},\ \bibinfo {pages} {040502} (\bibinfo
  {year} {2018})}\BibitemShut {NoStop}%
\bibitem [{\citenamefont {Goodfellow}\ \emph {et~al.}(2020)\citenamefont
  {Goodfellow}, \citenamefont {Pouget-Abadie}, \citenamefont {Mirza},
  \citenamefont {Xu}, \citenamefont {Warde-Farley}, \citenamefont {Ozair},
  \citenamefont {Courville},\ and\ \citenamefont
  {Bengio}}]{goodfellow2014generative}%
  \BibitemOpen
  \bibfield  {author} {\bibinfo {author} {\bibfnamefont {I.}~\bibnamefont
  {Goodfellow}}, \bibinfo {author} {\bibfnamefont {J.}~\bibnamefont
  {Pouget-Abadie}}, \bibinfo {author} {\bibfnamefont {M.}~\bibnamefont
  {Mirza}}, \bibinfo {author} {\bibfnamefont {B.}~\bibnamefont {Xu}}, \bibinfo
  {author} {\bibfnamefont {D.}~\bibnamefont {Warde-Farley}}, \bibinfo {author}
  {\bibfnamefont {S.}~\bibnamefont {Ozair}}, \bibinfo {author} {\bibfnamefont
  {A.}~\bibnamefont {Courville}},\ and\ \bibinfo {author} {\bibfnamefont
  {Y.}~\bibnamefont {Bengio}},\ }\href {https://doi.org/10.1145/3422622}
  {\bibfield  {journal} {\bibinfo  {journal} {Communications of the ACM}\
  }\textbf {\bibinfo {volume} {63}},\ \bibinfo {pages} {139–144} (\bibinfo
  {year} {2020})}\BibitemShut {NoStop}%
\bibitem [{\citenamefont {Zoufal}\ \emph {et~al.}(2019)\citenamefont {Zoufal},
  \citenamefont {Lucchi},\ and\ \citenamefont {Woerner}}]{zoufal2019quantum}%
  \BibitemOpen
  \bibfield  {author} {\bibinfo {author} {\bibfnamefont {C.}~\bibnamefont
  {Zoufal}}, \bibinfo {author} {\bibfnamefont {A.}~\bibnamefont {Lucchi}},\
  and\ \bibinfo {author} {\bibfnamefont {S.}~\bibnamefont {Woerner}},\ }\href
  {https://doi.org/10.1038/s41534-019-0223-2} {\bibfield  {journal} {\bibinfo
  {journal} {npj Quantum Information}\ }\textbf {\bibinfo {volume} {5}},\
  \bibinfo {pages} {1} (\bibinfo {year} {2019})}\BibitemShut {NoStop}%
\bibitem [{\citenamefont {Zeng}\ \emph {et~al.}(2019)\citenamefont {Zeng},
  \citenamefont {Wu}, \citenamefont {Liu}, \citenamefont {Wang},\ and\
  \citenamefont {Hu}}]{zeng2019learning}%
  \BibitemOpen
  \bibfield  {author} {\bibinfo {author} {\bibfnamefont {J.}~\bibnamefont
  {Zeng}}, \bibinfo {author} {\bibfnamefont {Y.}~\bibnamefont {Wu}}, \bibinfo
  {author} {\bibfnamefont {J.-G.}\ \bibnamefont {Liu}}, \bibinfo {author}
  {\bibfnamefont {L.}~\bibnamefont {Wang}},\ and\ \bibinfo {author}
  {\bibfnamefont {J.}~\bibnamefont {Hu}},\ }\href
  {https://doi.org/10.1103/PhysRevA.99.052306} {\bibfield  {journal} {\bibinfo
  {journal} {Physical Review A}\ }\textbf {\bibinfo {volume} {99}},\ \bibinfo
  {pages} {052306} (\bibinfo {year} {2019})}\BibitemShut {NoStop}%
\bibitem [{\citenamefont {Situ}\ \emph {et~al.}(2020)\citenamefont {Situ},
  \citenamefont {He}, \citenamefont {Wang}, \citenamefont {Li},\ and\
  \citenamefont {Zheng}}]{situ2020quantum}%
  \BibitemOpen
  \bibfield  {author} {\bibinfo {author} {\bibfnamefont {H.}~\bibnamefont
  {Situ}}, \bibinfo {author} {\bibfnamefont {Z.}~\bibnamefont {He}}, \bibinfo
  {author} {\bibfnamefont {Y.}~\bibnamefont {Wang}}, \bibinfo {author}
  {\bibfnamefont {L.}~\bibnamefont {Li}},\ and\ \bibinfo {author}
  {\bibfnamefont {S.}~\bibnamefont {Zheng}},\ }\href
  {https://doi.org/10.1016/j.ins.2020.05.127} {\bibfield  {journal} {\bibinfo
  {journal} {Information Sciences}\ }\textbf {\bibinfo {volume} {538}},\
  \bibinfo {pages} {193} (\bibinfo {year} {2020})}\BibitemShut {NoStop}%
\bibitem [{\citenamefont {Hu}\ \emph {et~al.}(2019)\citenamefont {Hu},
  \citenamefont {Wu}, \citenamefont {Cai}, \citenamefont {Ma}, \citenamefont
  {Mu}, \citenamefont {Xu}, \citenamefont {Wang}, \citenamefont {Song},
  \citenamefont {Deng}, \citenamefont {Zou} \emph {et~al.}}]{hu2019quantum}%
  \BibitemOpen
  \bibfield  {author} {\bibinfo {author} {\bibfnamefont {L.}~\bibnamefont
  {Hu}}, \bibinfo {author} {\bibfnamefont {S.-H.}\ \bibnamefont {Wu}}, \bibinfo
  {author} {\bibfnamefont {W.}~\bibnamefont {Cai}}, \bibinfo {author}
  {\bibfnamefont {Y.}~\bibnamefont {Ma}}, \bibinfo {author} {\bibfnamefont
  {X.}~\bibnamefont {Mu}}, \bibinfo {author} {\bibfnamefont {Y.}~\bibnamefont
  {Xu}}, \bibinfo {author} {\bibfnamefont {H.}~\bibnamefont {Wang}}, \bibinfo
  {author} {\bibfnamefont {Y.}~\bibnamefont {Song}}, \bibinfo {author}
  {\bibfnamefont {D.-L.}\ \bibnamefont {Deng}}, \bibinfo {author}
  {\bibfnamefont {C.-L.}\ \bibnamefont {Zou}}, \emph {et~al.},\ }\href
  {https://doi.org/10.1126/sciadv.aav2761} {\bibfield  {journal} {\bibinfo
  {journal} {Science advances}\ }\textbf {\bibinfo {volume} {5}},\ \bibinfo
  {pages} {eaav2761} (\bibinfo {year} {2019})}\BibitemShut {NoStop}%
\bibitem [{\citenamefont {Benedetti}\ \emph
  {et~al.}(2019{\natexlab{c}})\citenamefont {Benedetti}, \citenamefont {Grant},
  \citenamefont {Wossnig},\ and\ \citenamefont
  {Severini}}]{benedetti2019adversarial}%
  \BibitemOpen
  \bibfield  {author} {\bibinfo {author} {\bibfnamefont {M.}~\bibnamefont
  {Benedetti}}, \bibinfo {author} {\bibfnamefont {E.}~\bibnamefont {Grant}},
  \bibinfo {author} {\bibfnamefont {L.}~\bibnamefont {Wossnig}},\ and\ \bibinfo
  {author} {\bibfnamefont {S.}~\bibnamefont {Severini}},\ }\href
  {https://doi.org/10.1088/1367-2630/ab14b5} {\bibfield  {journal} {\bibinfo
  {journal} {New Journal of Physics}\ }\textbf {\bibinfo {volume} {21}},\
  \bibinfo {pages} {043023} (\bibinfo {year} {2019}{\natexlab{c}})}\BibitemShut
  {NoStop}%
\bibitem [{\citenamefont {Romero}\ and\ \citenamefont
  {Aspuru-Guzik}(2021)}]{romero2021variational}%
  \BibitemOpen
  \bibfield  {author} {\bibinfo {author} {\bibfnamefont {J.}~\bibnamefont
  {Romero}}\ and\ \bibinfo {author} {\bibfnamefont {A.}~\bibnamefont
  {Aspuru-Guzik}},\ }\href {https://doi.org/10.1002/qute.202000003} {\bibfield
  {journal} {\bibinfo  {journal} {Advanced Quantum Technologies}\ }\textbf
  {\bibinfo {volume} {4}},\ \bibinfo {pages} {2000003} (\bibinfo {year}
  {2021})}\BibitemShut {NoStop}%
\bibitem [{\citenamefont {Niu}\ \emph {et~al.}(2022)\citenamefont {Niu},
  \citenamefont {Zlokapa}, \citenamefont {Broughton}, \citenamefont {Boixo},
  \citenamefont {Mohseni}, \citenamefont {Smelyanskyi},\ and\ \citenamefont
  {Neven}}]{niu2021entangling}%
  \BibitemOpen
  \bibfield  {author} {\bibinfo {author} {\bibfnamefont {M.~Y.}\ \bibnamefont
  {Niu}}, \bibinfo {author} {\bibfnamefont {A.}~\bibnamefont {Zlokapa}},
  \bibinfo {author} {\bibfnamefont {M.}~\bibnamefont {Broughton}}, \bibinfo
  {author} {\bibfnamefont {S.}~\bibnamefont {Boixo}}, \bibinfo {author}
  {\bibfnamefont {M.}~\bibnamefont {Mohseni}}, \bibinfo {author} {\bibfnamefont
  {V.}~\bibnamefont {Smelyanskyi}},\ and\ \bibinfo {author} {\bibfnamefont
  {H.}~\bibnamefont {Neven}},\ }\href
  {https://doi.org/10.1103/PhysRevLett.128.220505} {\bibfield  {journal}
  {\bibinfo  {journal} {Physical Review Letters}\ }\textbf {\bibinfo {volume}
  {128}},\ \bibinfo {pages} {220505} (\bibinfo {year} {2022})}\BibitemShut
  {NoStop}%
\bibitem [{\citenamefont {Karras}\ \emph {et~al.}(2021)\citenamefont {Karras},
  \citenamefont {Laine},\ and\ \citenamefont {Aila}}]{karras2019style}%
  \BibitemOpen
  \bibfield  {author} {\bibinfo {author} {\bibfnamefont {T.}~\bibnamefont
  {Karras}}, \bibinfo {author} {\bibfnamefont {S.}~\bibnamefont {Laine}},\ and\
  \bibinfo {author} {\bibfnamefont {T.}~\bibnamefont {Aila}},\ }\href
  {https://doi.org/10.1109/TPAMI.2020.2970919} {\bibfield  {journal} {\bibinfo
  {journal} {IEEE Transactions on Pattern Analysis and Machine Intelligence}\
  }\textbf {\bibinfo {volume} {43}},\ \bibinfo {pages} {4217} (\bibinfo {year}
  {2021})}\BibitemShut {NoStop}%
\bibitem [{\citenamefont {Pérez-Salinas}\ \emph {et~al.}(2021)\citenamefont
  {Pérez-Salinas}, \citenamefont {Cruz-Martinez}, \citenamefont {Alhajri},\
  and\ \citenamefont {Carrazza}}]{P_rez_Salinas_2021}%
  \BibitemOpen
  \bibfield  {author} {\bibinfo {author} {\bibfnamefont {A.}~\bibnamefont
  {Pérez-Salinas}}, \bibinfo {author} {\bibfnamefont {J.}~\bibnamefont
  {Cruz-Martinez}}, \bibinfo {author} {\bibfnamefont {A.~A.}\ \bibnamefont
  {Alhajri}},\ and\ \bibinfo {author} {\bibfnamefont {S.}~\bibnamefont
  {Carrazza}},\ }\href {https://doi.org/10.1103/PhysRevD.103.034027} {\bibfield
   {journal} {\bibinfo  {journal} {Physical Review D}\ }\textbf {\bibinfo
  {volume} {103}},\ \bibinfo {pages} {034027} (\bibinfo {year}
  {2021})}\BibitemShut {NoStop}%
\bibitem [{\citenamefont {Guan}\ \emph {et~al.}(2021)\citenamefont {Guan},
  \citenamefont {Perdue}, \citenamefont {Pesah}, \citenamefont {Schuld},
  \citenamefont {Terashi}, \citenamefont {Vallecorsa},\ and\ \citenamefont
  {Vlimant}}]{Guan_2021}%
  \BibitemOpen
  \bibfield  {author} {\bibinfo {author} {\bibfnamefont {W.}~\bibnamefont
  {Guan}}, \bibinfo {author} {\bibfnamefont {G.}~\bibnamefont {Perdue}},
  \bibinfo {author} {\bibfnamefont {A.}~\bibnamefont {Pesah}}, \bibinfo
  {author} {\bibfnamefont {M.}~\bibnamefont {Schuld}}, \bibinfo {author}
  {\bibfnamefont {K.}~\bibnamefont {Terashi}}, \bibinfo {author} {\bibfnamefont
  {S.}~\bibnamefont {Vallecorsa}},\ and\ \bibinfo {author} {\bibfnamefont
  {J.-R.}\ \bibnamefont {Vlimant}},\ }\href
  {https://doi.org/10.1088/2632-2153/abc17d} {\bibfield  {journal} {\bibinfo
  {journal} {Machine Learning: Science and Technology}\ }\textbf {\bibinfo
  {volume} {2}},\ \bibinfo {pages} {011003} (\bibinfo {year}
  {2021})}\BibitemShut {NoStop}%
\bibitem [{\citenamefont {Chang}\ \emph
  {et~al.}(2021{\natexlab{a}})\citenamefont {Chang}, \citenamefont
  {Vallecorsa}, \citenamefont {Combarro},\ and\ \citenamefont
  {Carminati}}]{chang2021quantum}%
  \BibitemOpen
  \bibfield  {author} {\bibinfo {author} {\bibfnamefont {S.~Y.}\ \bibnamefont
  {Chang}}, \bibinfo {author} {\bibfnamefont {S.}~\bibnamefont {Vallecorsa}},
  \bibinfo {author} {\bibfnamefont {E.~F.}\ \bibnamefont {Combarro}},\ and\
  \bibinfo {author} {\bibfnamefont {F.}~\bibnamefont {Carminati}},\ }\href
  {https://doi.org/10.48550/arXiv.2101.11132} {\bibfield  {journal} {\bibinfo
  {journal} {arXiv preprint arXiv:2101.11132}\ } (\bibinfo {year}
  {2021}{\natexlab{a}})}\BibitemShut {NoStop}%
\bibitem [{\citenamefont {Chang}\ \emph
  {et~al.}(2021{\natexlab{b}})\citenamefont {Chang}, \citenamefont {Herbert},
  \citenamefont {Vallecorsa}, \citenamefont {Combarro},\ and\ \citenamefont
  {Duncan}}]{Chang_2021}%
  \BibitemOpen
  \bibfield  {author} {\bibinfo {author} {\bibfnamefont {S.~Y.}\ \bibnamefont
  {Chang}}, \bibinfo {author} {\bibfnamefont {S.}~\bibnamefont {Herbert}},
  \bibinfo {author} {\bibfnamefont {S.}~\bibnamefont {Vallecorsa}}, \bibinfo
  {author} {\bibfnamefont {E.~F.}\ \bibnamefont {Combarro}},\ and\ \bibinfo
  {author} {\bibfnamefont {R.}~\bibnamefont {Duncan}},\ }\href
  {https://doi.org/10.1051/epjconf/202125103050} {\bibfield  {journal}
  {\bibinfo  {journal} {EPJ Web of Conferences}\ }\textbf {\bibinfo {volume}
  {251}},\ \bibinfo {pages} {03050} (\bibinfo {year}
  {2021}{\natexlab{b}})}\BibitemShut {NoStop}%
\bibitem [{\citenamefont {Belis}\ \emph {et~al.}(2021)\citenamefont {Belis},
  \citenamefont {González-Castillo}, \citenamefont {Reissel}, \citenamefont
  {Vallecorsa}, \citenamefont {Combarro}, \citenamefont {Dissertori},\ and\
  \citenamefont {Reiter}}]{Belis_2021}%
  \BibitemOpen
  \bibfield  {author} {\bibinfo {author} {\bibfnamefont {V.}~\bibnamefont
  {Belis}}, \bibinfo {author} {\bibfnamefont {S.}~\bibnamefont
  {González-Castillo}}, \bibinfo {author} {\bibfnamefont {C.}~\bibnamefont
  {Reissel}}, \bibinfo {author} {\bibfnamefont {S.}~\bibnamefont {Vallecorsa}},
  \bibinfo {author} {\bibfnamefont {E.~F.}\ \bibnamefont {Combarro}}, \bibinfo
  {author} {\bibfnamefont {G.}~\bibnamefont {Dissertori}},\ and\ \bibinfo
  {author} {\bibfnamefont {F.}~\bibnamefont {Reiter}},\ }\href
  {https://doi.org/10.1051/epjconf/202125103070} {\bibfield  {journal}
  {\bibinfo  {journal} {EPJ Web of Conferences}\ }\textbf {\bibinfo {volume}
  {251}},\ \bibinfo {pages} {03070} (\bibinfo {year} {2021})}\BibitemShut
  {NoStop}%
\bibitem [{\citenamefont {Khattak}\ \emph {et~al.}(2022)\citenamefont
  {Khattak}, \citenamefont {Vallecorsa}, \citenamefont {Carminati},\ and\
  \citenamefont {Khan}}]{khattak2021fast}%
  \BibitemOpen
  \bibfield  {author} {\bibinfo {author} {\bibfnamefont {G.~R.}\ \bibnamefont
  {Khattak}}, \bibinfo {author} {\bibfnamefont {S.}~\bibnamefont {Vallecorsa}},
  \bibinfo {author} {\bibfnamefont {F.}~\bibnamefont {Carminati}},\ and\
  \bibinfo {author} {\bibfnamefont {G.~M.}\ \bibnamefont {Khan}},\ }\href
  {https://doi.org/10.1140/epjc/s10052-022-10258-4} {\bibfield  {journal}
  {\bibinfo  {journal} {The European Physical Journal C}\ }\textbf {\bibinfo
  {volume} {82}},\ \bibinfo {pages} {1} (\bibinfo {year} {2022})}\BibitemShut
  {NoStop}%
\bibitem [{\citenamefont {Baldi}\ \emph {et~al.}(2021)\citenamefont {Baldi},
  \citenamefont {Blecher}, \citenamefont {Butter}, \citenamefont {Collado},
  \citenamefont {Howard}, \citenamefont {Keilbach}, \citenamefont {Plehn},
  \citenamefont {Kasieczka},\ and\ \citenamefont {Whiteson}}]{baldi2021gan}%
  \BibitemOpen
  \bibfield  {author} {\bibinfo {author} {\bibfnamefont {P.}~\bibnamefont
  {Baldi}}, \bibinfo {author} {\bibfnamefont {L.}~\bibnamefont {Blecher}},
  \bibinfo {author} {\bibfnamefont {A.}~\bibnamefont {Butter}}, \bibinfo
  {author} {\bibfnamefont {J.}~\bibnamefont {Collado}}, \bibinfo {author}
  {\bibfnamefont {J.~N.}\ \bibnamefont {Howard}}, \bibinfo {author}
  {\bibfnamefont {F.}~\bibnamefont {Keilbach}}, \bibinfo {author}
  {\bibfnamefont {T.}~\bibnamefont {Plehn}}, \bibinfo {author} {\bibfnamefont
  {G.}~\bibnamefont {Kasieczka}},\ and\ \bibinfo {author} {\bibfnamefont
  {D.}~\bibnamefont {Whiteson}},\ }\href
  {https://doi.org/10.48550/arXiv.2012.11944} {\bibfield  {journal} {\bibinfo
  {journal} {arXiv preprint arXiv:2012.11944}\ } (\bibinfo {year}
  {2021})}\BibitemShut {NoStop}%
\bibitem [{\citenamefont {Backes}\ \emph {et~al.}(2021)\citenamefont {Backes},
  \citenamefont {Butter}, \citenamefont {Plehn},\ and\ \citenamefont
  {Winterhalder}}]{Backes_2021}%
  \BibitemOpen
  \bibfield  {author} {\bibinfo {author} {\bibfnamefont {M.}~\bibnamefont
  {Backes}}, \bibinfo {author} {\bibfnamefont {A.}~\bibnamefont {Butter}},
  \bibinfo {author} {\bibfnamefont {T.}~\bibnamefont {Plehn}},\ and\ \bibinfo
  {author} {\bibfnamefont {R.}~\bibnamefont {Winterhalder}},\ }\href
  {https://doi.org/10.21468/SciPostPhys.10.4.089} {\bibfield  {journal}
  {\bibinfo  {journal} {SciPost Physics}\ }\textbf {\bibinfo {volume} {10}},\
  \bibinfo {pages} {89} (\bibinfo {year} {2021})}\BibitemShut {NoStop}%
\bibitem [{\citenamefont {Butter}\ and\ \citenamefont
  {Plehn}(2022)}]{butter2020generative}%
  \BibitemOpen
  \bibfield  {author} {\bibinfo {author} {\bibfnamefont {A.}~\bibnamefont
  {Butter}}\ and\ \bibinfo {author} {\bibfnamefont {T.}~\bibnamefont {Plehn}},\
  }in\ \href {https://doi.org/10.1142/9789811234033_0007} {\emph {\bibinfo
  {booktitle} {Artificial Intelligence For High Energy Physics}}}\ (\bibinfo
  {publisher} {World Scientific},\ \bibinfo {year} {2022})\ pp.\ \bibinfo
  {pages} {191--240}\BibitemShut {NoStop}%
\bibitem [{\citenamefont {Butter}\ \emph {et~al.}(2021)\citenamefont {Butter},
  \citenamefont {Diefenbacher}, \citenamefont {Kasieczka}, \citenamefont
  {Nachman},\ and\ \citenamefont {Plehn}}]{Butter_2021}%
  \BibitemOpen
  \bibfield  {author} {\bibinfo {author} {\bibfnamefont {A.}~\bibnamefont
  {Butter}}, \bibinfo {author} {\bibfnamefont {S.}~\bibnamefont
  {Diefenbacher}}, \bibinfo {author} {\bibfnamefont {G.}~\bibnamefont
  {Kasieczka}}, \bibinfo {author} {\bibfnamefont {B.}~\bibnamefont {Nachman}},\
  and\ \bibinfo {author} {\bibfnamefont {T.}~\bibnamefont {Plehn}},\ }\href
  {https://doi.org/10.21468/SciPostPhys.10.6.139} {\bibfield  {journal}
  {\bibinfo  {journal} {SciPost Physics}\ }\textbf {\bibinfo {volume} {10}},\
  \bibinfo {pages} {139} (\bibinfo {year} {2021})}\BibitemShut {NoStop}%
\bibitem [{\citenamefont {Butter}\ \emph {et~al.}(2020)\citenamefont {Butter},
  \citenamefont {Plehn},\ and\ \citenamefont {Winterhalder}}]{Butter_2020}%
  \BibitemOpen
  \bibfield  {author} {\bibinfo {author} {\bibfnamefont {A.}~\bibnamefont
  {Butter}}, \bibinfo {author} {\bibfnamefont {T.}~\bibnamefont {Plehn}},\ and\
  \bibinfo {author} {\bibfnamefont {R.}~\bibnamefont {Winterhalder}},\ }\href
  {https://doi.org/10.21468/SciPostPhysCore.3.2.009} {\bibfield  {journal}
  {\bibinfo  {journal} {SciPost Physics Core}\ }\textbf {\bibinfo {volume}
  {3}},\ \bibinfo {pages} {9} (\bibinfo {year} {2020})}\BibitemShut {NoStop}%
\bibitem [{\citenamefont {Bellagente}\ \emph {et~al.}(2020)\citenamefont
  {Bellagente}, \citenamefont {Butter}, \citenamefont {Kasieczka},
  \citenamefont {Plehn},\ and\ \citenamefont {Winterhalder}}]{Bellagente_2020}%
  \BibitemOpen
  \bibfield  {author} {\bibinfo {author} {\bibfnamefont {M.}~\bibnamefont
  {Bellagente}}, \bibinfo {author} {\bibfnamefont {A.}~\bibnamefont {Butter}},
  \bibinfo {author} {\bibfnamefont {G.}~\bibnamefont {Kasieczka}}, \bibinfo
  {author} {\bibfnamefont {T.}~\bibnamefont {Plehn}},\ and\ \bibinfo {author}
  {\bibfnamefont {R.}~\bibnamefont {Winterhalder}},\ }\href
  {https://doi.org/10.21468/SciPostPhys.8.4.070} {\bibfield  {journal}
  {\bibinfo  {journal} {SciPost Physics}\ }\textbf {\bibinfo {volume} {8}},\
  \bibinfo {pages} {70} (\bibinfo {year} {2020})}\BibitemShut {NoStop}%
\bibitem [{\citenamefont {Butter}\ \emph {et~al.}(2019)\citenamefont {Butter},
  \citenamefont {Plehn},\ and\ \citenamefont {Winterhalder}}]{Butter_2019}%
  \BibitemOpen
  \bibfield  {author} {\bibinfo {author} {\bibfnamefont {A.}~\bibnamefont
  {Butter}}, \bibinfo {author} {\bibfnamefont {T.}~\bibnamefont {Plehn}},\ and\
  \bibinfo {author} {\bibfnamefont {R.}~\bibnamefont {Winterhalder}},\ }\href
  {https://doi.org/10.21468/SciPostPhys.7.6.075} {\bibfield  {journal}
  {\bibinfo  {journal} {SciPost Physics}\ }\textbf {\bibinfo {volume} {7}},\
  \bibinfo {pages} {75} (\bibinfo {year} {2019})}\BibitemShut {NoStop}%
\bibitem [{\citenamefont {Efthymiou}\ \emph
  {et~al.}(2021{\natexlab{a}})\citenamefont {Efthymiou}, \citenamefont
  {Ramos-Calderer}, \citenamefont {Bravo-Prieto}, \citenamefont
  {P{\'e}rez-Salinas}, \citenamefont {Garc{\'\i}a-Mart{\'\i}n}, \citenamefont
  {Garcia-Saez}, \citenamefont {Latorre},\ and\ \citenamefont
  {Carrazza}}]{efthymiou2020qibo}%
  \BibitemOpen
  \bibfield  {author} {\bibinfo {author} {\bibfnamefont {S.}~\bibnamefont
  {Efthymiou}}, \bibinfo {author} {\bibfnamefont {S.}~\bibnamefont
  {Ramos-Calderer}}, \bibinfo {author} {\bibfnamefont {C.}~\bibnamefont
  {Bravo-Prieto}}, \bibinfo {author} {\bibfnamefont {A.}~\bibnamefont
  {P{\'e}rez-Salinas}}, \bibinfo {author} {\bibfnamefont {D.}~\bibnamefont
  {Garc{\'\i}a-Mart{\'\i}n}}, \bibinfo {author} {\bibfnamefont
  {A.}~\bibnamefont {Garcia-Saez}}, \bibinfo {author} {\bibfnamefont {J.~I.}\
  \bibnamefont {Latorre}},\ and\ \bibinfo {author} {\bibfnamefont
  {S.}~\bibnamefont {Carrazza}},\ }\href
  {https://doi.org/10.1088/2058-9565/ac39f5} {\bibfield  {journal} {\bibinfo
  {journal} {Quantum Science and Technology}\ }\textbf {\bibinfo {volume}
  {7}},\ \bibinfo {pages} {015018} (\bibinfo {year}
  {2021}{\natexlab{a}})}\BibitemShut {NoStop}%
\bibitem [{\citenamefont {Efthymiou}\ \emph
  {et~al.}(2021{\natexlab{b}})\citenamefont {Efthymiou}, \citenamefont
  {Carrazza}, \citenamefont {Ramos}, \citenamefont {bpcarlos}, \citenamefont
  {AdrianPerezSalinas}, \citenamefont {García-Martín}, \citenamefont {Paul},
  \citenamefont {Serrano},\ and\ \citenamefont
  {atomicprinter}}]{stavros_efthymiou_2021_5088103}%
  \BibitemOpen
  \bibfield  {author} {\bibinfo {author} {\bibfnamefont {S.}~\bibnamefont
  {Efthymiou}}, \bibinfo {author} {\bibfnamefont {S.}~\bibnamefont {Carrazza}},
  \bibinfo {author} {\bibfnamefont {S.}~\bibnamefont {Ramos}}, \bibinfo
  {author} {\bibnamefont {bpcarlos}}, \bibinfo {author} {\bibnamefont
  {AdrianPerezSalinas}}, \bibinfo {author} {\bibfnamefont {D.}~\bibnamefont
  {García-Martín}}, \bibinfo {author} {\bibnamefont {Paul}}, \bibinfo
  {author} {\bibfnamefont {J.}~\bibnamefont {Serrano}},\ and\ \bibinfo {author}
  {\bibnamefont {atomicprinter}},\ }\href
  {https://doi.org/10.5281/zenodo.5088103} {\bibinfo {title} {qiboteam/qibo:
  Qibo 0.1.6-rc1}} (\bibinfo {year} {2021}{\natexlab{b}})\BibitemShut {NoStop}%
\bibitem [{\citenamefont {Abadi}\ \emph {et~al.}(2015)\citenamefont {Abadi},
  \citenamefont {Agarwal}, \citenamefont {Barham}, \citenamefont {Brevdo},
  \citenamefont {Chen}, \citenamefont {Citro}, \citenamefont {Corrado},
  \citenamefont {Davis}, \citenamefont {Dean}, \citenamefont {Devin} \emph
  {et~al.}}]{tensorflow2015-whitepaper}%
  \BibitemOpen
  \bibfield  {author} {\bibinfo {author} {\bibfnamefont {M.}~\bibnamefont
  {Abadi}}, \bibinfo {author} {\bibfnamefont {A.}~\bibnamefont {Agarwal}},
  \bibinfo {author} {\bibfnamefont {P.}~\bibnamefont {Barham}}, \bibinfo
  {author} {\bibfnamefont {E.}~\bibnamefont {Brevdo}}, \bibinfo {author}
  {\bibfnamefont {Z.}~\bibnamefont {Chen}}, \bibinfo {author} {\bibfnamefont
  {C.}~\bibnamefont {Citro}}, \bibinfo {author} {\bibfnamefont {G.~S.}\
  \bibnamefont {Corrado}}, \bibinfo {author} {\bibfnamefont {A.}~\bibnamefont
  {Davis}}, \bibinfo {author} {\bibfnamefont {J.}~\bibnamefont {Dean}},
  \bibinfo {author} {\bibfnamefont {M.}~\bibnamefont {Devin}}, \emph {et~al.},\
  }\href {https://www.tensorflow.org/} {\bibinfo {title} {{TensorFlow}:
  Large-scale machine learning on heterogeneous systems}} (\bibinfo {year}
  {2015}),\ \bibinfo {note} {software available from
  tensorflow.org}\BibitemShut {NoStop}%
\bibitem [{\citenamefont {afrancis heplat}\ \emph {et~al.}(2021)\citenamefont
  {afrancis heplat}, \citenamefont {Bravo-Prieto}, \citenamefont {Carrazza},
  \citenamefont {Cè}, \citenamefont {Baglio},\ and\ \citenamefont {d-m
  grabowska}}]{afrancis_heplat_2021_5567077}%
  \BibitemOpen
  \bibfield  {author} {\bibinfo {author} {\bibnamefont {afrancis heplat}},
  \bibinfo {author} {\bibfnamefont {C.}~\bibnamefont {Bravo-Prieto}}, \bibinfo
  {author} {\bibfnamefont {S.}~\bibnamefont {Carrazza}}, \bibinfo {author}
  {\bibfnamefont {M.}~\bibnamefont {Cè}}, \bibinfo {author} {\bibfnamefont
  {J.}~\bibnamefont {Baglio}},\ and\ \bibinfo {author} {\bibnamefont {d-m
  grabowska}},\ }\href {https://doi.org/10.5281/zenodo.5567077} {\bibinfo
  {title} {Qti-th/style-qgan: v1.0.0}} (\bibinfo {year} {2021})\BibitemShut
  {NoStop}%
\bibitem [{\citenamefont {Zeiler}(2012)}]{zeiler2012adadelta}%
  \BibitemOpen
  \bibfield  {author} {\bibinfo {author} {\bibfnamefont {M.~D.}\ \bibnamefont
  {Zeiler}},\ }\href {https://doi.org/10.48550/arXiv.1212.5701} {\bibfield
  {journal} {\bibinfo  {journal} {arXiv preprint arXiv:1212.5701}\ } (\bibinfo
  {year} {2012})}\BibitemShut {NoStop}%
\bibitem [{\citenamefont {Ostaszewski}\ \emph {et~al.}(2021)\citenamefont
  {Ostaszewski}, \citenamefont {Grant},\ and\ \citenamefont
  {Benedetti}}]{ostaszewski2021structure}%
  \BibitemOpen
  \bibfield  {author} {\bibinfo {author} {\bibfnamefont {M.}~\bibnamefont
  {Ostaszewski}}, \bibinfo {author} {\bibfnamefont {E.}~\bibnamefont {Grant}},\
  and\ \bibinfo {author} {\bibfnamefont {M.}~\bibnamefont {Benedetti}},\ }\href
  {https://doi.org/10.22331/q-2021-01-28-391} {\bibfield  {journal} {\bibinfo
  {journal} {Quantum}\ }\textbf {\bibinfo {volume} {5}},\ \bibinfo {pages}
  {391} (\bibinfo {year} {2021})}\BibitemShut {NoStop}%
\bibitem [{\citenamefont {Kullback}\ and\ \citenamefont
  {Leibler}(1951)}]{kullback1951information}%
  \BibitemOpen
  \bibfield  {author} {\bibinfo {author} {\bibfnamefont {S.}~\bibnamefont
  {Kullback}}\ and\ \bibinfo {author} {\bibfnamefont {R.~A.}\ \bibnamefont
  {Leibler}},\ }\href {https://doi.org/10.1214/aoms/1177729694} {\bibfield
  {journal} {\bibinfo  {journal} {The Annals of Mathematical Statistics}\
  }\textbf {\bibinfo {volume} {22}},\ \bibinfo {pages} {79} (\bibinfo {year}
  {1951})}\BibitemShut {NoStop}%
\bibitem [{\citenamefont {Frid-Adar}\ \emph {et~al.}(2018)\citenamefont
  {Frid-Adar}, \citenamefont {Klang}, \citenamefont {Amitai}, \citenamefont
  {Goldberger},\ and\ \citenamefont {Greenspan}}]{frid2018synthetic}%
  \BibitemOpen
  \bibfield  {author} {\bibinfo {author} {\bibfnamefont {M.}~\bibnamefont
  {Frid-Adar}}, \bibinfo {author} {\bibfnamefont {E.}~\bibnamefont {Klang}},
  \bibinfo {author} {\bibfnamefont {M.}~\bibnamefont {Amitai}}, \bibinfo
  {author} {\bibfnamefont {J.}~\bibnamefont {Goldberger}},\ and\ \bibinfo
  {author} {\bibfnamefont {H.}~\bibnamefont {Greenspan}},\ }in\ \href
  {https://doi.org/10.1109/ISBI.2018.8363576} {\emph {\bibinfo {booktitle}
  {2018 IEEE 15th International Symposium on Biomedical Imaging (ISBI 2018)}}}\
  (\bibinfo {year} {2018})\ pp.\ \bibinfo {pages} {289--293}\BibitemShut
  {NoStop}%
\bibitem [{\citenamefont {dos Santos~Tanaka}\ and\ \citenamefont
  {Aranha}(2019)}]{tanaka2019data}%
  \BibitemOpen
  \bibfield  {author} {\bibinfo {author} {\bibfnamefont {F.~H.~K.}\
  \bibnamefont {dos Santos~Tanaka}}\ and\ \bibinfo {author} {\bibfnamefont
  {C.}~\bibnamefont {Aranha}},\ }\href
  {https://doi.org/10.48550/arXiv.1904.09135} {\bibfield  {journal} {\bibinfo
  {journal} {arXiv preprint arXiv:1904.09135}\ } (\bibinfo {year}
  {2019})}\BibitemShut {NoStop}%
\bibitem [{\citenamefont {Alwall}\ \emph {et~al.}(2014)\citenamefont {Alwall},
  \citenamefont {Frederix}, \citenamefont {Frixione}, \citenamefont {Hirschi},
  \citenamefont {Maltoni}, \citenamefont {Mattelaer}, \citenamefont {Shao},
  \citenamefont {Stelzer}, \citenamefont {Torrielli},\ and\ \citenamefont
  {Zaro}}]{Alwall:2014hca}%
  \BibitemOpen
  \bibfield  {author} {\bibinfo {author} {\bibfnamefont {J.}~\bibnamefont
  {Alwall}}, \bibinfo {author} {\bibfnamefont {R.}~\bibnamefont {Frederix}},
  \bibinfo {author} {\bibfnamefont {S.}~\bibnamefont {Frixione}}, \bibinfo
  {author} {\bibfnamefont {V.}~\bibnamefont {Hirschi}}, \bibinfo {author}
  {\bibfnamefont {F.}~\bibnamefont {Maltoni}}, \bibinfo {author} {\bibfnamefont
  {O.}~\bibnamefont {Mattelaer}}, \bibinfo {author} {\bibfnamefont {H.~S.}\
  \bibnamefont {Shao}}, \bibinfo {author} {\bibfnamefont {T.}~\bibnamefont
  {Stelzer}}, \bibinfo {author} {\bibfnamefont {P.}~\bibnamefont {Torrielli}},\
  and\ \bibinfo {author} {\bibfnamefont {M.}~\bibnamefont {Zaro}},\ }\href
  {https://doi.org/10.1007/JHEP07(2014)079} {\bibfield  {journal} {\bibinfo
  {journal} {Journal of High Energy Physics}\ }\textbf {\bibinfo {volume}
  {07}},\ \bibinfo {pages} {079} (\bibinfo {year} {2014})}\BibitemShut
  {NoStop}%
\bibitem [{\citenamefont {Frederix}\ \emph {et~al.}(2018)\citenamefont
  {Frederix}, \citenamefont {Frixione}, \citenamefont {Hirschi}, \citenamefont
  {Pagani}, \citenamefont {Shao},\ and\ \citenamefont
  {Zaro}}]{Frederix:2018nkq}%
  \BibitemOpen
  \bibfield  {author} {\bibinfo {author} {\bibfnamefont {R.}~\bibnamefont
  {Frederix}}, \bibinfo {author} {\bibfnamefont {S.}~\bibnamefont {Frixione}},
  \bibinfo {author} {\bibfnamefont {V.}~\bibnamefont {Hirschi}}, \bibinfo
  {author} {\bibfnamefont {D.}~\bibnamefont {Pagani}}, \bibinfo {author}
  {\bibfnamefont {H.~S.}\ \bibnamefont {Shao}},\ and\ \bibinfo {author}
  {\bibfnamefont {M.}~\bibnamefont {Zaro}},\ }\href
  {https://doi.org/10.1007/JHEP07(2018)185} {\bibfield  {journal} {\bibinfo
  {journal} {Journal of High Energy Physics}\ }\textbf {\bibinfo {volume}
  {07}},\ \bibinfo {pages} {185} (\bibinfo {year} {2018})}\BibitemShut
  {NoStop}%
\bibitem [{\citenamefont {Yeo}\ and\ \citenamefont
  {Johnson}(2000)}]{yeo2000new}%
  \BibitemOpen
  \bibfield  {author} {\bibinfo {author} {\bibfnamefont {I.-K.}\ \bibnamefont
  {Yeo}}\ and\ \bibinfo {author} {\bibfnamefont {R.~A.}\ \bibnamefont
  {Johnson}},\ }\href {https://doi.org/10.1093/biomet/87.4.954} {\bibfield
  {journal} {\bibinfo  {journal} {Biometrika}\ }\textbf {\bibinfo {volume}
  {87}},\ \bibinfo {pages} {954} (\bibinfo {year} {2000})}\BibitemShut
  {NoStop}%
\bibitem [{\citenamefont {Pedregosa}\ \emph {et~al.}(2011)\citenamefont
  {Pedregosa}, \citenamefont {Varoquaux}, \citenamefont {Gramfort},
  \citenamefont {Michel}, \citenamefont {Thirion}, \citenamefont {Grisel},
  \citenamefont {Blondel}, \citenamefont {Prettenhofer}, \citenamefont {Weiss},
  \citenamefont {Dubourg}, \citenamefont {Vanderplas}, \citenamefont {Passos},
  \citenamefont {Cournapeau}, \citenamefont {Brucher}, \citenamefont {Perrot},\
  and\ \citenamefont {Duchesnay}}]{scikit-learn}%
  \BibitemOpen
  \bibfield  {author} {\bibinfo {author} {\bibfnamefont {F.}~\bibnamefont
  {Pedregosa}}, \bibinfo {author} {\bibfnamefont {G.}~\bibnamefont
  {Varoquaux}}, \bibinfo {author} {\bibfnamefont {A.}~\bibnamefont {Gramfort}},
  \bibinfo {author} {\bibfnamefont {V.}~\bibnamefont {Michel}}, \bibinfo
  {author} {\bibfnamefont {B.}~\bibnamefont {Thirion}}, \bibinfo {author}
  {\bibfnamefont {O.}~\bibnamefont {Grisel}}, \bibinfo {author} {\bibfnamefont
  {M.}~\bibnamefont {Blondel}}, \bibinfo {author} {\bibfnamefont
  {P.}~\bibnamefont {Prettenhofer}}, \bibinfo {author} {\bibfnamefont
  {R.}~\bibnamefont {Weiss}}, \bibinfo {author} {\bibfnamefont
  {V.}~\bibnamefont {Dubourg}}, \bibinfo {author} {\bibfnamefont
  {J.}~\bibnamefont {Vanderplas}}, \bibinfo {author} {\bibfnamefont
  {A.}~\bibnamefont {Passos}}, \bibinfo {author} {\bibfnamefont
  {D.}~\bibnamefont {Cournapeau}}, \bibinfo {author} {\bibfnamefont
  {M.}~\bibnamefont {Brucher}}, \bibinfo {author} {\bibfnamefont
  {M.}~\bibnamefont {Perrot}},\ and\ \bibinfo {author} {\bibfnamefont
  {E.}~\bibnamefont {Duchesnay}},\ }\href
  {https://dl.acm.org/doi/10.5555/1953048.2078195} {\bibfield  {journal}
  {\bibinfo  {journal} {Journal of Machine Learning Research}\ }\textbf
  {\bibinfo {volume} {12}},\ \bibinfo {pages} {2825–2830} (\bibinfo {year}
  {2011})}\BibitemShut {NoStop}%
\bibitem [{\citenamefont {Aleksandrowicz}\ \emph {et~al.}(2019)\citenamefont
  {Aleksandrowicz}, \citenamefont {Alexander}, \citenamefont {Barkoutsos},
  \citenamefont {Bello}, \citenamefont {Ben-Haim}, \citenamefont {Bucher},
  \citenamefont {Cabrera-Hernández}, \citenamefont {Carballo-Franquis},
  \citenamefont {Chen}, \citenamefont {Chen} \emph
  {et~al.}}]{gadi_aleksandrowicz_2019_2562111}%
  \BibitemOpen
  \bibfield  {author} {\bibinfo {author} {\bibfnamefont {G.}~\bibnamefont
  {Aleksandrowicz}}, \bibinfo {author} {\bibfnamefont {T.}~\bibnamefont
  {Alexander}}, \bibinfo {author} {\bibfnamefont {P.}~\bibnamefont
  {Barkoutsos}}, \bibinfo {author} {\bibfnamefont {L.}~\bibnamefont {Bello}},
  \bibinfo {author} {\bibfnamefont {Y.}~\bibnamefont {Ben-Haim}}, \bibinfo
  {author} {\bibfnamefont {D.}~\bibnamefont {Bucher}}, \bibinfo {author}
  {\bibfnamefont {F.~J.}\ \bibnamefont {Cabrera-Hernández}}, \bibinfo {author}
  {\bibfnamefont {J.}~\bibnamefont {Carballo-Franquis}}, \bibinfo {author}
  {\bibfnamefont {A.}~\bibnamefont {Chen}}, \bibinfo {author} {\bibfnamefont
  {C.-F.}\ \bibnamefont {Chen}}, \emph {et~al.},\ }\href
  {https://doi.org/10.5281/zenodo.2562111} {\bibinfo {title} {{Qiskit: An
  Open-source Framework for Quantum Computing}}} (\bibinfo {year}
  {2019})\BibitemShut {NoStop}%
\end{thebibliography}%

\vspace*{0.5cm}

\appendix

\section{Noise simulation on IBM~Q device}
\label{sec:appendixnoise}

We have performed a noise simulation of our style-qGAN on an IBM~Q device, taking as a device baseline the {\tt ibmq\_santiago}
5-qubit Falcon r4L quantum processor that we have used for our runs on real IBM~Q hardware.

The noise model takes into account the readout error probability of each qubit (mean value of the probability of reading $|1\rangle$ while being in the state $|0\rangle$, and the probability of reading $|0\rangle$ while being in the state $|1\rangle$), the relaxation time constants of each qubit (relaxation time and dephasing time), the gate error probability of each basis gate, and the gate length (timing of the gate) of each basis gate. The values are taken from the calibration information of the selected device for the noise simulation. Note that this calibration is performed at regular intervals. The generation of $10^5$ samples on the actual machine took about one week, implying that the calibration parameters may have varied significantly during the full run.

\begin{figure*}

  \hspace{1.5em}\includegraphics[width=0.295\textwidth]{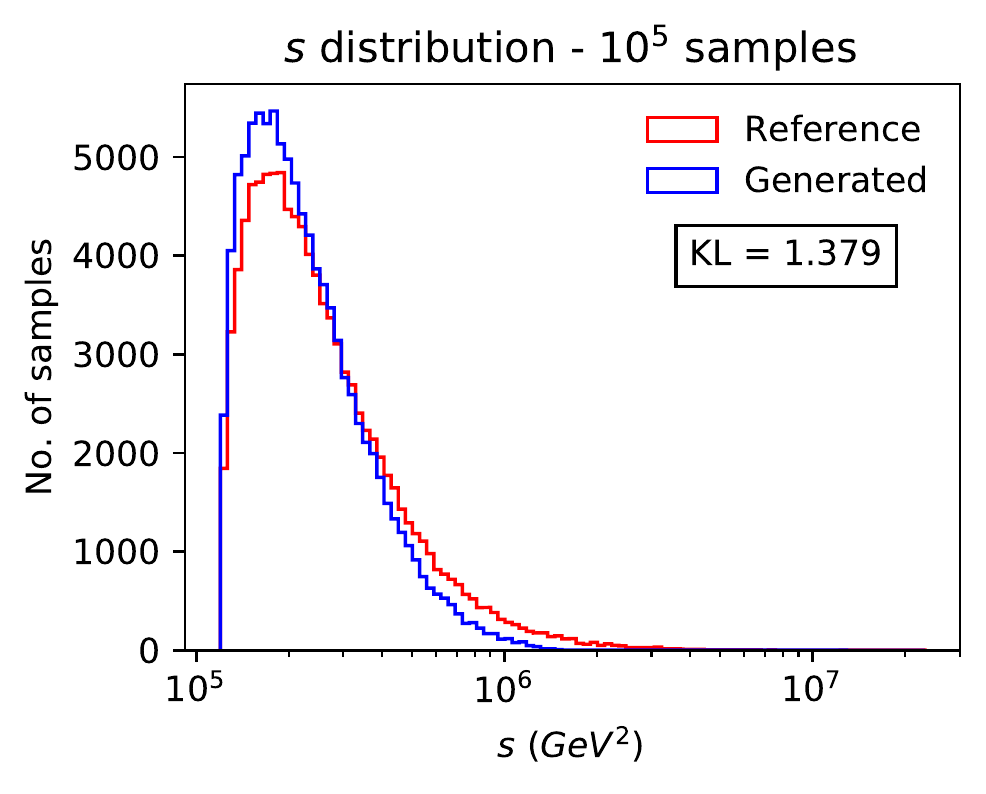}%
  \includegraphics[width=0.295\textwidth]{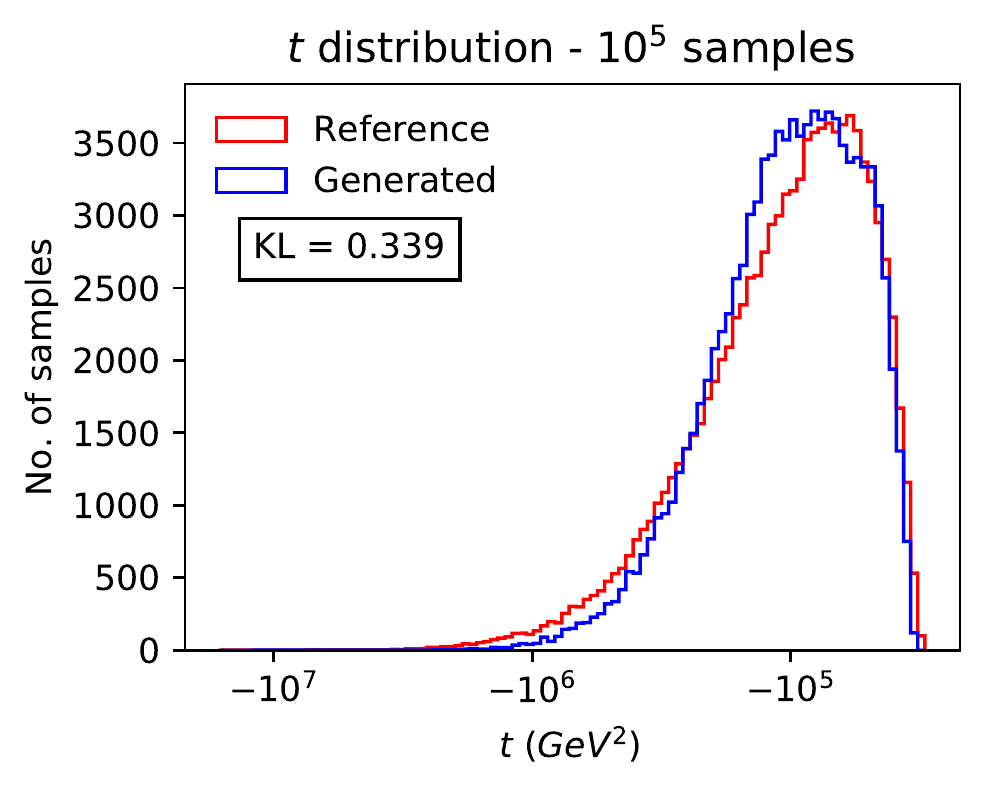}%
  \includegraphics[width=0.295\textwidth]{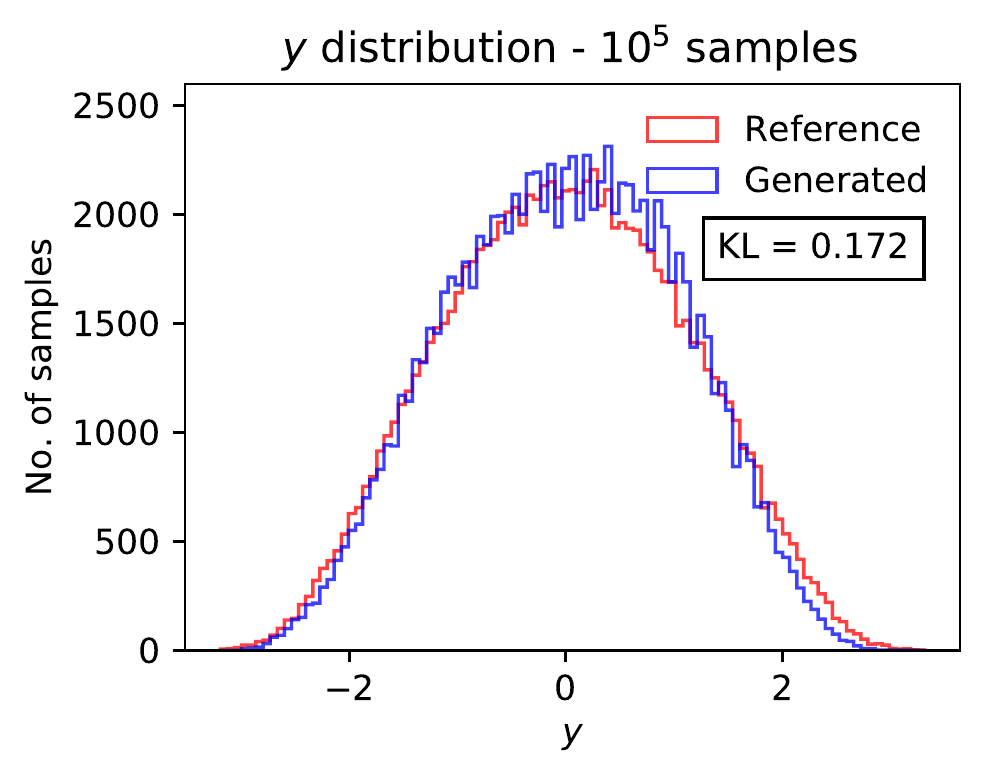}

  \hspace{0.7em}\includegraphics[width=0.328\textwidth]{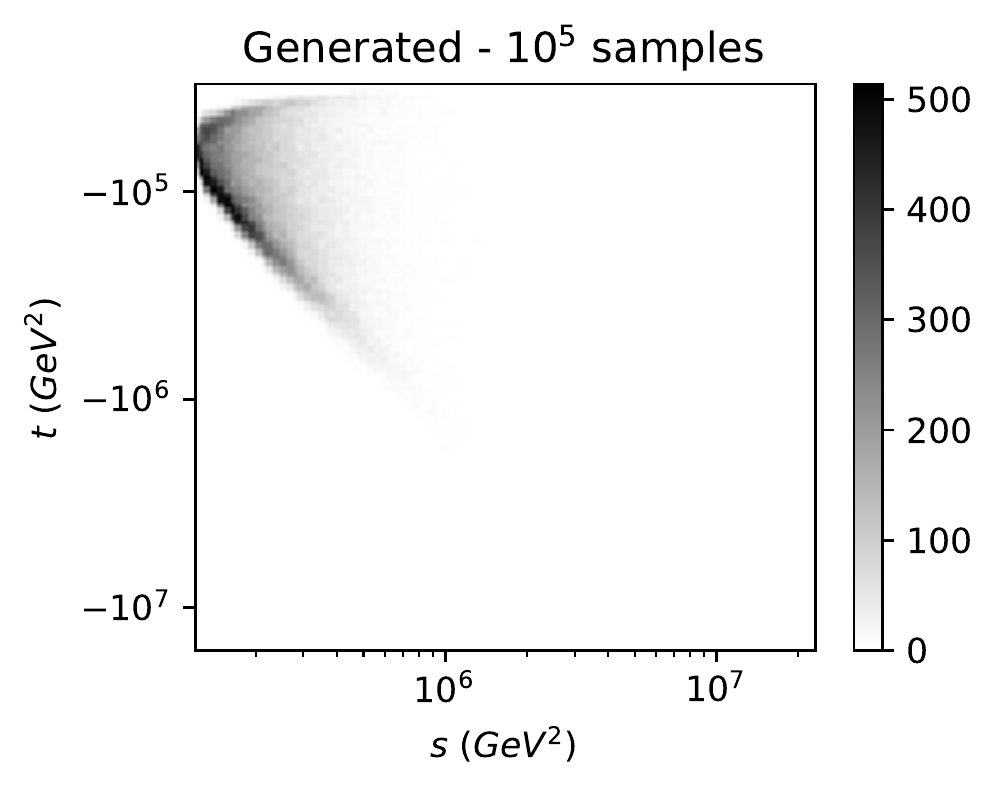}%
  \includegraphics[width=0.303\textwidth]{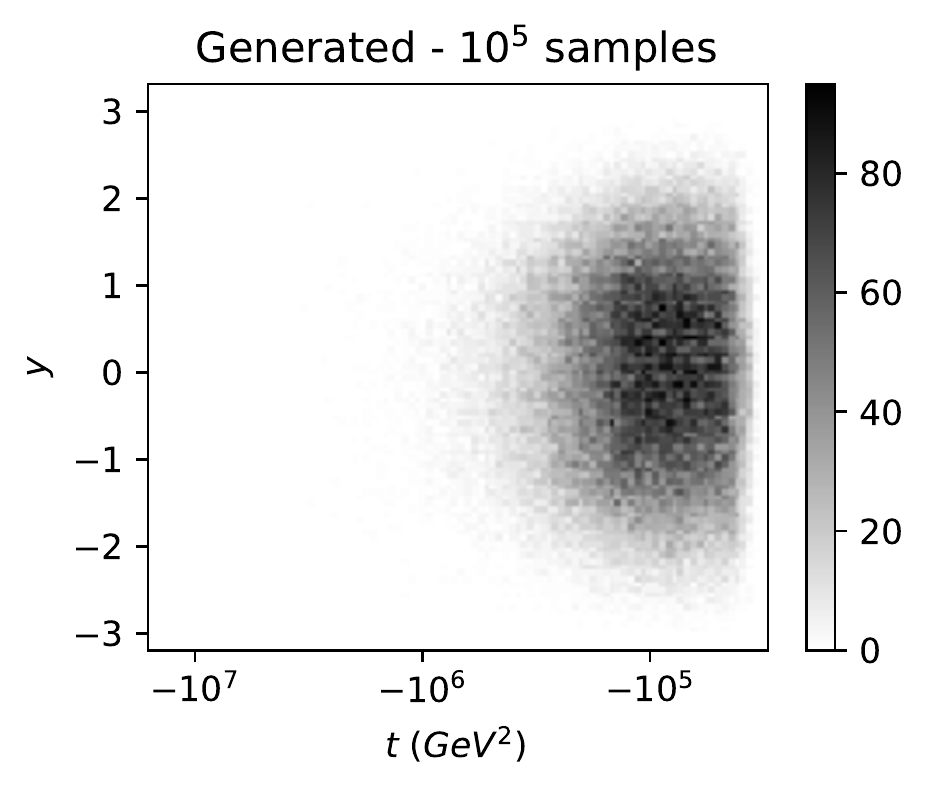}%
  \includegraphics[width=0.318\textwidth]{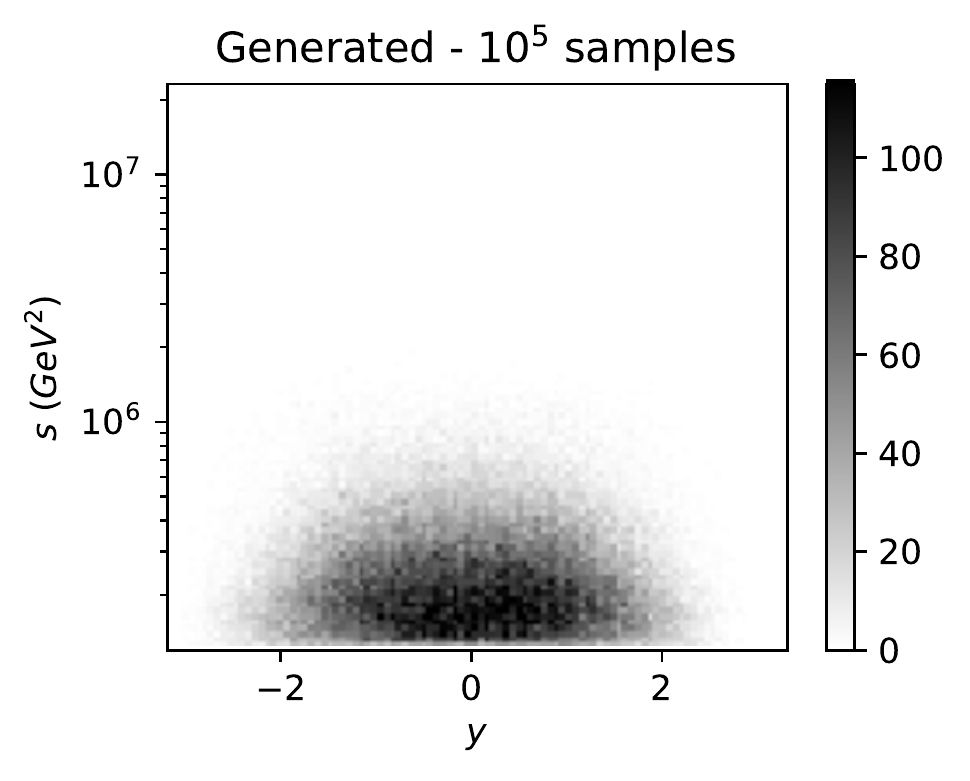}

  \hspace{0.8em}\includegraphics[width=0.315\textwidth]{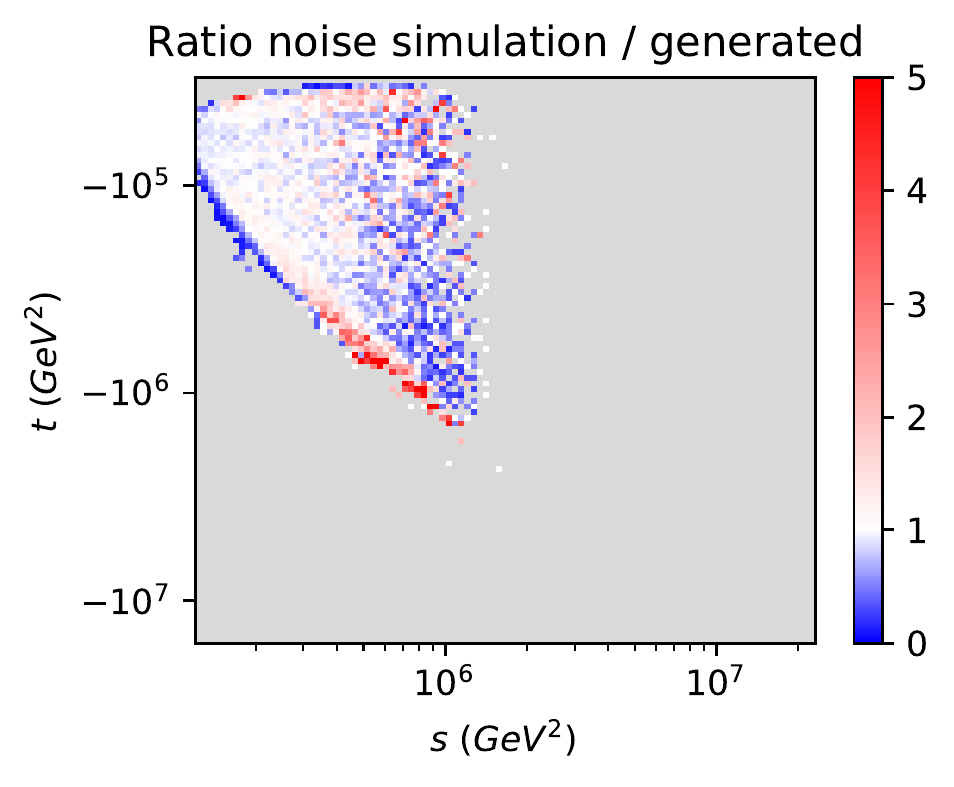}%
  \includegraphics[width=0.299\textwidth]{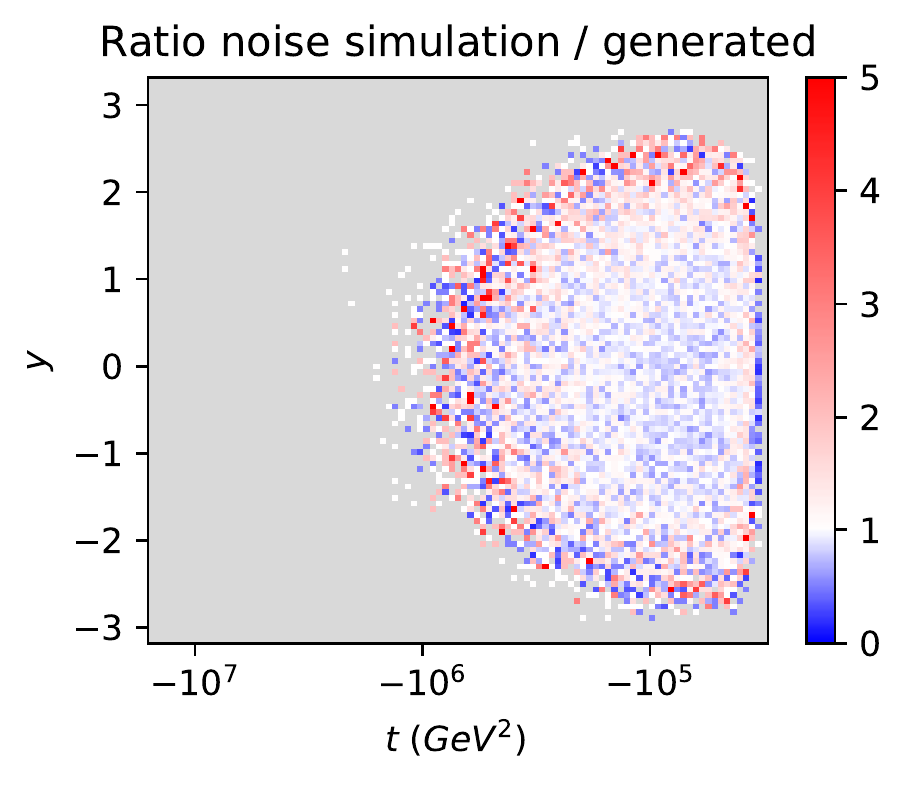}%
  \includegraphics[width=0.310\textwidth]{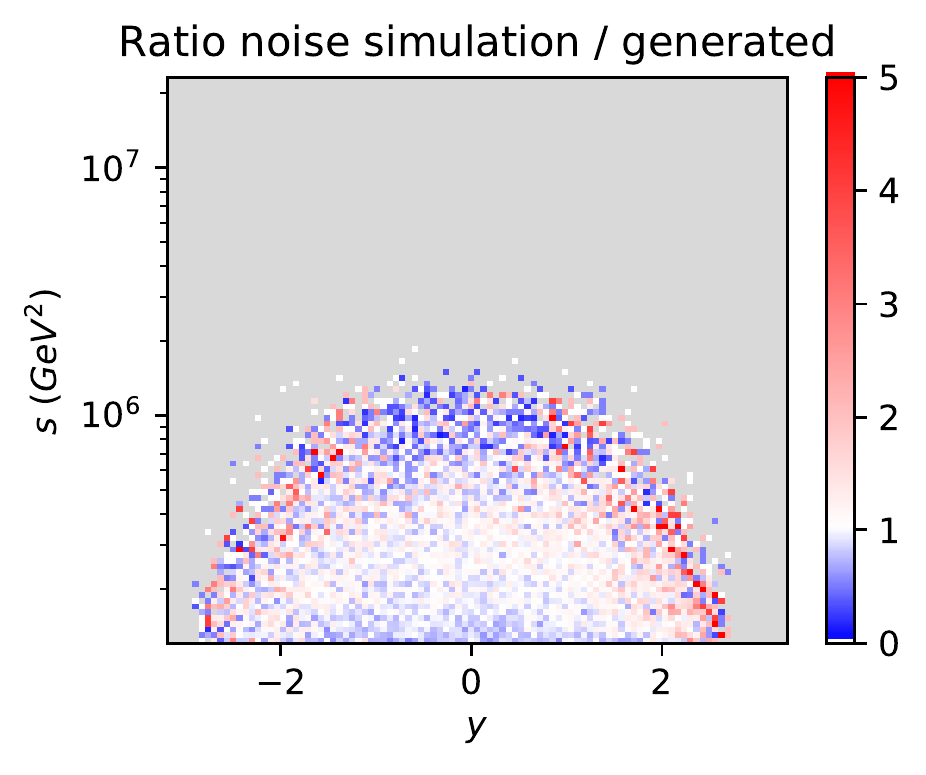}

  \caption{\label{fig:ibmnoise}
    Marginal samples distributions for the physical observables $s,t,y$
    in $pp\rightarrow t\bar{t}$ production at the LHC using the style-qGAN generator model in
    a noise simulation of {\tt ibmq\_santiago} device (top row), trained with $10^4$ samples (top row),
    together with the corresponding two-dimensional sampling projections (middle row) and the ratio to the reference underlying
    prior MC distribution (bottom row). Note that we choose a grey background for the plots at the bottom
    row to highlight when the reference and generated samples are identical.}
\end{figure*}

We show in Figure~\ref{fig:ibmnoise} the result of our noise simulation. The KL distances displayed in the top row are comparable to the KL distances reported in Figure~\ref{fig:ibm}. We also compare our noise simulation to the run on actual IBM~Q hardware, the latter being reported in Section~\ref{sec:deployment}; this is shown in the bottom row. The plots show a significant amount of white points, signaling that the noise simulation seems to capture most of the errors induced by running on actual quantum hardware and that errors beyond the parameters reported in the previous paragraph are subdominant.

\begin{figure*}

  \hspace{0.9em}\includegraphics[width=0.322\textwidth]{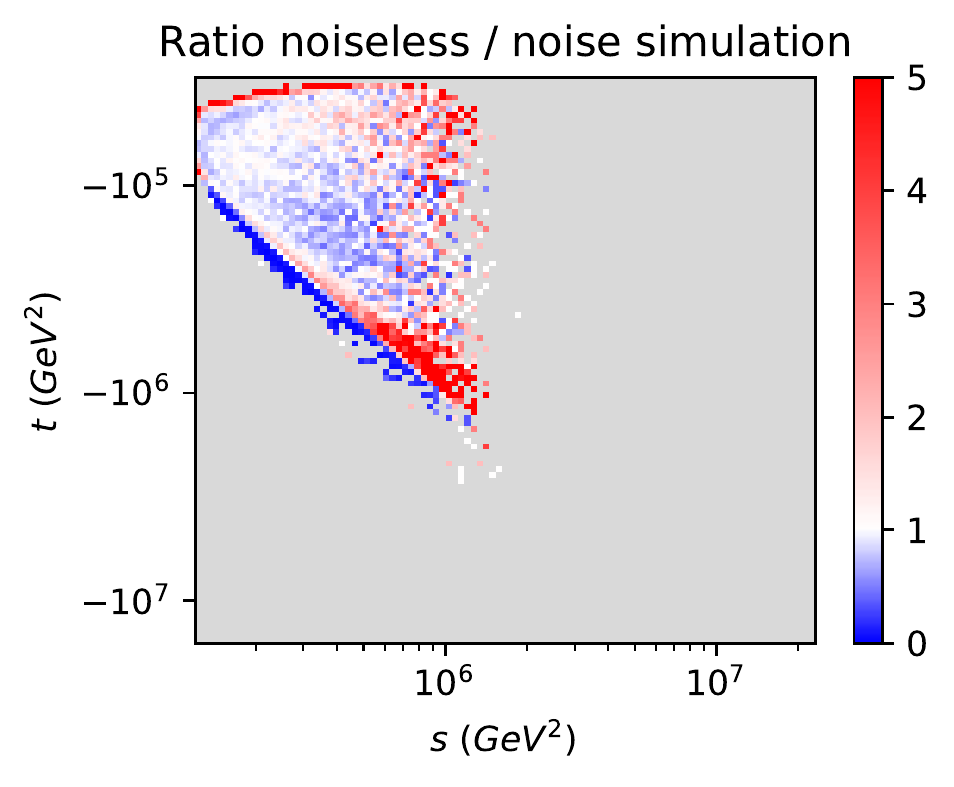}%
  \includegraphics[width=0.305\textwidth]{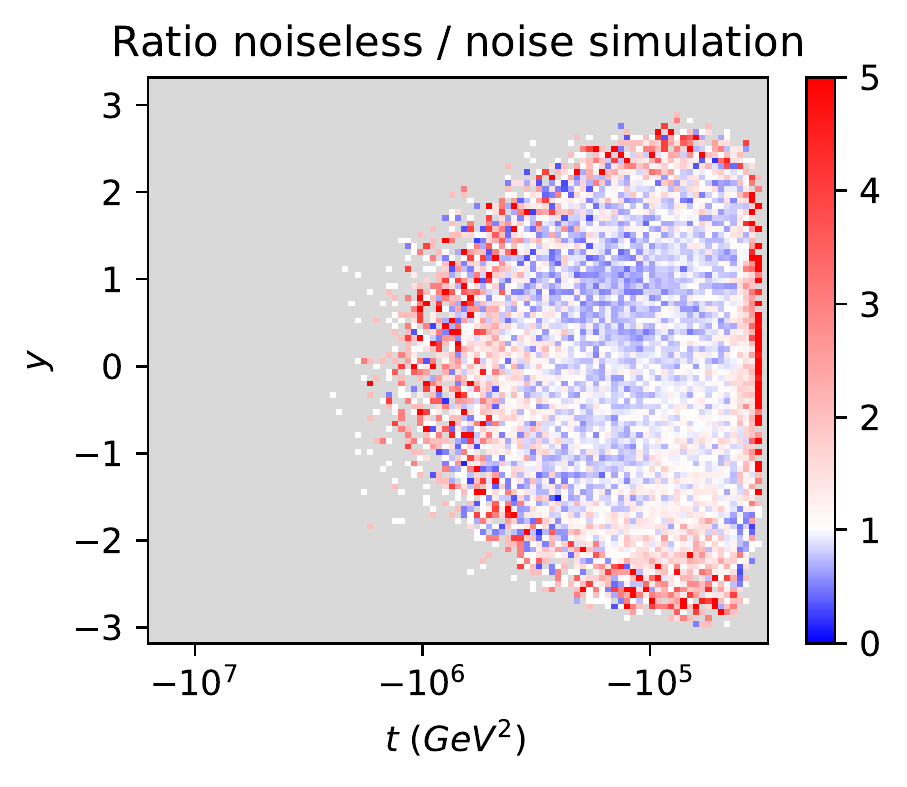}%
  \includegraphics[width=0.315\textwidth]{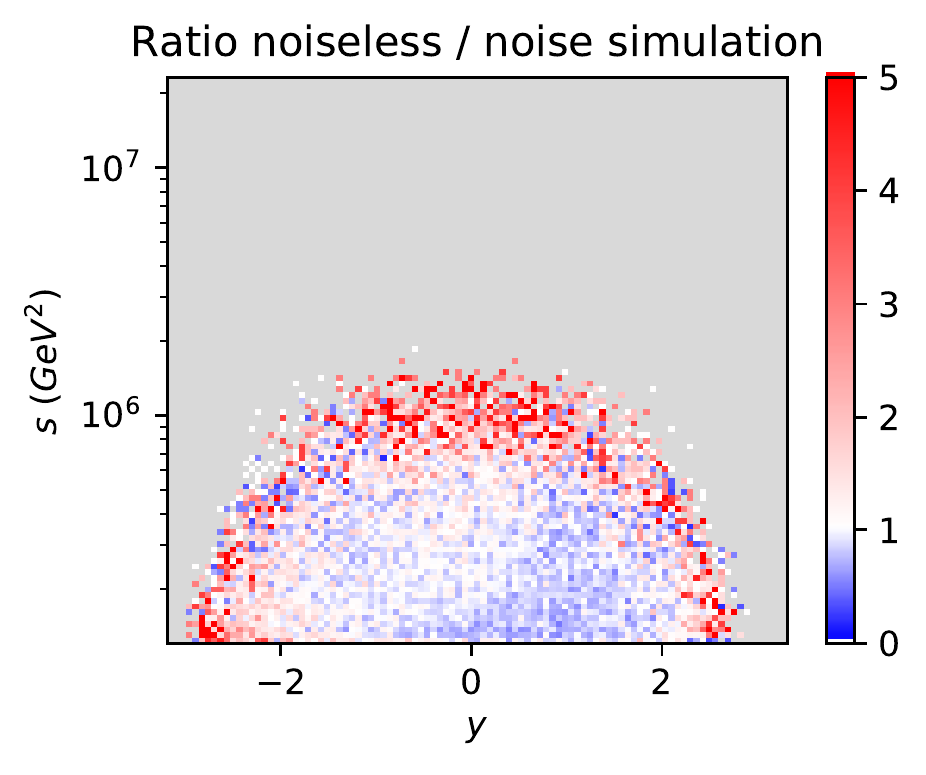}

  \caption{\label{fig:ibmnoise2}Ratio of two-dimensional sampling projections using the noise simulation of {\tt ibmq\_santiago} device to the
    corresponding noiseless simulation.}
\end{figure*}

We also compare our noise simulation to the noiseless simulation.
The results shown in Figure~\ref{fig:ibmnoise2} indicate that while the noise has an impact, as expected, there are
still many points close to the ratio of one; therefore the style-qGAN still performs fairly well in a noisy environment. This has led us
to believe that running on actual quantum hardware gives good results.

\end{document}